\documentclass[12pt,preprint]{aastex}

\newcommand\hms{h~~m~~s}
\newcommand\dms{$~~~^{\circ}~~~^{\prime}~~~^{{\prime}{\prime}}$}

\received{}
\accepted{}
\journalid{}{}
\articleid{}{}

\slugcomment{Astronomical Journal, submitted}

\shortauthors{Landolt}
\shorttitle{Southern $UBVRI$ Standard Stars}

\begin{document}

\title{$UBVRI$ PHOTOMETRIC STANDARD STARS AROUND \\ THE SKY AT $-50$ DEGREES DECLINATION}

\author{Arlo U.~Landolt\altaffilmark{1}}
\affil{Department of Physics \& Astronomy, Louisiana State University, Baton Rouge, LA 70803-4001}
\email{landolt@phys.lsu.edu}

\altaffiltext{1}{Visiting Astronomer, Cerro Tololo Inter-American Observatory, National Optical 
Astronomical Observatories, which are operated by the Association of Universities for Research in 
Astronomy, under contract with the National Science Foundation.}

\begin{abstract}
\label{sec:abstract}

$UBVRI$ photoelectric observations have been made of 109 stars around the sky, and centered more or 
less at $-50$ degrees declination.  The majority of the stars fall in the magnitude range 
$10.4<V<15.5$, and in the color index range $-0.33<(B-V)<+1.66$.  These new broad-band photometric 
standard stars average 16.4 measures each from data taken on 116 different nights over a period of 
four years.  Similar data are tabulated for 19 stars of interest, but which were not observed often 
enough to make them well-defined standard stars.

\end{abstract}

\keywords{stars: standard --- photometry: broad-band --- photometry: standardization}

\section{Introduction} 
\label{sec:introduction}

Accurate and readily accessible standard star sequences are necessary for the calibration of 
intensity and color data obtained for objects projected against the celestial sphere.  Toward this 
end, this author has published sequences around the sky centered on the celestial equator 
\citep{Landolt1973, Landolt1983, Landolt1992}.

Far southern declination photometric sequences, including Cousins early work, as summarized by 
\citet{Menzies1980}, were published by \citet{Graham1982}, \citet{Menzies1989}, and 
\cite{Kilkenny1998}.  A more complete history may be found in \citet{Landolt2007}.  $UBVRI$ 
photometry for a number of spectrophotometric standard stars has appeared in 
\citet{LandoltUomoto2007}.

\section{Observations}
\label{sec:observations}

The CTIO 1.5-m telescope together with a GaAs photomultiplier was assigned on 168 nights for the 
program in the time interval 1998-2001.  Useful data were obtained on 116 whole or partial different 
nights, wherein 55\% of the possible observing hours were photometric.

These broad-band $UBVRI$ photometric observations all were obtained with the same RCA 31034A-02 
photomultiplier mounted in KPNO cold box \#53.  The photomultiplier was operated at $-1600\,V$. 
The sensitivity function of the photomultiplier was unavailable.  Data for the same brand 
photomultiplier has been tabulated in \citet{Landolt1992}, Appendix B, Table 11, Figures 52-54.

The filter set was CTIO's $UBVRI$ filter set \#3.  Information describing the composition of that 
filter set may be found in \citet[][Table III]{Landolt1983}.  The transmission characteristics of 
those filters are tabulated in \citet[][Appendix B]{Landolt1992}.

Between 20 and 25 $UBVRI$ standard stars, as defined by \citet{Landolt1992}, were observed each night 
together with the program stars.  A night's observations began and ended with a group of four or five 
standard stars.  Similar groups were observed periodically throughout the night.  Each of these 
groups contained stars closely spaced on the sky, and possessing as wide a color range as possible.  
A more complete outline of the author's observing philosophy has been given in \citet{Landolt2007}.

A complete data set for a star consisted of a series of measures: $VBURIIRUBV$.  A 
$14.0^{\prime\prime}$ diaphragm was used throughout the observing program.  The integration or 
counting time depended upon the faintness of a particular star.  The counting time never was less 
than ten seconds per filter, and was as long as $60\,s$ for the faintest stars.  Data reduction 
procedures followed the precepts outlined by \citet{SchulteCrawford1961} and by \citet{Landolt2007}.

Extinction coefficients were calculated from three or four standard stars possessing a range in color 
index that were followed from near the meridian over to an air mass of 2.1, or so.  Each night's data 
were reduced using the primary extinction coefficients derived from that night, whenever possible.  
Average secondary extinction coefficients for a given run were used.  Examples of the range in 
extinction coefficients which an observer in fact encounters have been tabulated in 
\citet{Landolt2007}.  Such tabulations should remind any observer of the perils in using mean 
extinction coefficients.

The final computer printout for each night's reductions contained the magnitude and color indices for 
each of the standard stars.  Since the time of observation was recorded for each measurement, it was 
possible to plot the residuals in the $V$ magnitude and in the different color indices for each 
standard star against Universal Time for a given night.  These plots permitted small corrections to 
be made to all program star measures.  The corrections usually were less than a few hundredths of a 
magnitude.  Such corrections took into account small changes in both atmospheric and instrumental 
conditions that occurred during the course of a night's observations.

\section{Discussion}
\label{sec:discussion}

A total of 128 stars, distributed around the sky, and more or less centered at $-50$ degrees 
declination, comprised this program. The data were reduced night by night.  The results were tied 
into the $UBVRI$ photometric system defined by \citet{Landolt1992}.

A check on the accuracy of the magnitude and color index transformations was made via a comparison of 
the magnitudes and color indices of the stars from \citet{Landolt1992} that were used as standard 
stars in this paper, with the magnitudes and color indices of these same standard stars obtained 
during this current program.  The comparisons, the delta quantities, were in the sense of data from 
this program {\it minus} the corresponding magnitudes and color indices from \citet{Landolt1992}.

\begin{figure} 
\plotone{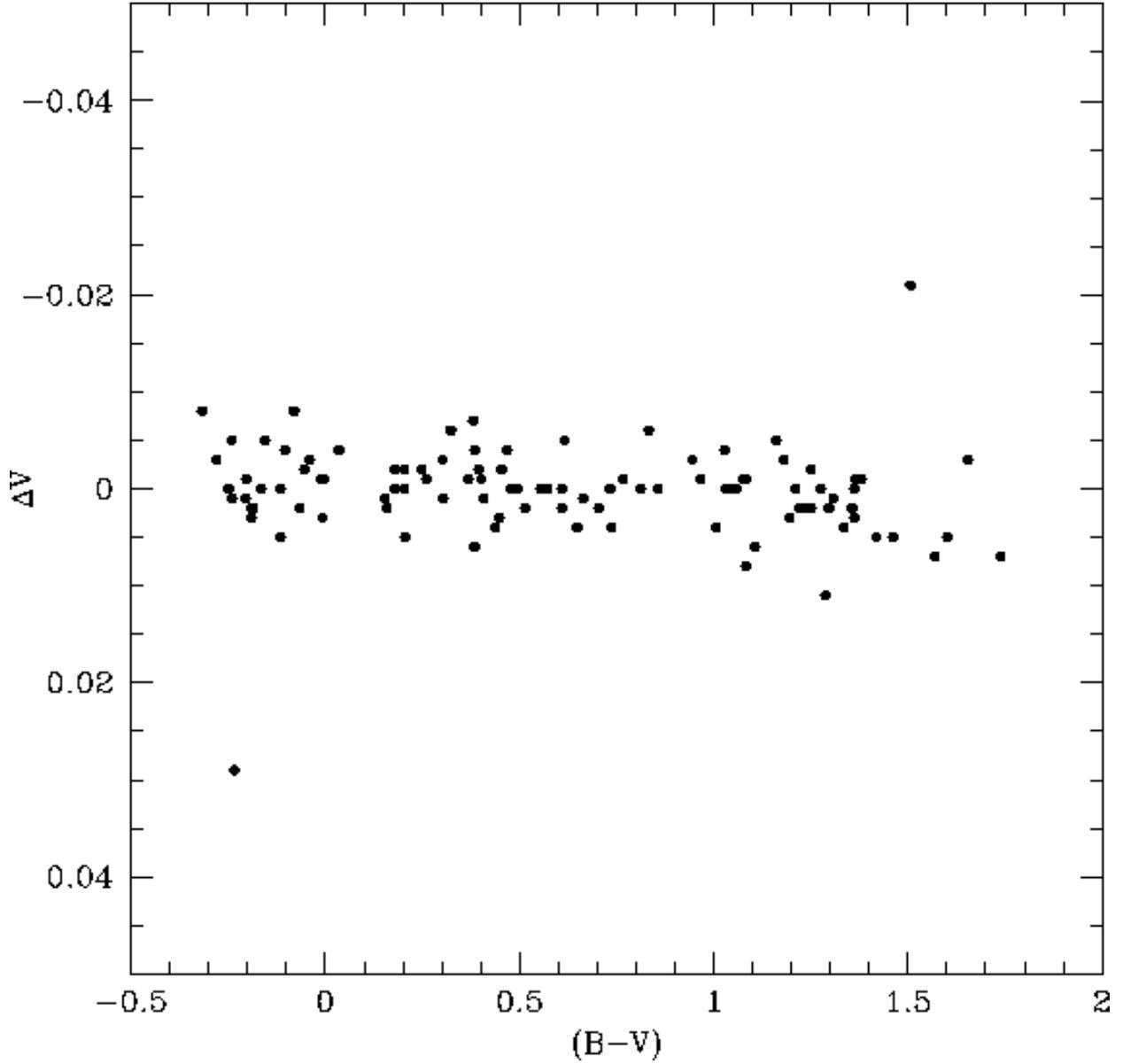} 
\caption{Comparison of the $V$ magnitudes tied into \citet{Landolt1992} standard stars as a function 
of the \citet{Landolt1992} equatorial standard star's $(B-V)$ color indices.  The two discrepant 
points are the white dwarf Feige~108 and the M-dwarf G163-51 (see text).}
\label{fig:figure1}
\end{figure}

The $V$ magnitude for the standard star Feige~108 shows an appreciable deviation in Figure 
\ref{fig:figure1} from the past.  A suspected variable star number, NSV~26050, has been 
assigned based on a comment in \citet{Bergeron1984}.

The $V$ magnitude for the standard star G~163-51 shows an approximate 0.02~mag difference from 
the past.  No suspected variable star number has been assigned, insofar as is known.

\begin{figure} 
\plotone{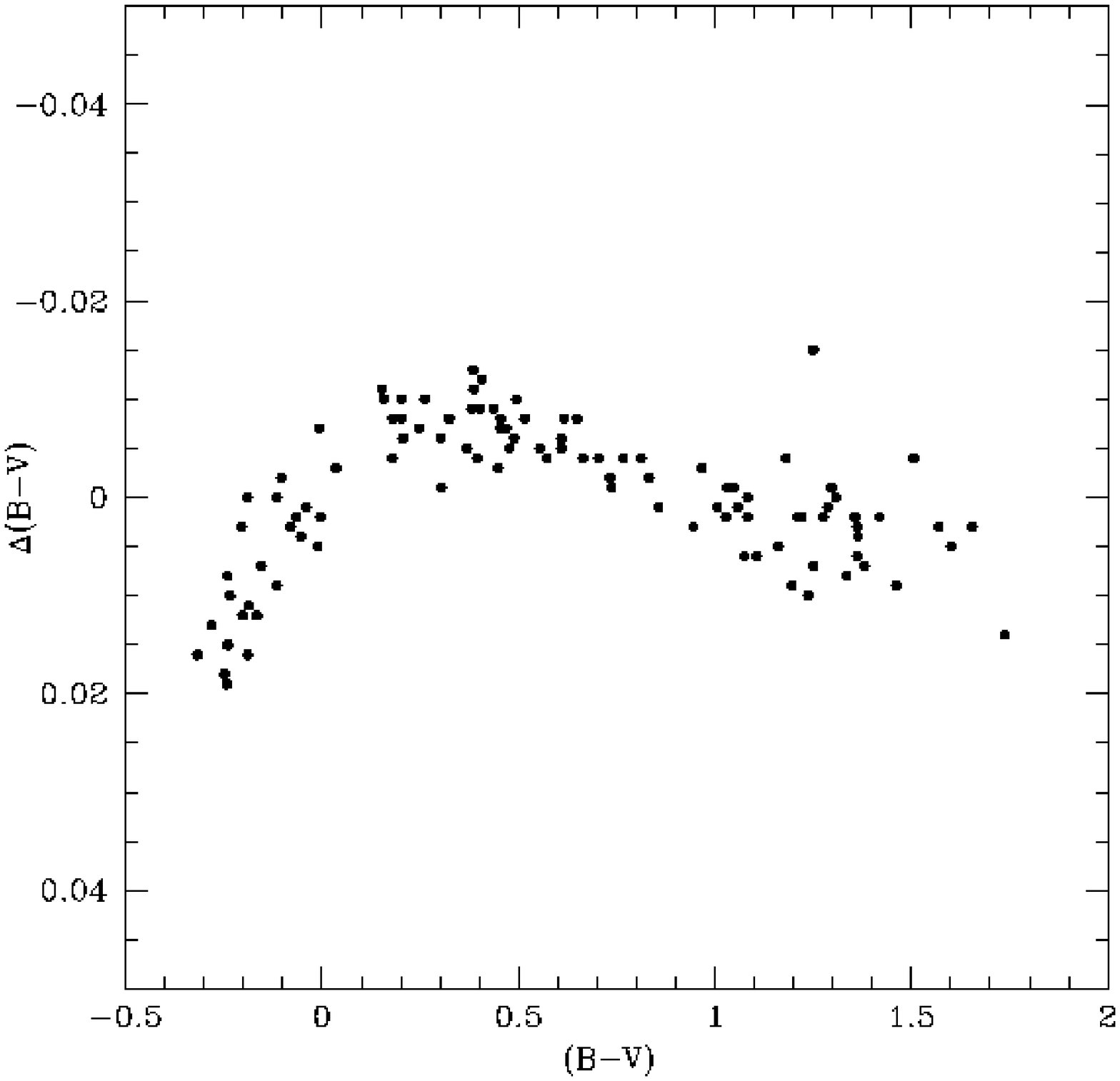} 
\caption{Comparison of the $(B-V)$ magnitudes tied into \citet{Landolt1992} standard stars as a function 
of the \citet{Landolt1992} equatorial standard star's $(B-V)$ color indices.}
\label{fig:figure2}
\end{figure}

\begin{figure} 
\plotone{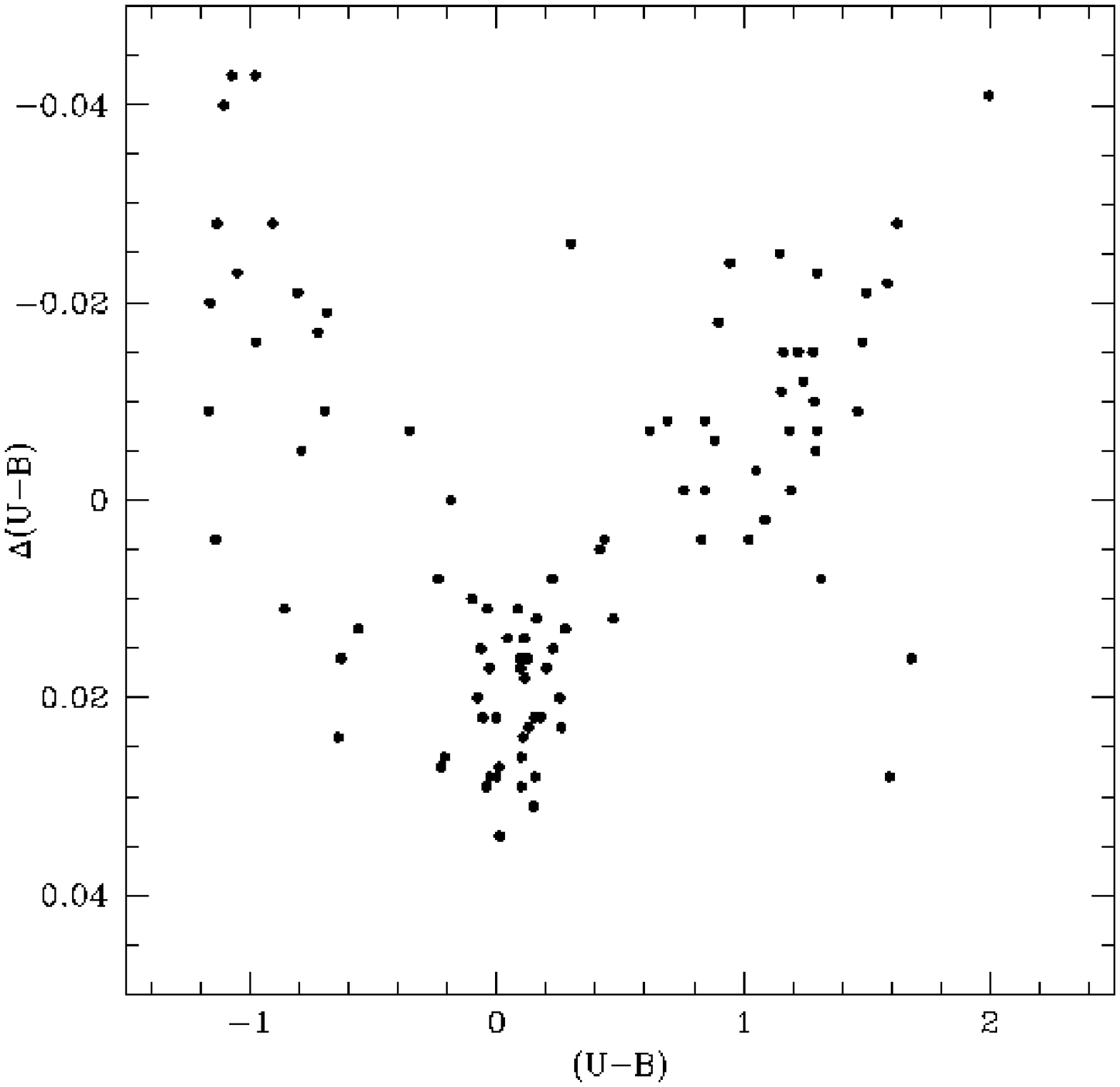} 
\caption{Comparison of the $(U-B)$ magnitudes tied into \citet{Landolt1992} standard stars as a function 
of the \citet{Landolt1992} equatorial standard star's $(U-B)$ color indices.}
\label{fig:figure3}
\end{figure}

\begin{figure} 
\plotone{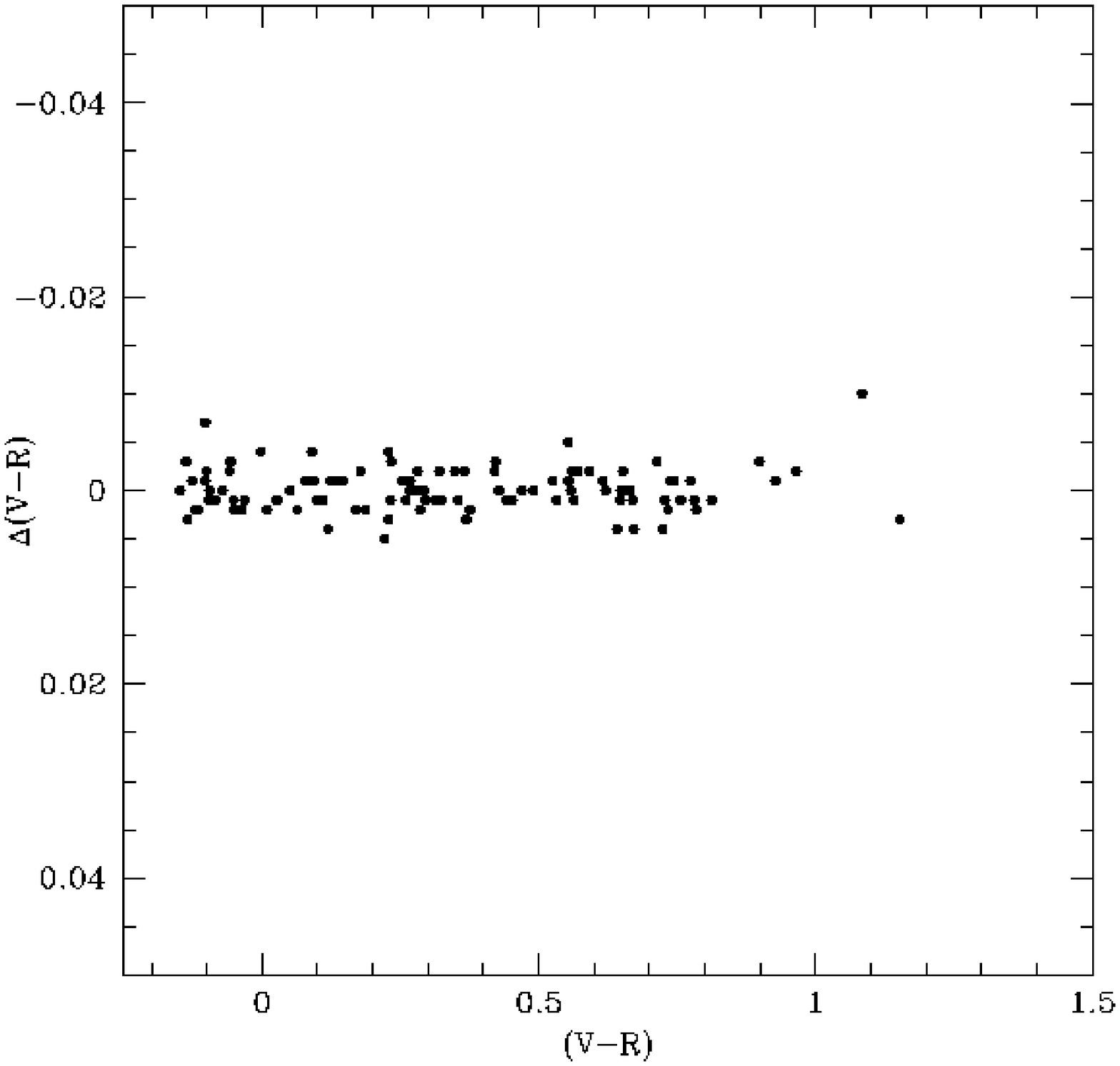} 
\caption{Comparison of the $(V-R)$ magnitudes tied into \citet{Landolt1992} standard stars as a function 
of the \citet{Landolt1992} equatorial standard star's $(V-R)$ color indices.}
\label{fig:figure4}
\end{figure}

\begin{figure} 
\plotone{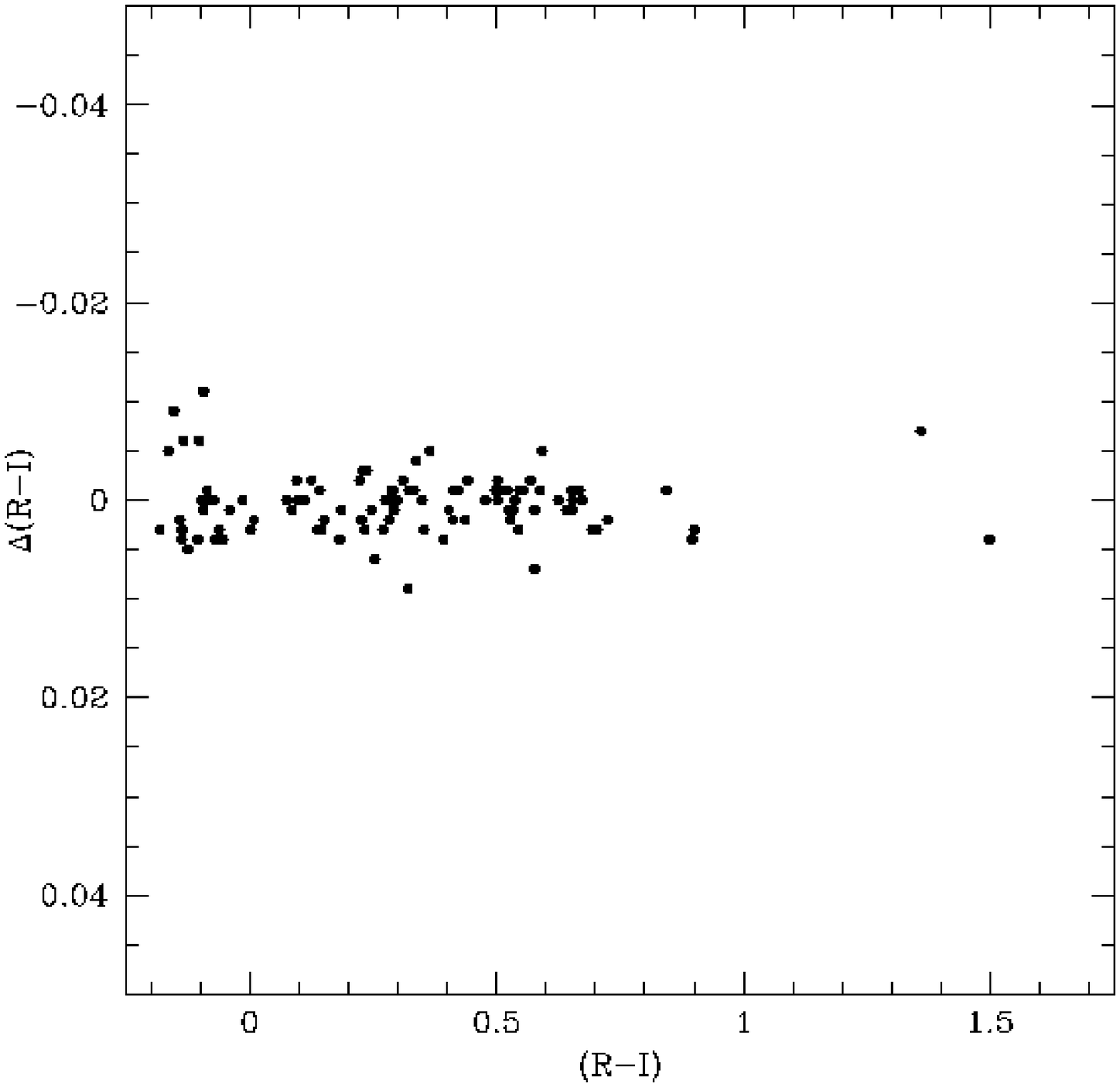} 
\caption{Comparison of the $(R-I)$ magnitudes tied into \citet{Landolt1992} standard stars as a function 
of the \citet{Landolt1992} equatorial standard star's $(R-I)$ color indices.}
\label{fig:figure5}
\end{figure}

\begin{figure} 
\plotone{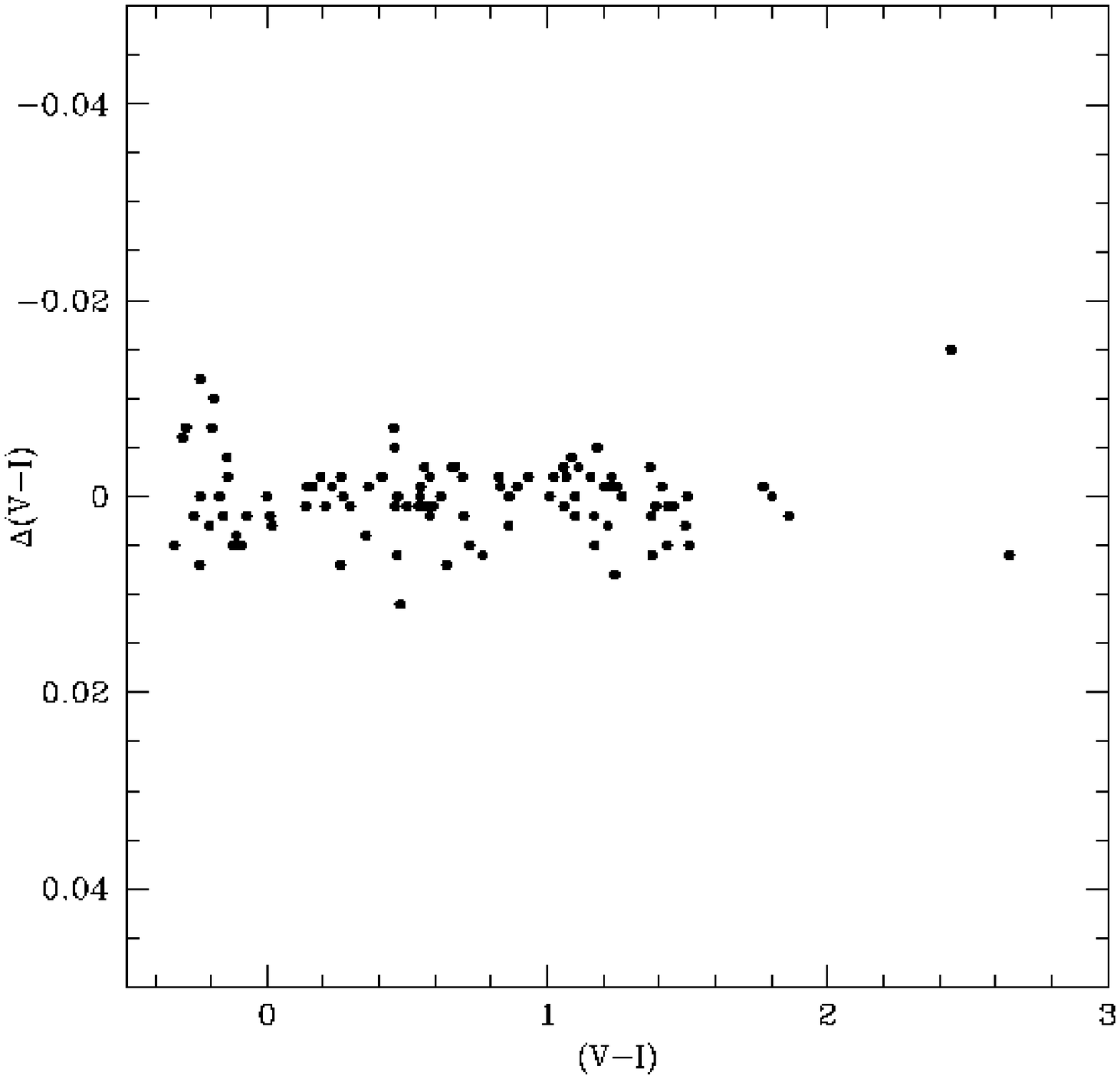} 
\caption{Comparison of the $(V-I)$ magnitudes tied into \citet{Landolt1992} standard stars as a function 
of the \citet{Landolt1992} equatorial standard star's $(V-I)$ color indices.}
\label{fig:figure6}
\end{figure}

Figures \ref{fig:figure1} - \ref{fig:figure6} illustrate the plots of the delta quantities on the 
ordinates versus the color indices on the abscissas.  Nonlinearities are apparent in the figures.  
Inspection of each figure allowed the nonlinear ``break points" to be chosen.  They are indicated 
below in association with the appropriate nonlinear transformation relation, which were derived by 
least-squares fitting from the data appearing in Figures \ref{fig:figure1} - \ref{fig:figure6}.

The non-linear transformation relations had the form, where a subscript ``c" indicates ``catalog" 
and subscript ``obs" indicates ``observed", as follows:

\begin{eqnarray}
(B-V)_{c}&=&+0.00127 + 1.06192(B-V)_{obs}~~~~~~~~~~~~~~~~~~~~~~~~(B-V)<+0.1, \nonumber \\
& & \pm 0.00154 \pm 0.00880 \nonumber \\
& & \nonumber \\
(B-V)_{c}&=&+0.01091 + 0.99029(B-V)_{obs}~~~~~~~~~~~~~~~~~~~+0.1<(B-V)<+1.0, \nonumber \\
& & \pm	0.00106 \pm 0.00202 \nonumber \\
& & \nonumber \\
(B-V)_{c}&=&+0.00781 + 0.99172(B-V)_{obs}~~~~~~~~~~~~~~~~~~~~~~~~(B-V)>+1.0, \nonumber \\
& & \pm 0.00624 \pm 0.00485 \nonumber \\
& & \nonumber \\
(U-B)_{c}&=&-0.03057 + 0.94734(U-B)_{obs}~~~~~~~~~~~~~~~~~~~~~~~~(U-B)<-0.2, \nonumber \\
& & \pm 0.00923 \pm 0.01053 \nonumber \\
& & \nonumber \\
(U-B)_{c}&=&-0.02134 + 1.02212(U-B)_{obs}~~~~~~~~~~~~~~~~~~~-0.2<(U-B)<+0.5, \nonumber \\
& & \pm 0.00157 \pm 0.00948 \nonumber \\
& & \nonumber \\
(U-B)_{c}&=&-0.01422 + 1.02218(U-B)_{obs}~~~~~~~~~~~~~~~~~~~~~~~~(U-B)>+0.5, \nonumber \\
& & \pm 0.01110 \pm 0.00868 \nonumber \\
& & \nonumber \\
V_{c}&=&+0.00068 - 0.00536(B-V)_{c} + V_{obs}~~~~~~~~~~~~~~~~~~~~(B-V)<+0.1, \nonumber \\
& & \pm 0.00129 \pm 0.00750 \nonumber \\
& & \nonumber \\
V_{c}&=&+0.00002 + 0.00123(B-V)_{c} + V_{obs}~~~~~~~~~~~~~~~+0.1<(B-V)<+1.0, \nonumber \\
& & \pm 0.00114 \pm 0.00216 \nonumber \\
& & \nonumber \\
V_{c}&=&-0.00244 + 0.00106(B-V)_{c} + V_{obs}~~~~~~~~~~~~~~~~~~~~(B-V)>+1.0, \nonumber \\
& & \pm 0.00633 \pm 0.00515 \nonumber \\
& & \nonumber \\
(V-R)_{c}&=&+0.00030 + 1.00034(V-R)_{obs}~~~~~~~~~~~~~~~~~~~~~~~~(V-R)<+0.1, \nonumber \\
& & \pm 0.00052 \pm 0.00588 \nonumber \\
& & \nonumber \\
(V-R)_{c}&=&-0.00052 + 1.00150(V-R)_{obs}~~~~~~~~~~~~~~~~~~~+0.1<(V-R)<+0.5, \nonumber \\
& & \pm 0.00101 \pm 0.00334 \nonumber\\
& & \nonumber \\
(V-R)_{c}&=&-0.00387 + 1.00617(V-R)_{obs}~~~~~~~~~~~~~~~~~~~~~~~~(V-R)>+0.5, \nonumber \\
& & \pm 0.00237 \pm 0.00335 \nonumber \\
& & \nonumber \\
(R-I)_{c}&=&-0.00129 + 0.99863(R-I)_{obs}~~~~~~~~~~~~~~~~~~~~~~~~(R-I)<+0.1, \nonumber \\
& & \pm 0.00098 \pm 0.00963 \nonumber \\
& & \nonumber \\
(R-I)_{c}&=&-0.00118 + 1.00367(R-I)_{obs}~~~~~~~~~~~~~~~~~~~+0.1<(R-I)<+0.5, \nonumber \\
& & \pm 0.00114 \pm 0.00369 \nonumber \\
& & \nonumber \\
(R-I)_{c}&=&-0.00254 + 1.00343(R-I)_{obs}~~~~~~~~~~~~~~~~~~~~~~~~(R-I)>+0.5, \nonumber \\
& & \pm 0.00185 \pm 0.00277 \nonumber \\
& & \nonumber \\
(V-I)_{c}&=&-0.00215 + 0.98744(V-I)_{obs}~~~~~~~~~~~~~~~~~~~~~~~~(V-I)<+0.1, \nonumber \\
& & \pm 0.00289 \pm 0.01413 \nonumber \\
& & \nonumber \\
(V-I)_{c}&=&+0.00014 + 0.99962(V-I)_{obs}~~~~~~~~~~~~~~~~~~~+0.1<(V-I)<+1.0, \nonumber \\
& & \pm 0.00123 \pm 0.00219 \nonumber \\
& & \nonumber \\
(V-I)_{c}&=&-0.00587 + 1.00431(V-I)_{obs}~~~~~~~~~~~~~~~~~~~~~~~~(V-I)>+1.0. \nonumber \\
& & \pm 0.00310 \pm 0.00227 \nonumber \\
& & \nonumber \\
\nonumber
\end{eqnarray}

\begin{figure}
\plotone{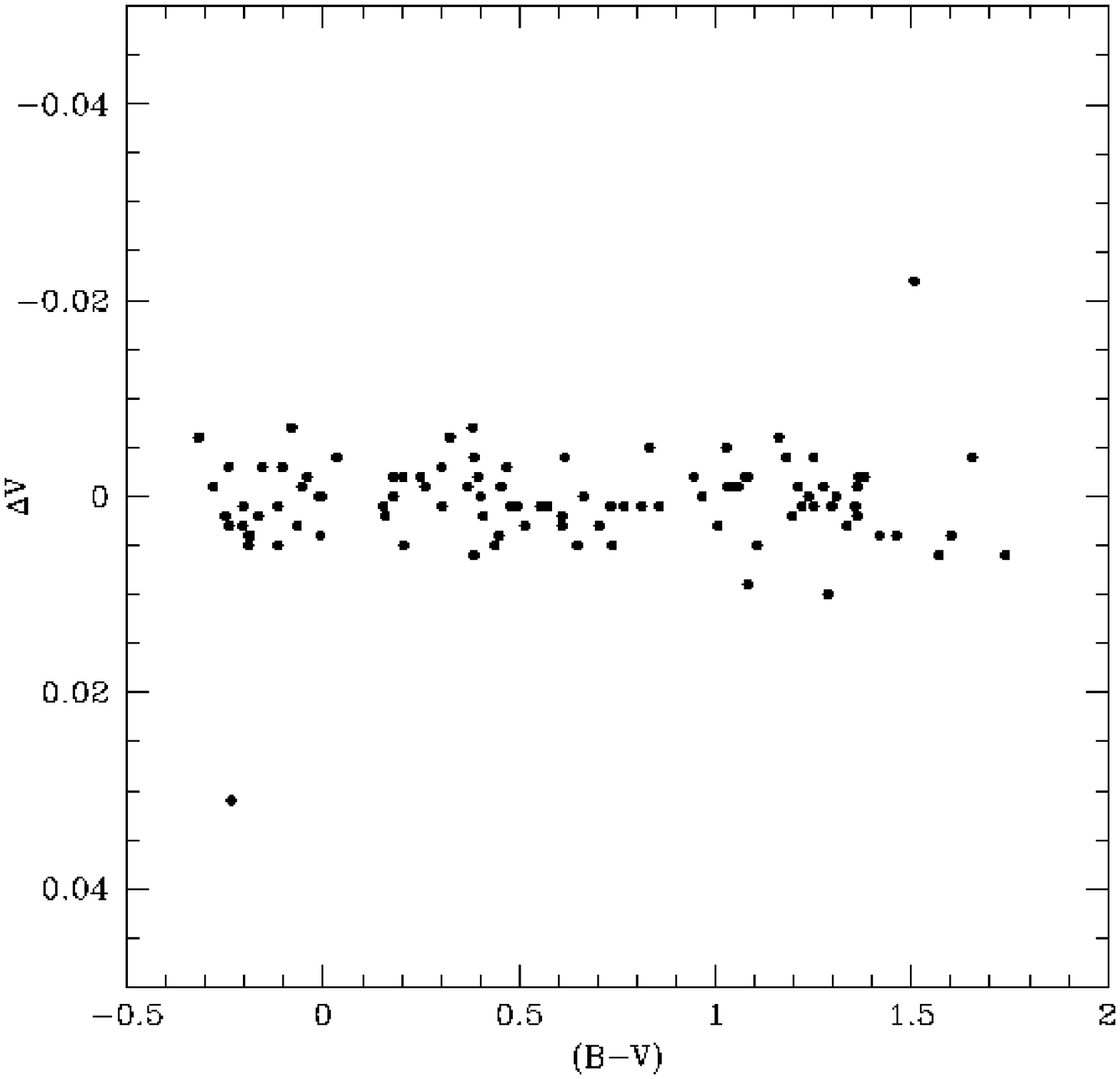}
\caption{Comparison of the $V$ magnitudes after removal of the transformation nonlinearity as a 
function of the \citet{Landolt1992} equatorial standard star's $(B-V)$ color index.}
\label{fig:figure7}
\end{figure}

\begin{figure}
\plotone{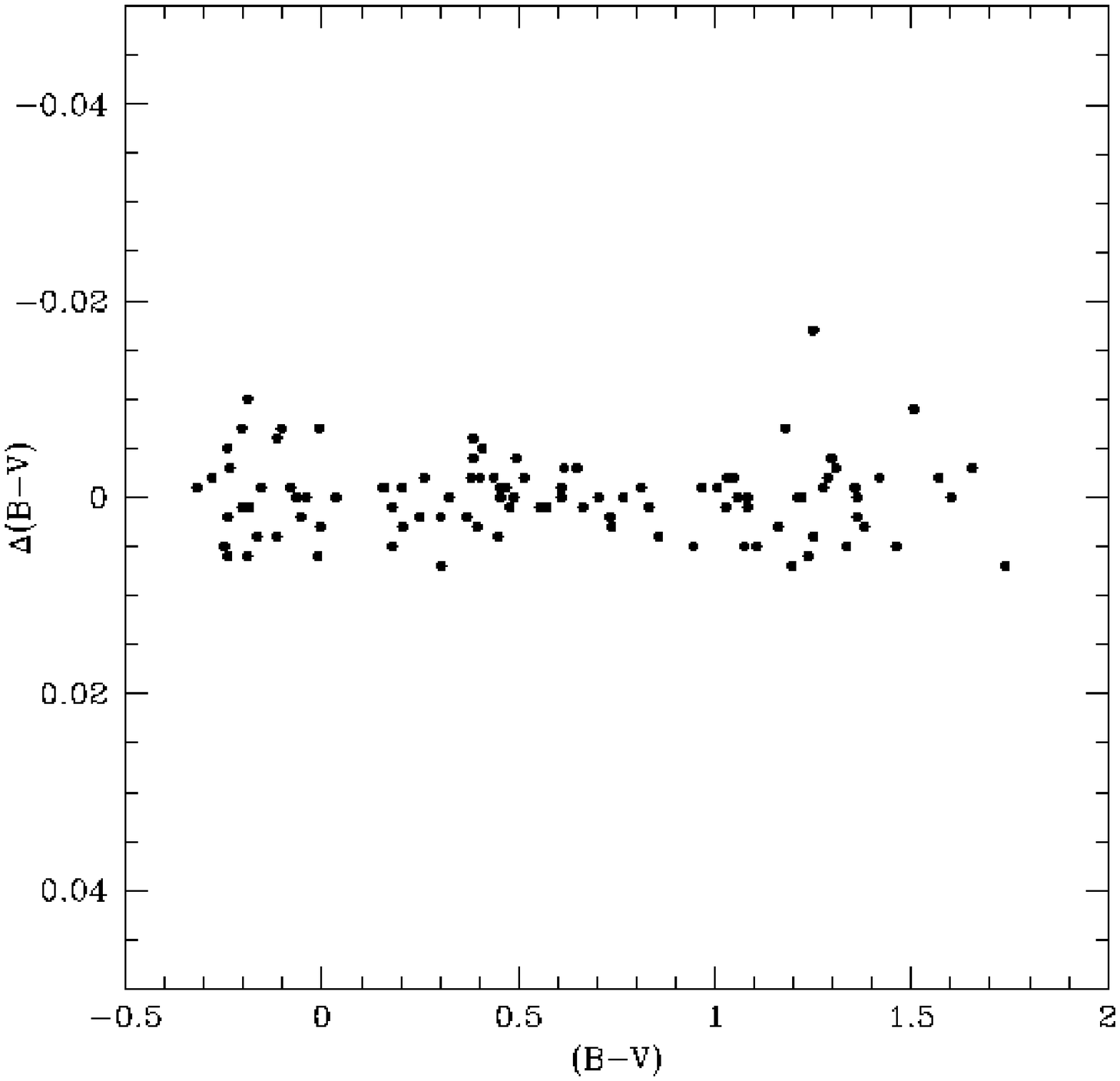}
\caption{Comparison of the $(B-V)$ magnitudes after removal of the transformation nonlinearity as a 
function of the \citet{Landolt1992} equatorial standard star's $(B-V)$ color index.}
\label{fig:figure8}
\end{figure}

\begin{figure}
\plotone{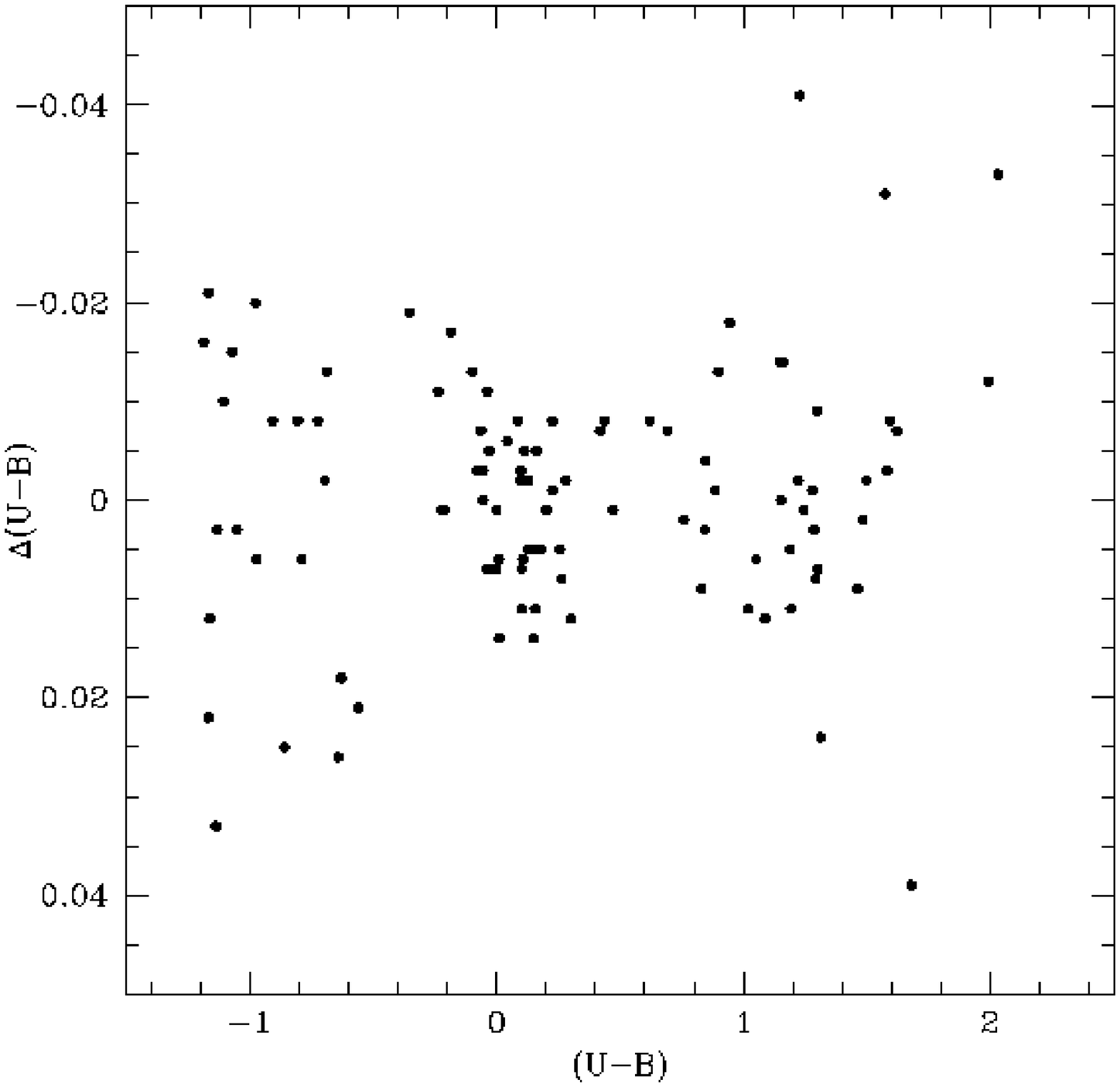}
\caption{Comparison of the $(U-B)$ magnitudes after removal of the transformation nonlinearity as a 
function of the \citet{Landolt1992} equatorial standard star's $(U-B)$ color index.}
\label{fig:figure9}
\end{figure}

\begin{figure}
\plotone{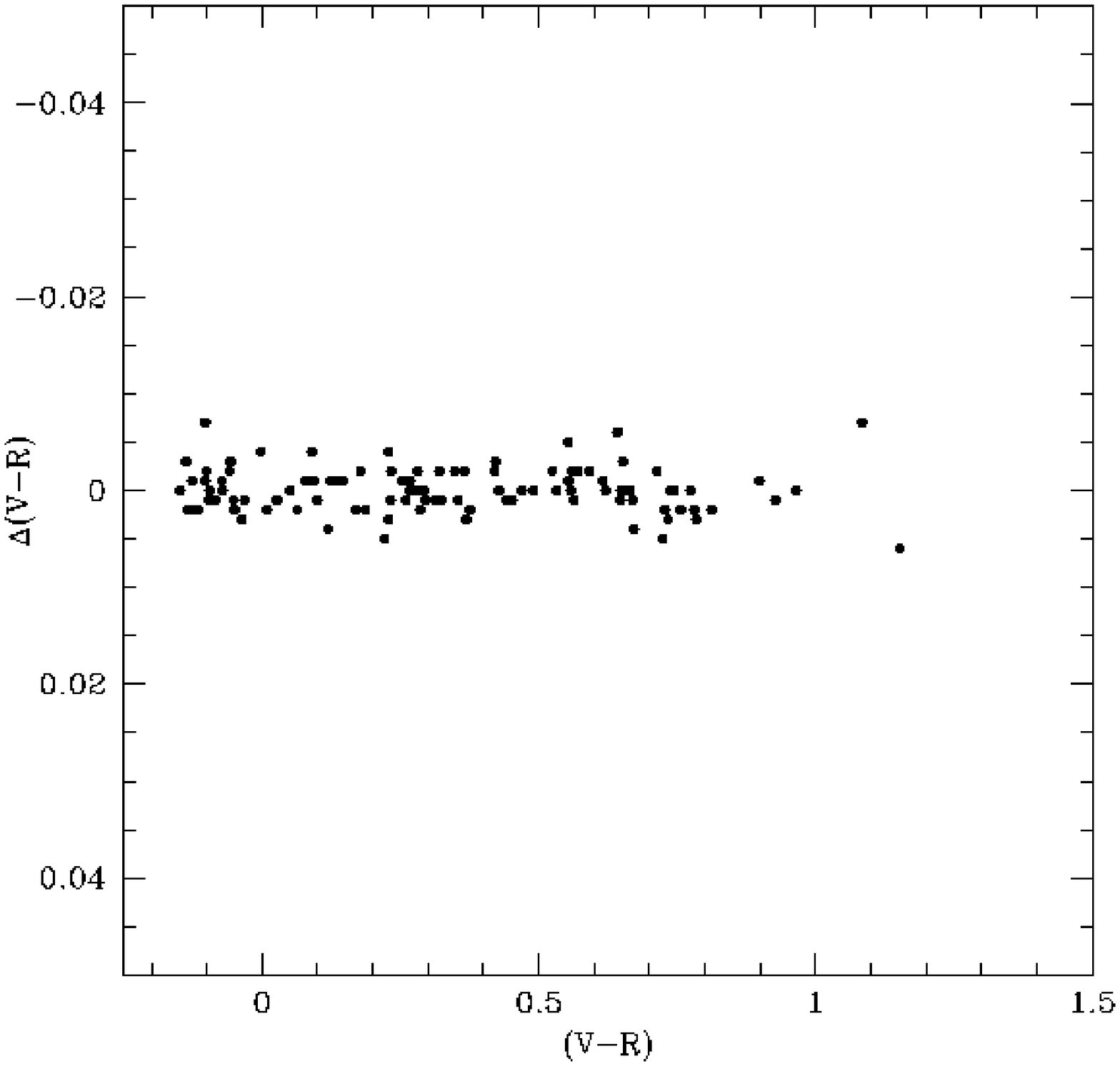}
\caption{Comparison of the $(V-R)$ magnitudes after removal of the transformation nonlinearity as a 
function of the \citet{Landolt1992} equatorial standard star's $(V-R)$ color index.}
\label{fig:figure10}
\end{figure}

\begin{figure}
\plotone{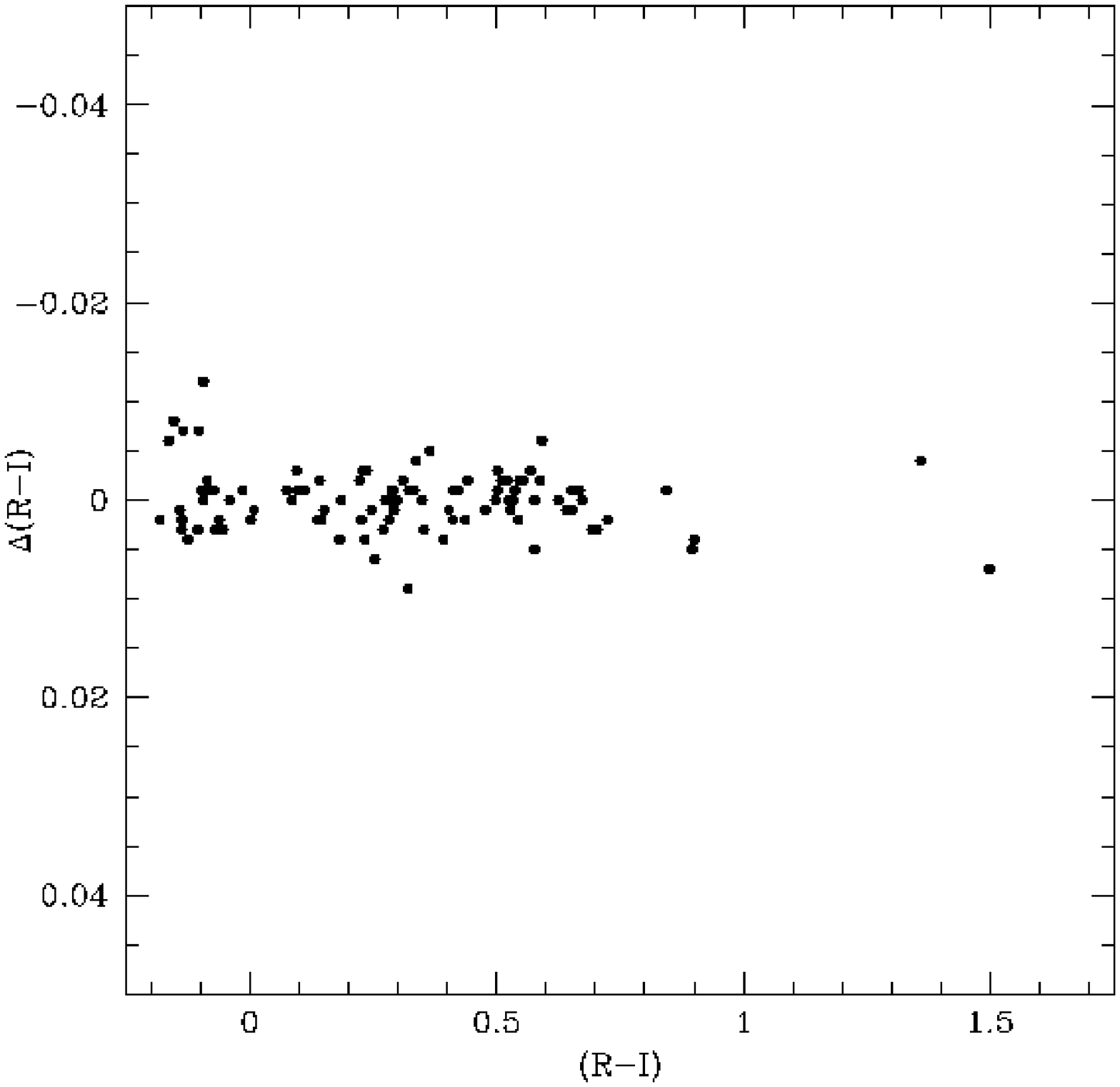}
\caption{Comparison of the $(R-I)$ magnitudes after removal of the transformation nonlinearity as a 
function of the \citet{Landolt1992} equatorial standard star's $(R-I)$ color index.}
\label{fig:figure11}
\end{figure}

\begin{figure}
\plotone{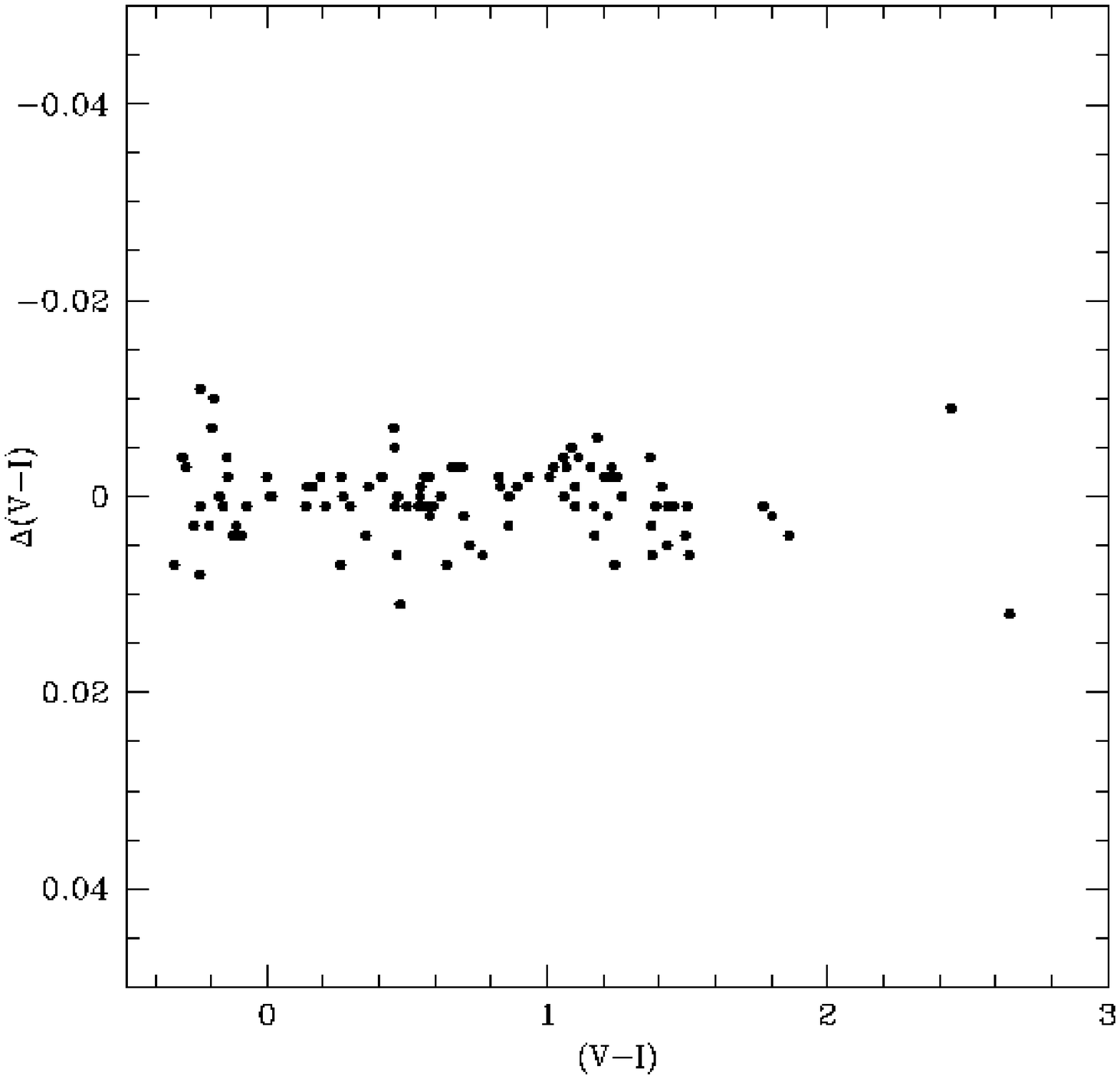}
\caption{Comparison of the $(V-I)$ magnitudes after removal of the transformation nonlinearity as a 
function of the \citet{Landolt1992} equatorial standard star's $(V-I)$ color index.}
\label{fig:figure12}
\end{figure}

After the above relations were applied to the recovered magnitudes and color indices of the standard 
stars used in this project, the data were on the broadband $UBVRI$ photometric system defined by the 
standard stars in \citet{Landolt1992}.  Next the standard star magnitudes and color indices, now 
corrected for nonlinear transformation, once again were compared to the published values in the sense 
of corrected values minus published magnitudes and color indices.  The fact that the nonlinear 
effects were corrected successfully is illustrated in Figures \ref{fig:figure7} - \ref{fig:figure12}.  
Therefore, the data in this paper have been transformed to the photometric system defined in 
\citet{Landolt1992}.

\begin{figure}
\plotone{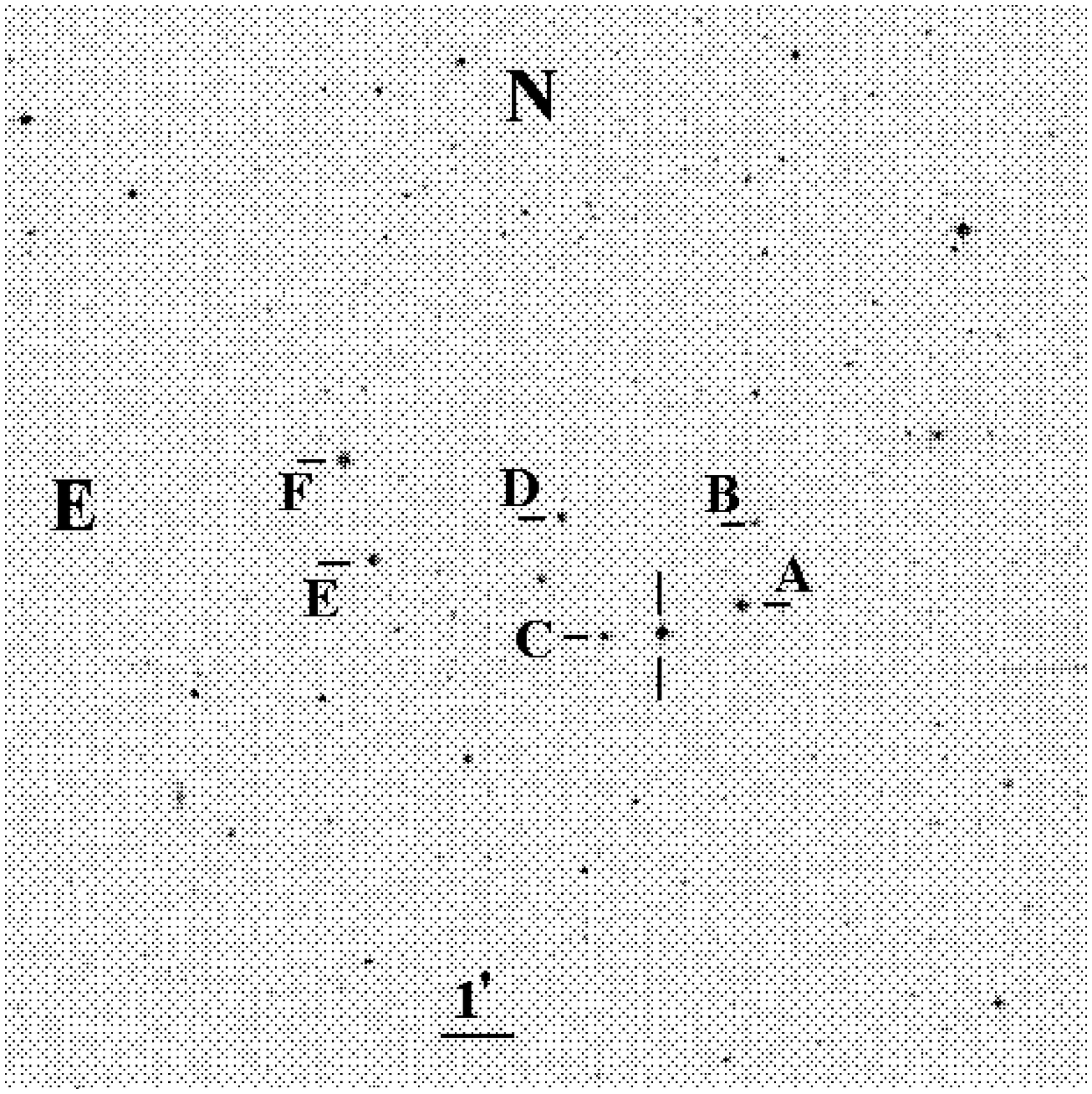}
\caption{Field, $15^{\prime}$ on a side, of the sequence in the vicinity of the star JL~163.}
\label{fig:figure13}
\end{figure}

\begin{figure}
\plotone{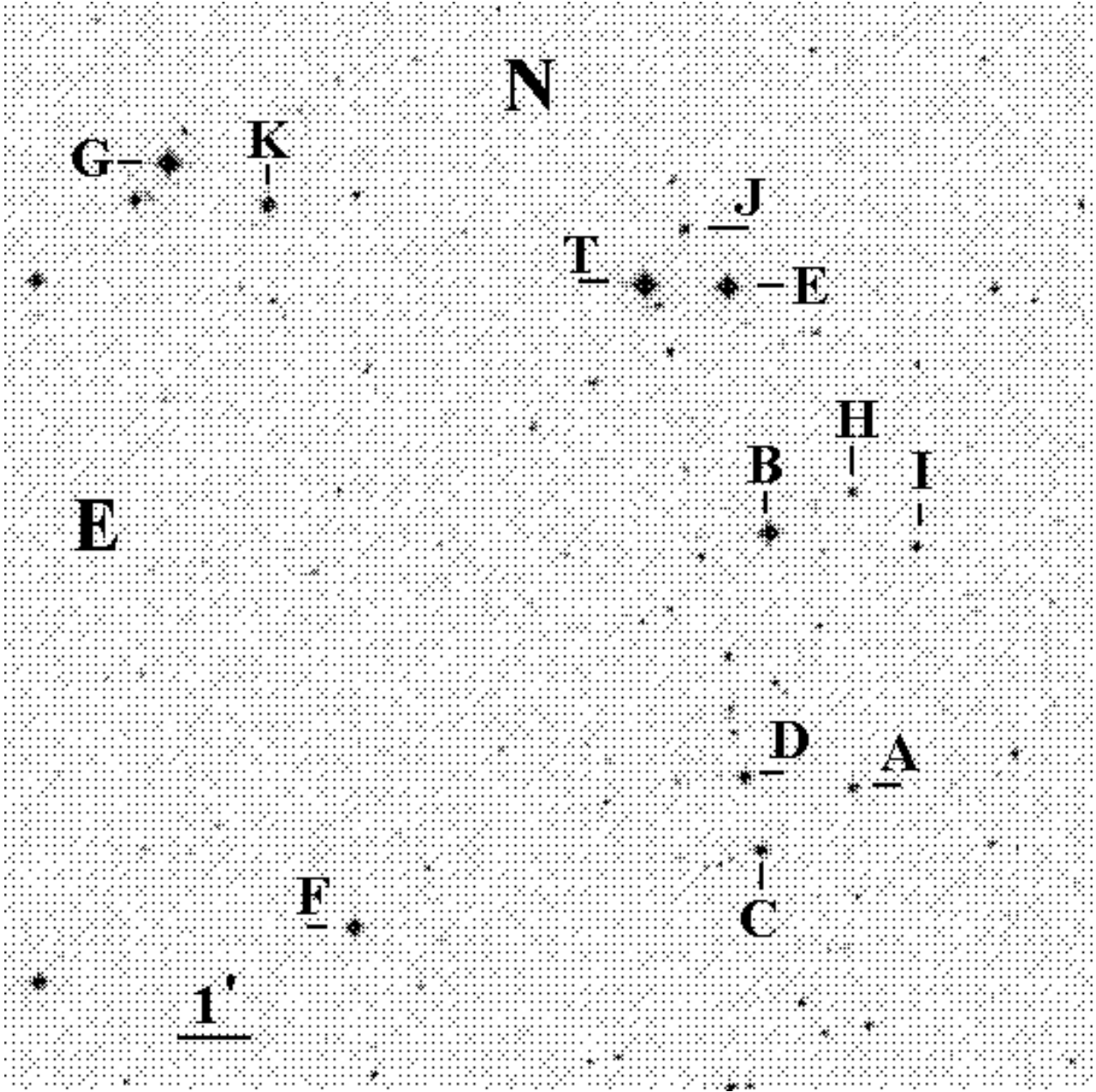}
\caption{Field, $15^{\prime}$ on a side, of the sequence in the vicinity of the Mira variable star T~Phe, 
marked as ``T" in the figure.}
\label{fig:figure14}
\end{figure}

\begin{figure}
\plotone{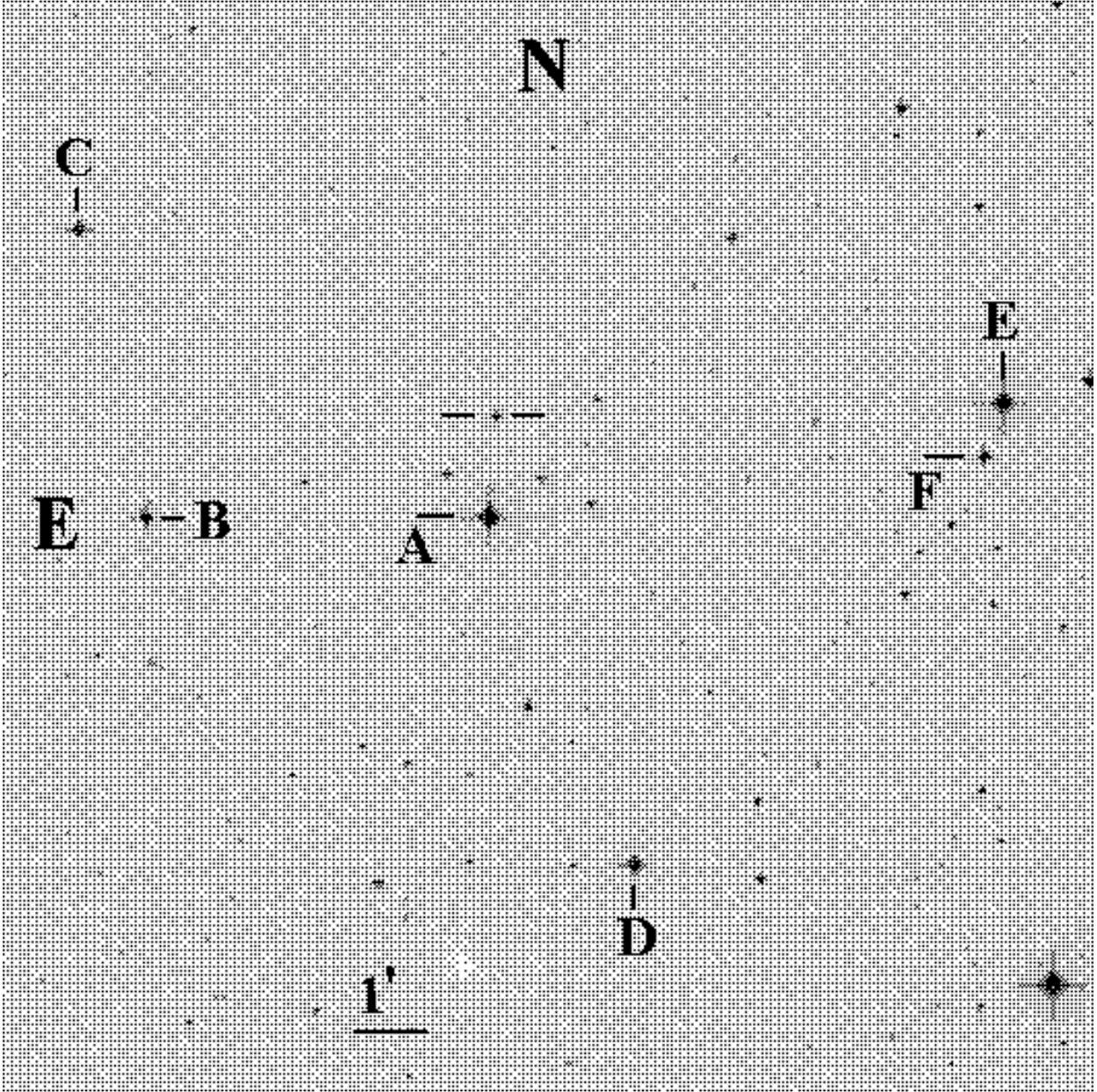}
\caption{Field, $15^{\prime}$ on a side, of the sequence in the vicinity of the star MCT~0401-4017.}
\label{fig:figure15}
\end{figure}

\begin{figure}
\plotone{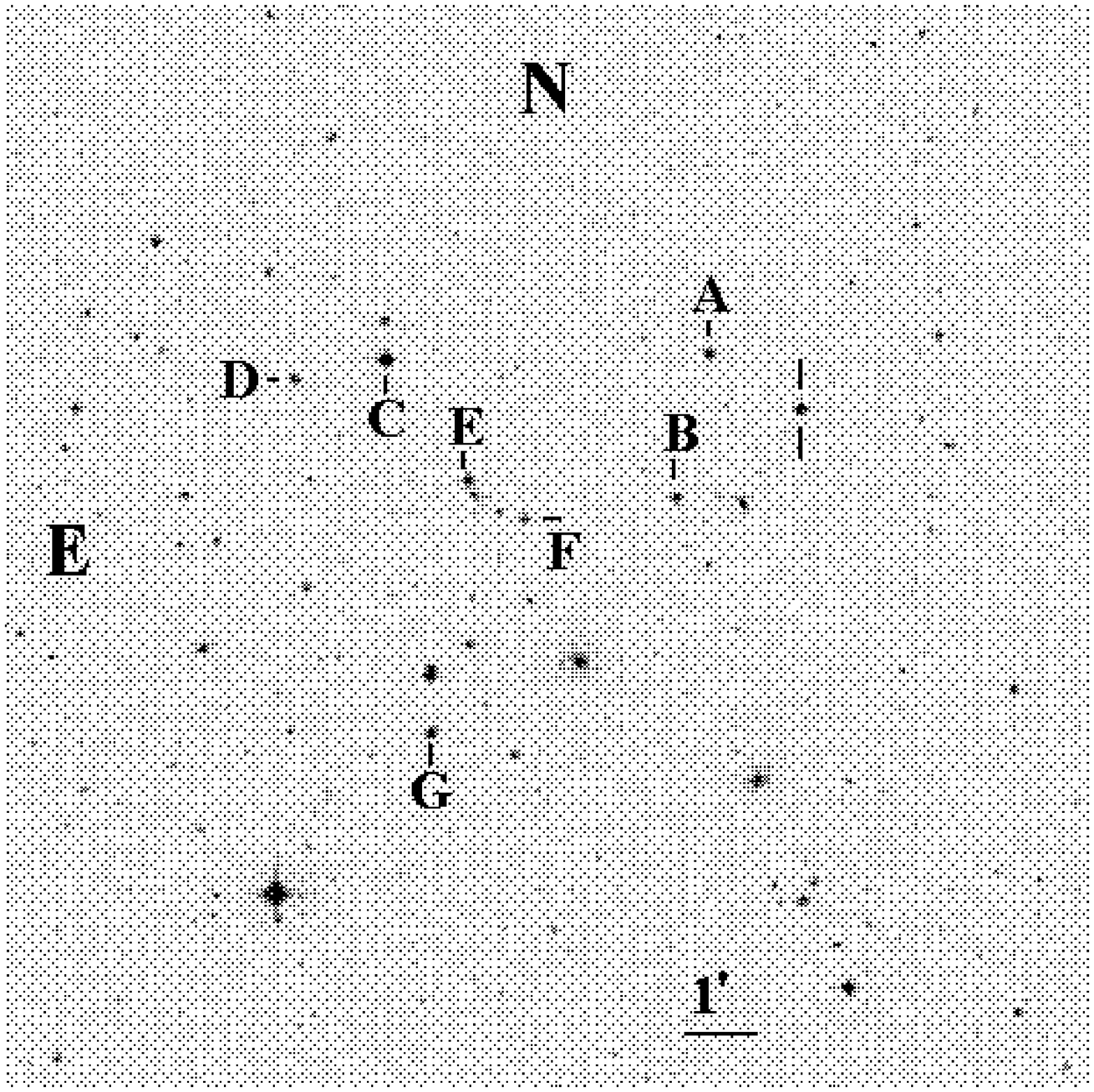}
\caption{Field, $15^{\prime}$ on a side, of the sequence in the vicinity of the star LB~1735.}
\label{fig:figure16}
\end{figure}

\begin{figure}
\plotone{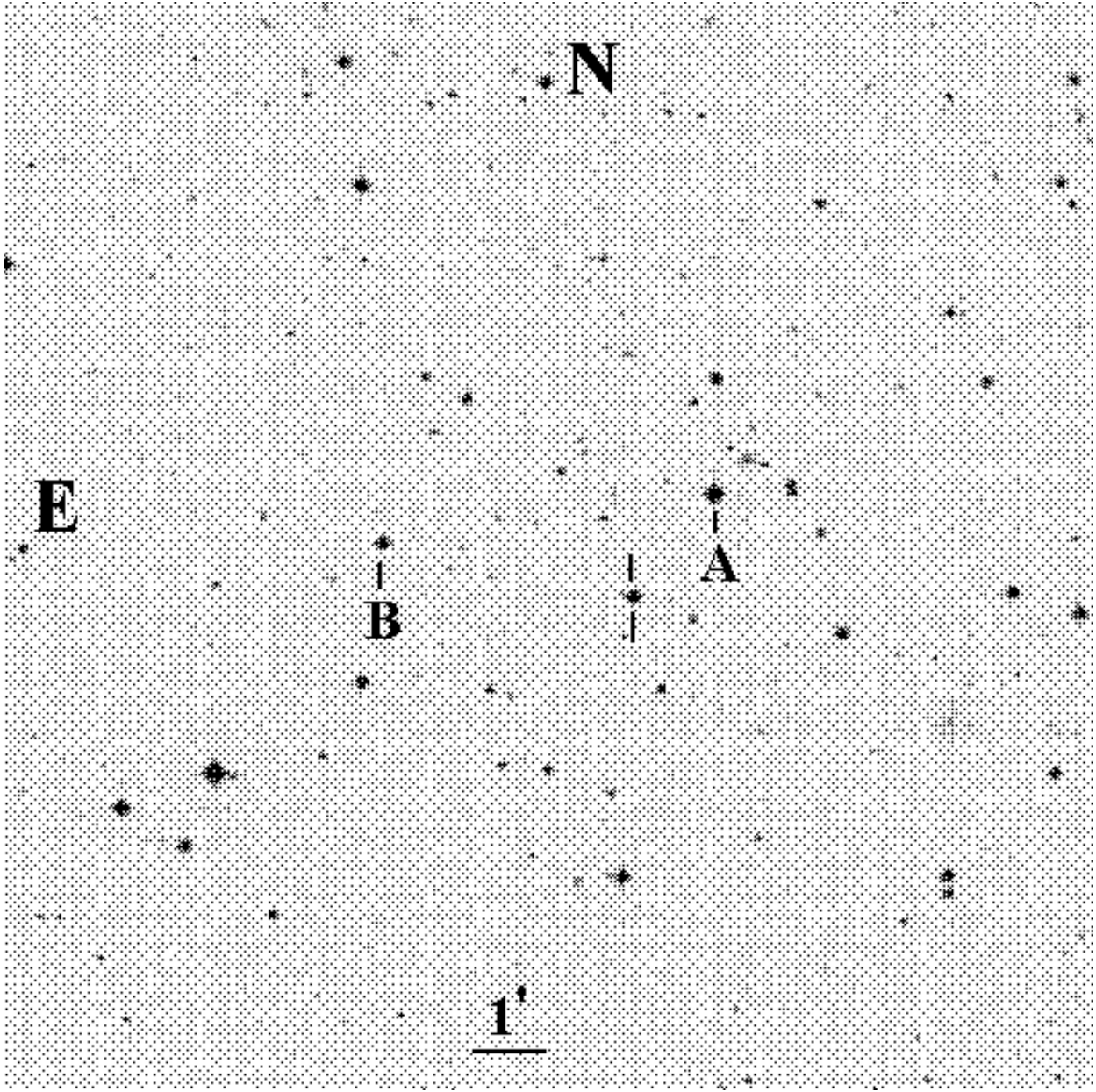}
\caption{Field, $15^{\prime}$ on a side, of the sequence in the vicinity of the star MCT~0436-4616.}
\label{fig:figure17}
\end{figure}

\begin{figure}
\plotone{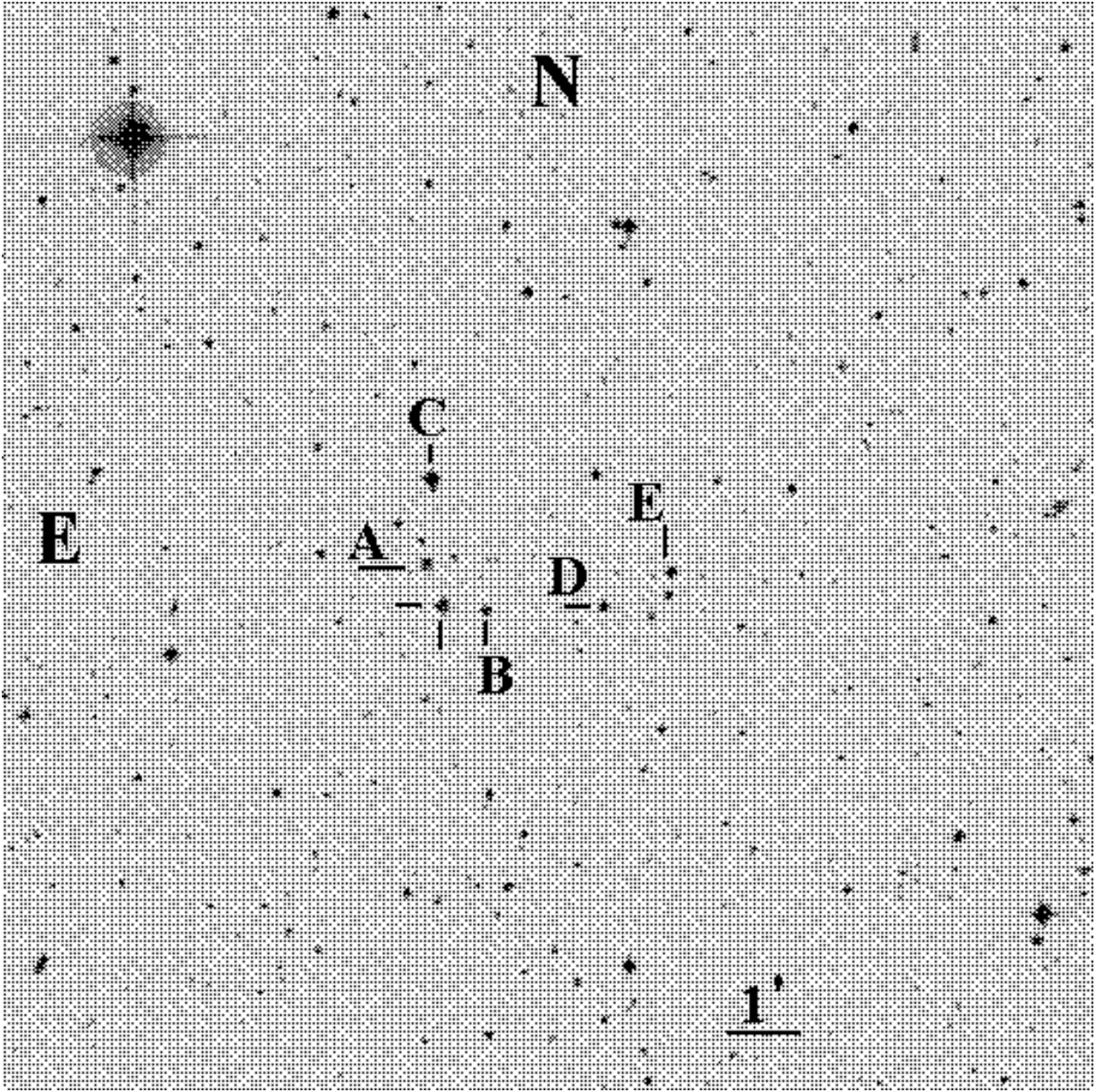}
\caption{Field, $15^{\prime}$ on a side, of the sequence in the vicinity of the star MCT~0550-4911.}
\label{fig:figure18}
\end{figure}

\begin{figure}
\plotone{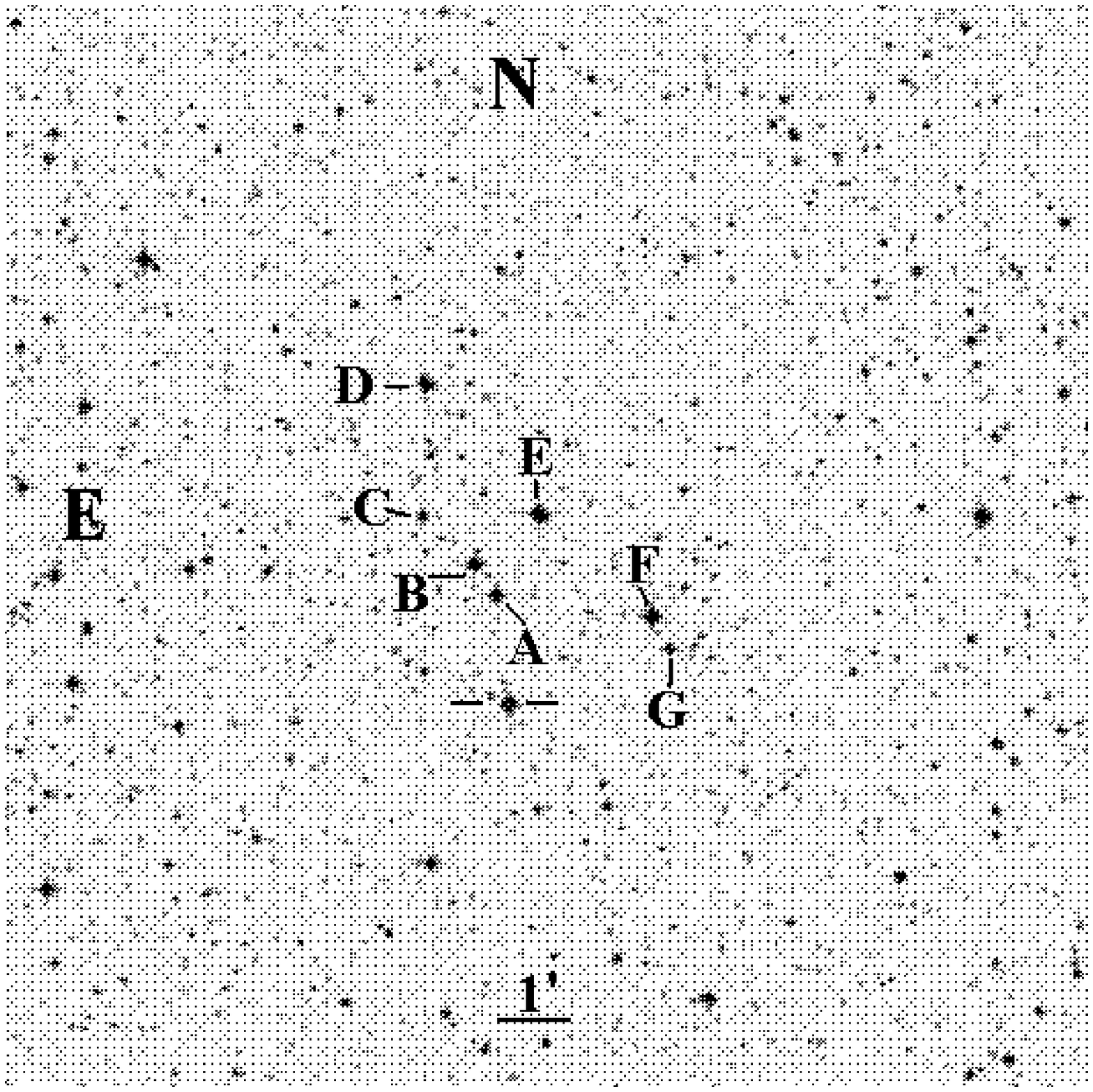}
\caption{Field, $15^{\prime}$ on a side, of the sequence in the vicinity of the star LSS~982.}
\label{fig:figure19}
\end{figure}

\begin{figure}
\plotone{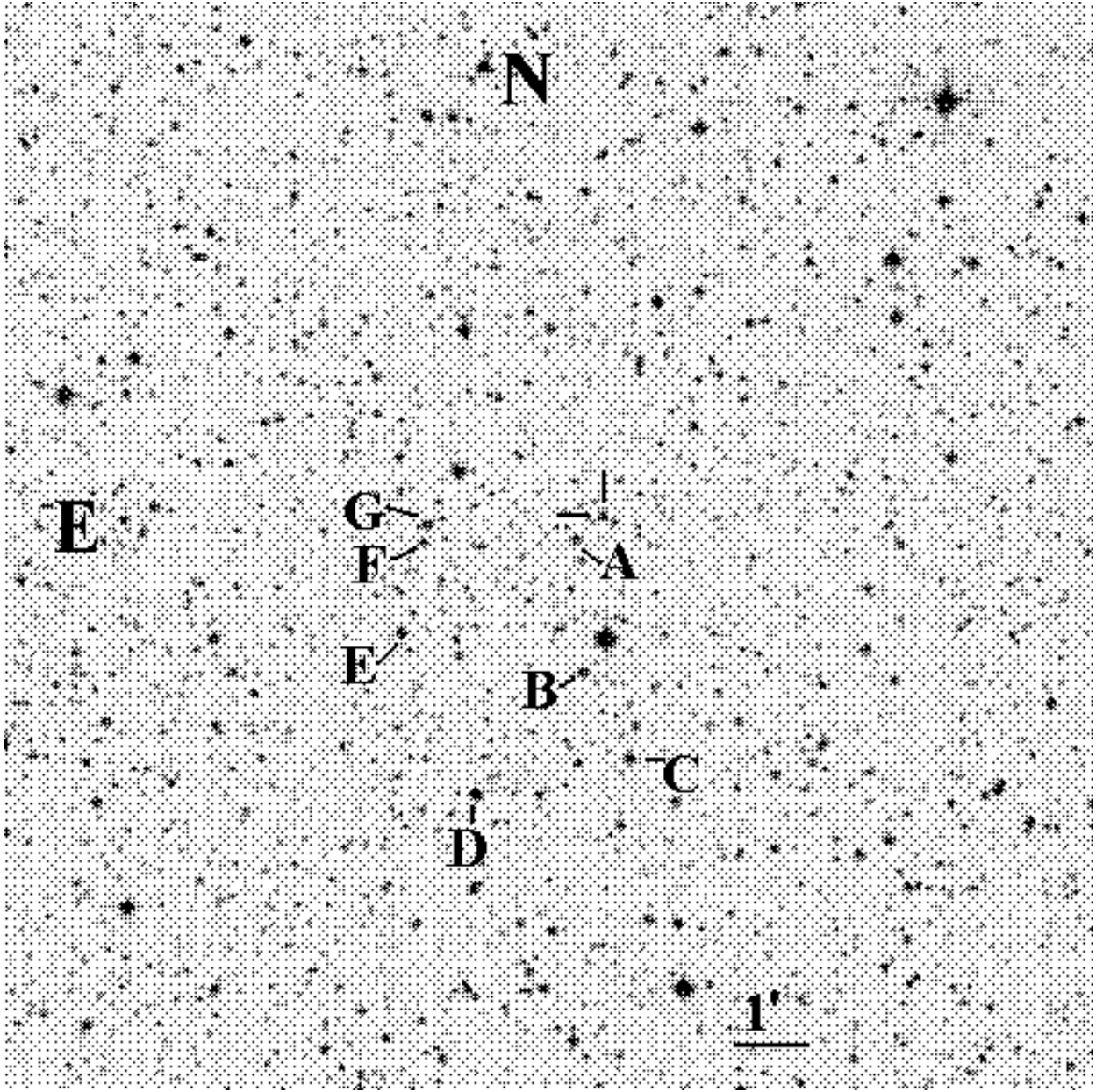}
\caption{Field, $15^{\prime}$ on a side, of the sequence in the vicinity of the star WD~0830-535.}
\label{fig:figure20}
\end{figure}

\begin{figure}
\plotone{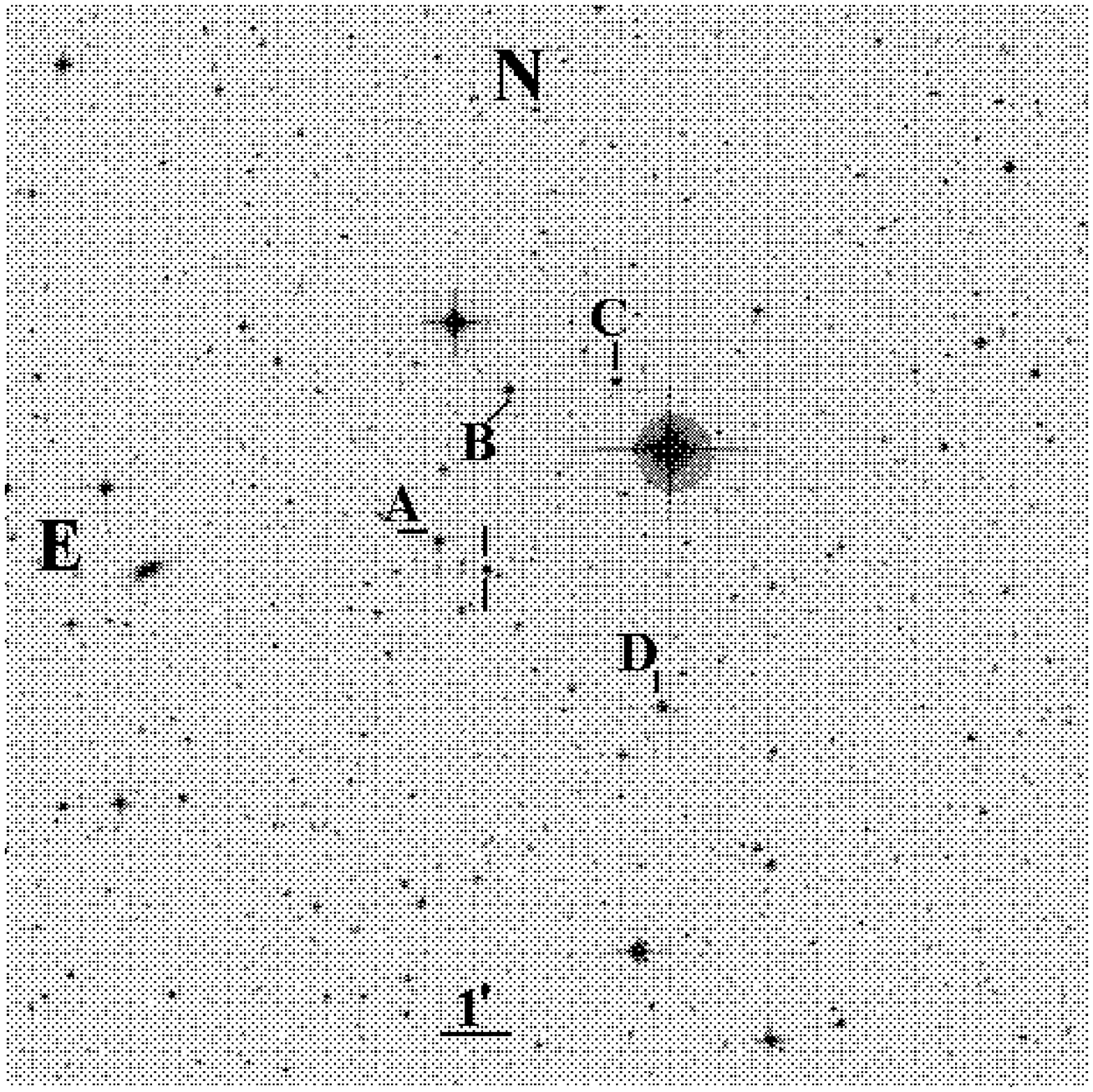}
\caption{Field, $15^{\prime}$ on a side, of the sequence in the vicinity of the star WD~1056-384.}
\label{fig:figure21}
\end{figure}

\begin{figure}
\plotone{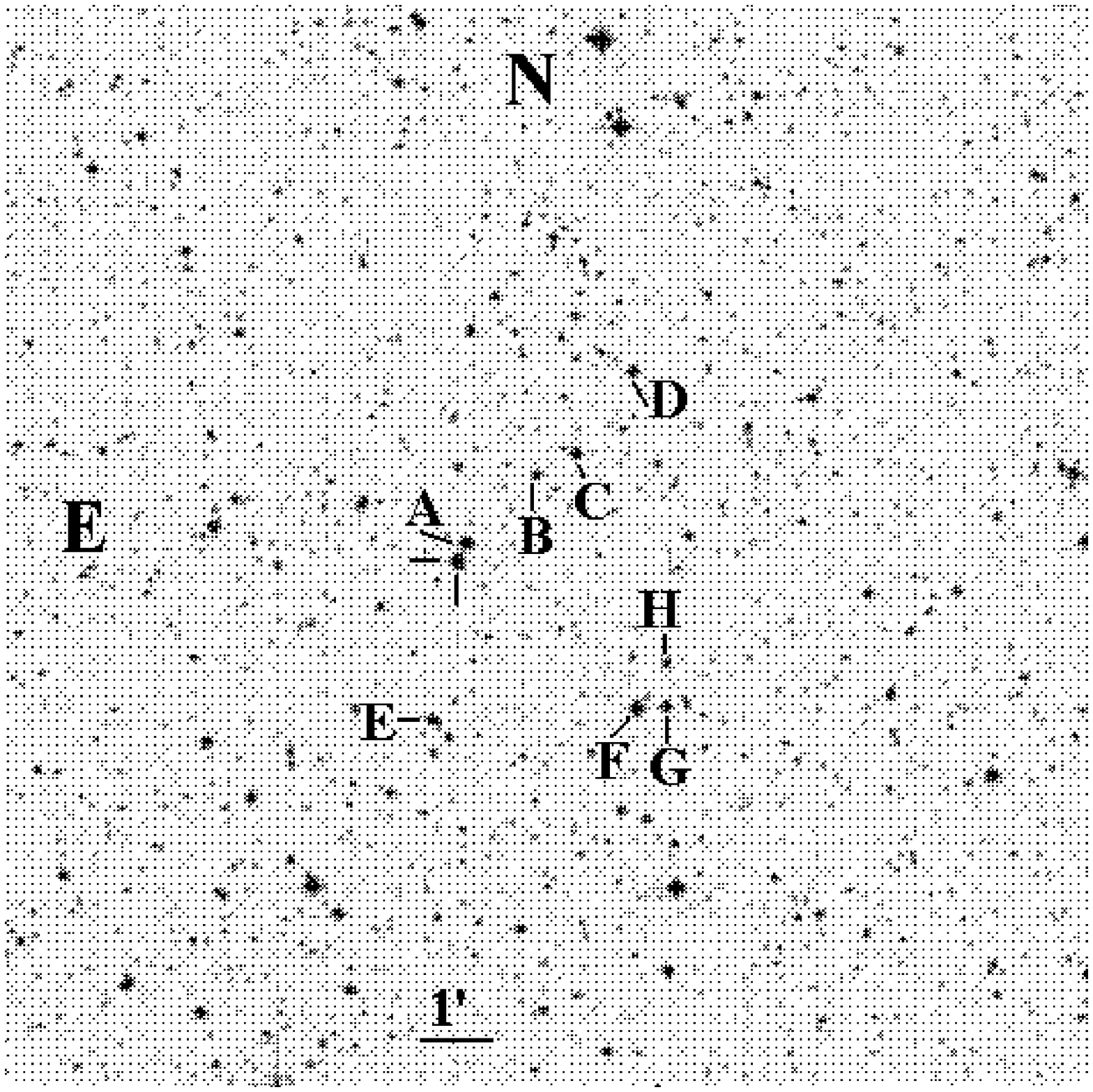}
\caption{Field, $15^{\prime}$ on a side, of the sequence in the vicinity of the star WD~1153-484.}
\label{fig:figure22}
\end{figure}

\begin{figure}
\plotone{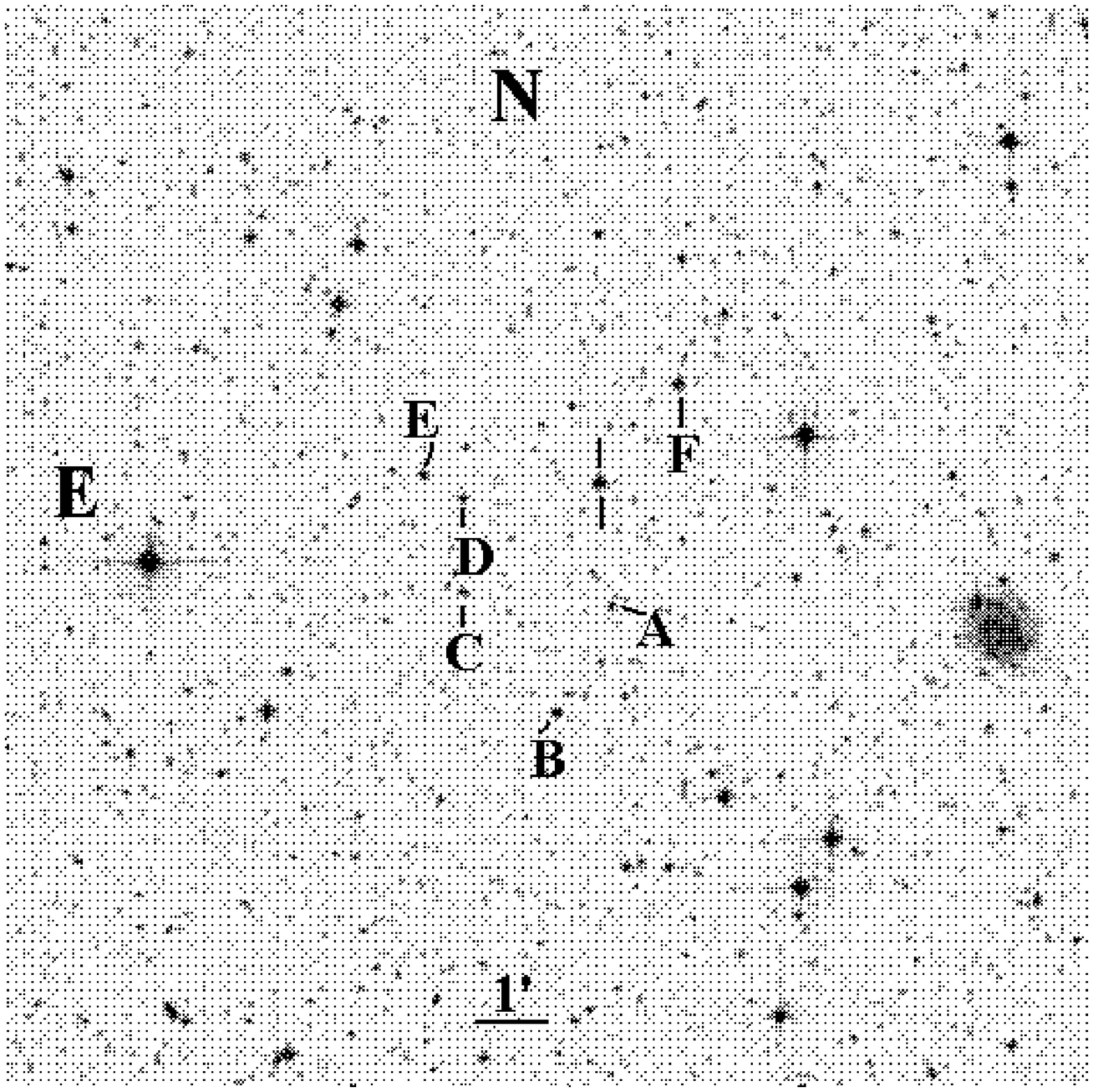}
\caption{Field, $15^{\prime}$ on a side, of the sequence in the vicinity of the star LSE~44.}
\label{fig:figure23}
\end{figure}

\begin{figure}
\plotone{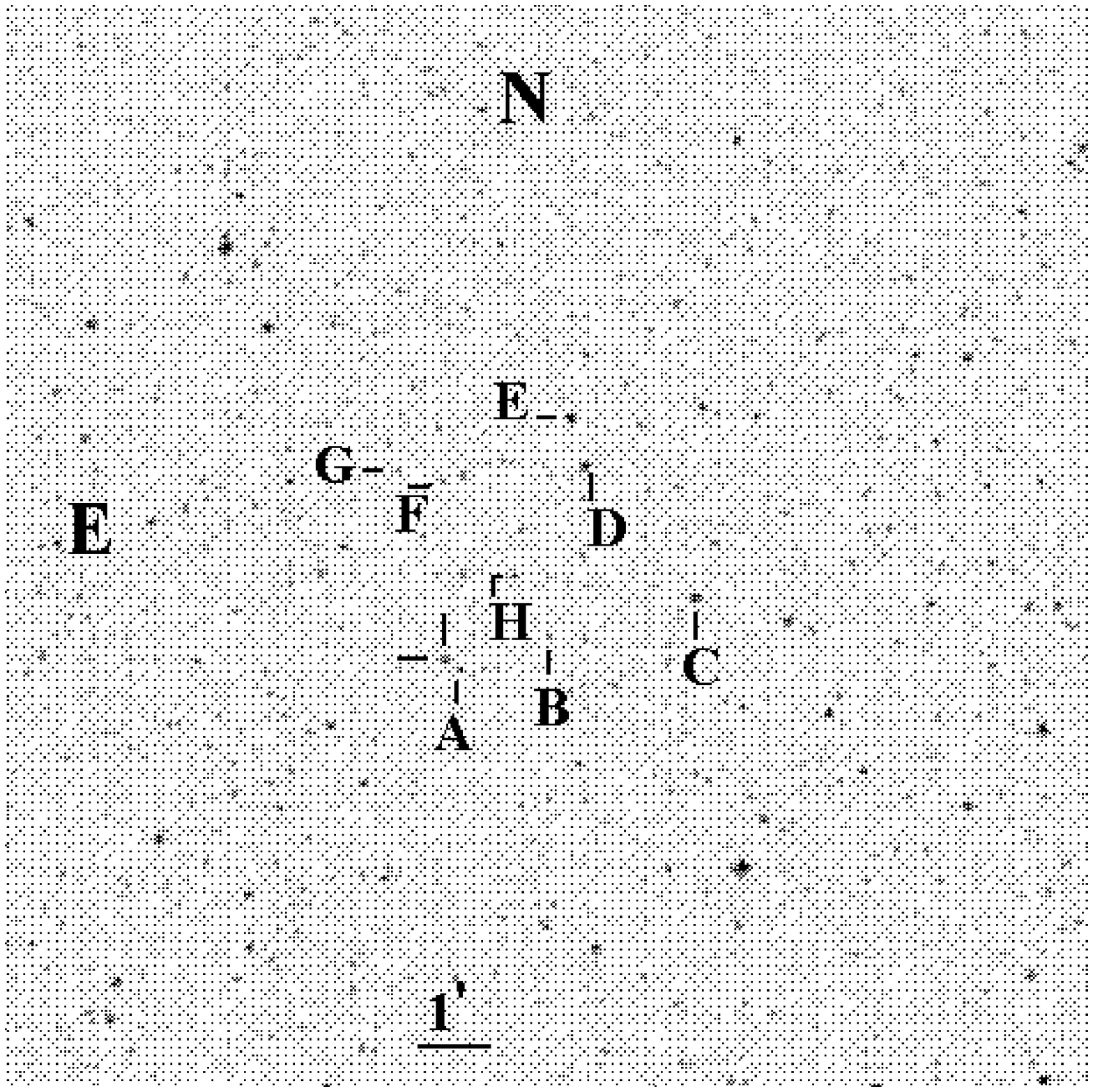}
\caption{Field, $15^{\prime}$ on a side, of the sequence in the vicinity of the star LSE~259.}
\label{fig:figure24}
\end{figure}

\begin{figure}
\plotone{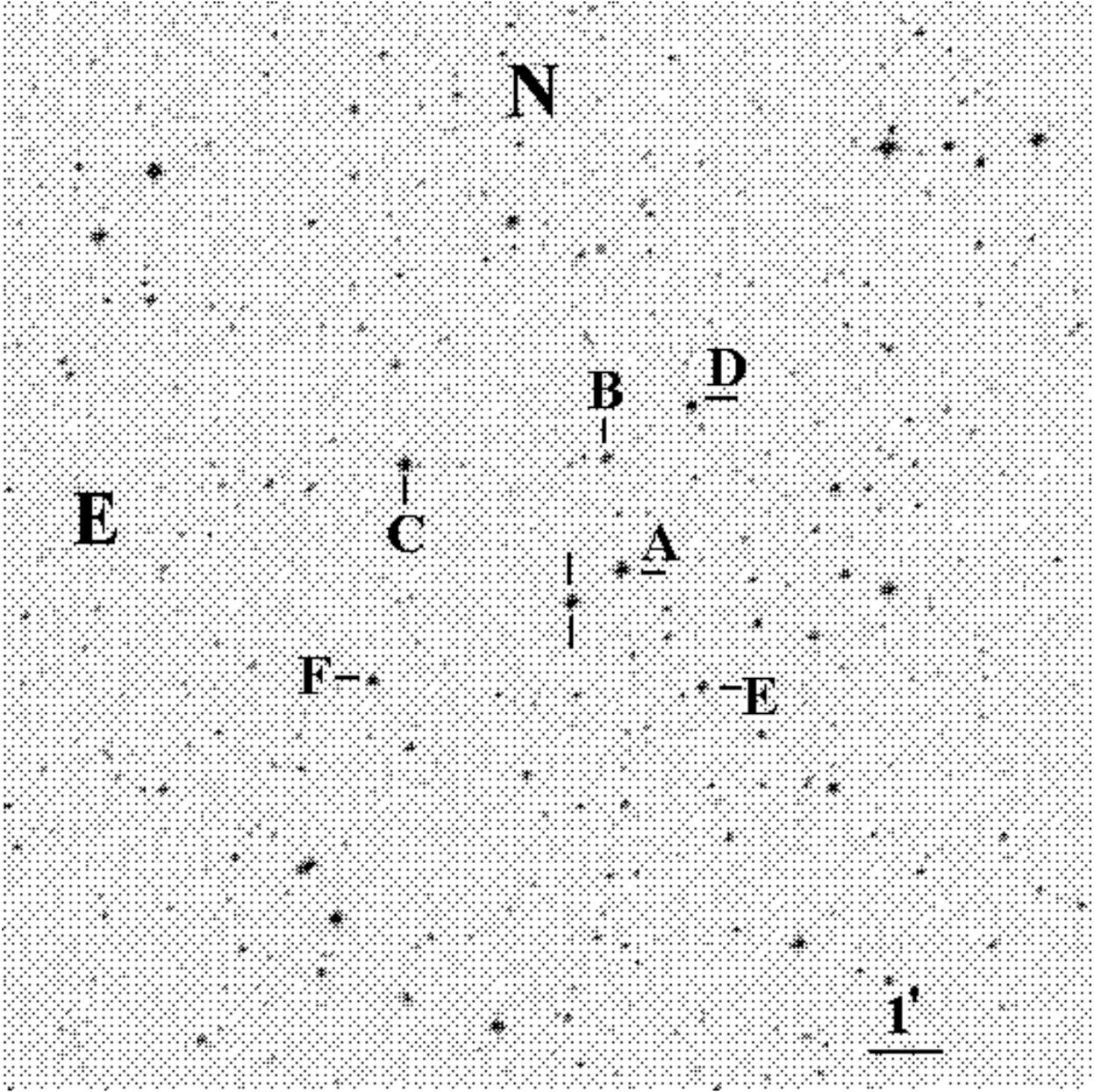}
\caption{Field, $15^{\prime}$ on a side, of the sequence in the vicinity of the star MCT~2019-4339.}
\label{fig:figure25}
\end{figure}

\begin{figure}
\plotone{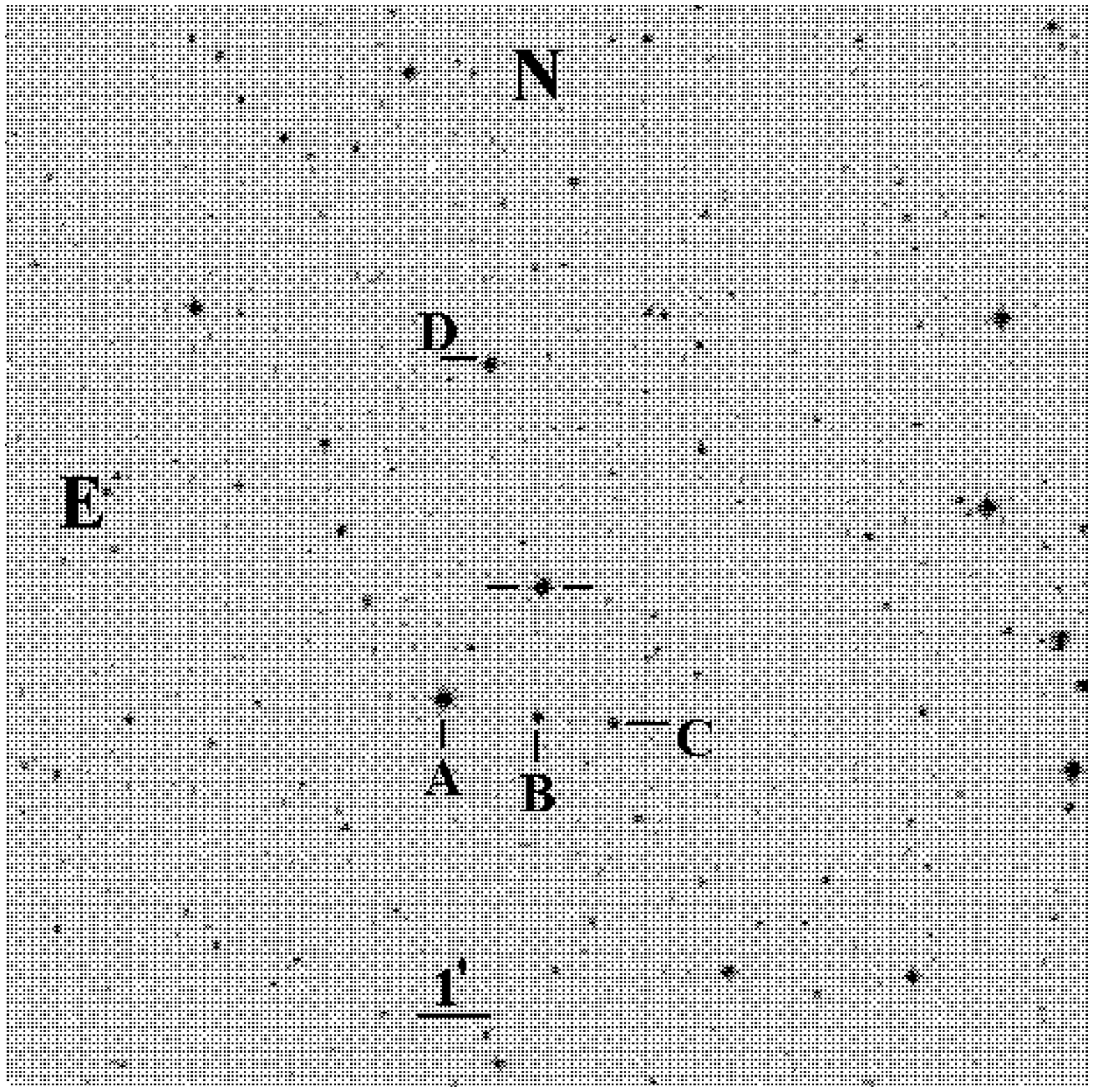}
\caption{Field, $15^{\prime}$ on a side, of the sequence in the vicinity of the star JL~82.}
\label{fig:figure26}
\end{figure}

\begin{figure}
\plotone{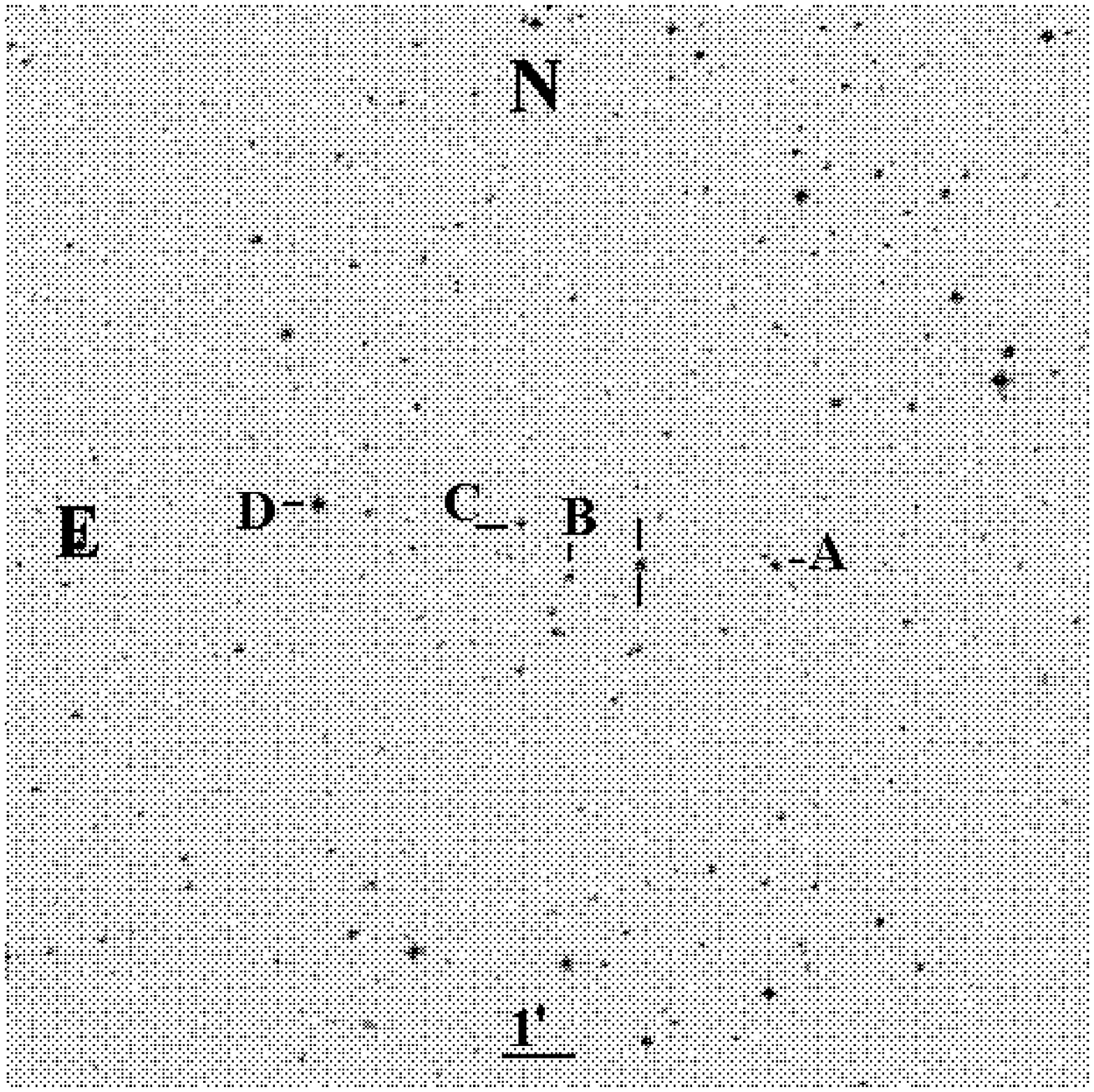}
\caption{Field, $15^{\prime}$ on a side, of the sequence in the vicinity of the star JL~117.}
\label{fig:figure27}
\end{figure}

The final magnitudes and color indices for the new standard stars resulting from this program are 
tabulated in Table \ref{tab:table1}.  Each star was observed an average of 16.4 times on an average 
9 nights.  Finding charts are provided via Figures \ref{fig:figure13} - \ref{fig:figure27}.  The 
coordinates in Table \ref{tab:table1} were taken from the UCAC2 catalog \citep{Zacharias2004} when 
possible.  Positions for stars not in the UCAC2 catalog were taken from the 2MASS Point Source 
Catalog which coordinates came from the Two Micron All Sky survey \citep[2MASS;][]{Skrutskie2006}.

Columns (4)-(9) in Table \ref{tab:table1} give the final magnitude and color indices in the 
$UBVRI$ photometric system as defined by \citet{Landolt1992}.  Column (10) indicates the number of 
times {\it n} that each star was observed.  Column (11) gives the number of nights {\it m} that 
each star was observed.  The numbers in columns (4)-(9) are mean magnitudes and color indices.  
Hence, the errors tabulated in columns (12)-(17) are mean errors of the mean magnitude and color 
indices \citep[see][p. 450]{Landolt1983}.  

The stars in the WD~1153-484 sequence are of interest because they encompass a good range in 
color.  However, there are insufficient measures, on average, for the stars to make the sequence 
as robust as one would like.  The number of measures is fewer than had been anticipated because 
during the reduction and analysis, two nights of data had to be rejected apparently due to thin 
cirrus.

The numerical size of the average mean error of a single observation of a $V$ magnitude or a 
color index for the 109 new standard stars in Table \ref{tab:table1} is given in the second 
column of Table \ref{tab:table2}. The last column shows the average mean error of the mean 
observed magnitude or color index.  Errors in the second column for a single observation are 
as large as they are for $(U-B)$, $(R-I)$, and $(V-I)$ since red stars are faint in $U$ and 
blue stars are faint in $I$.

While accurate coordinates for individual stars are necessary in many circumstances, modern area 
detectors need knowledge of the coordinate center of a photometric sequence.  Table 
\ref{tab:table3} provides the coordinate centers for the new $UBVRI$ photometric sequences 
listed in Table \ref{tab:table1}.  The field name is based on a blue star chosen from the 
literature, or recommended by a colleague, with the exception of the T$\,$Phe field.  T$\,$Phe 
is a Mira variable (= HD~2725 = CD-47~131 = CPD-47~50 = GSC~08024-01000). The current T$\,$Phe 
sequence has been enlarged from the sequence initially published in \citet{Landolt1992}.

\begin{figure}
\plotone{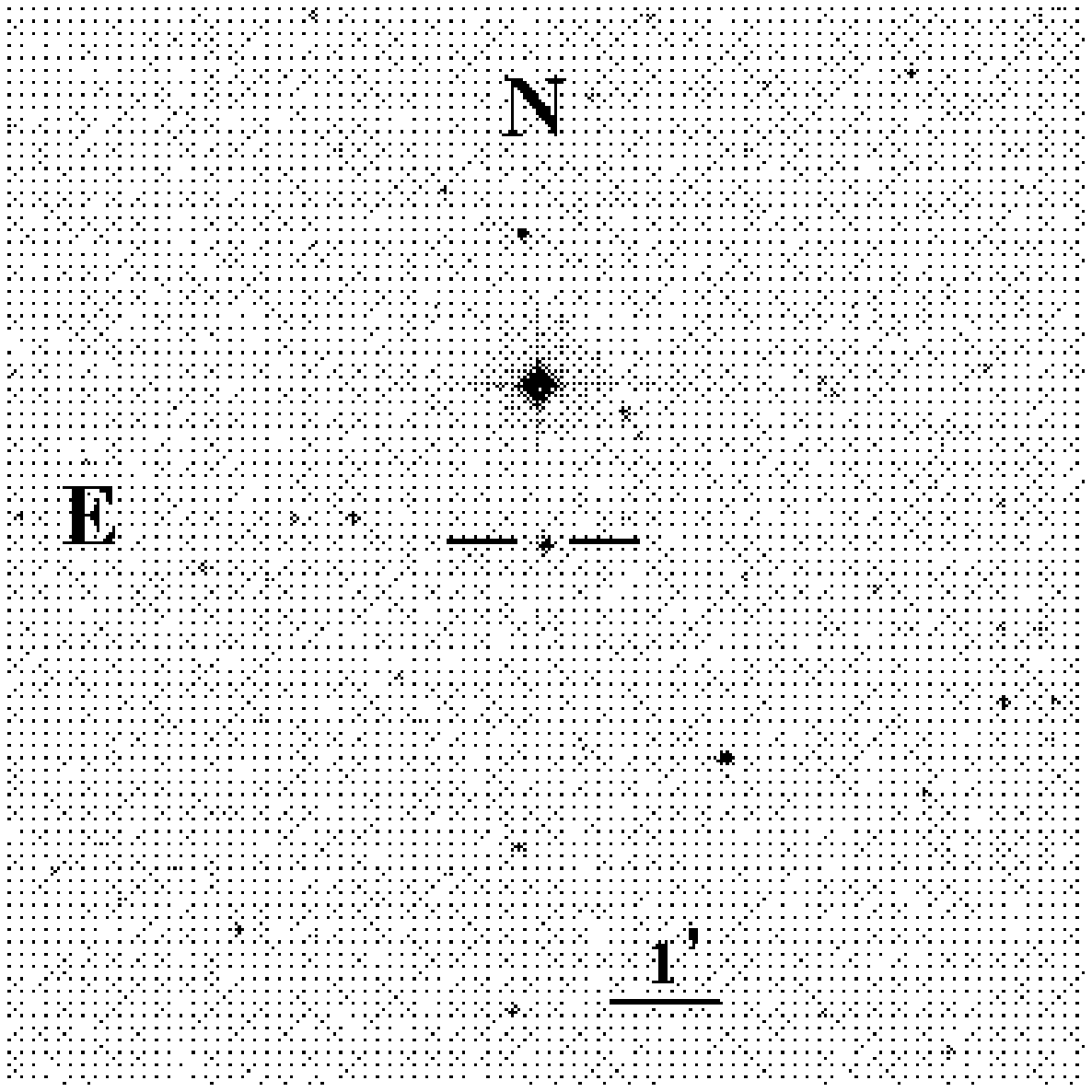}
\caption{Field, $10^{\prime}$ on a side, of the star JL~166.}
\label{fig:figure28}
\end{figure}

\begin{figure}
\plotone{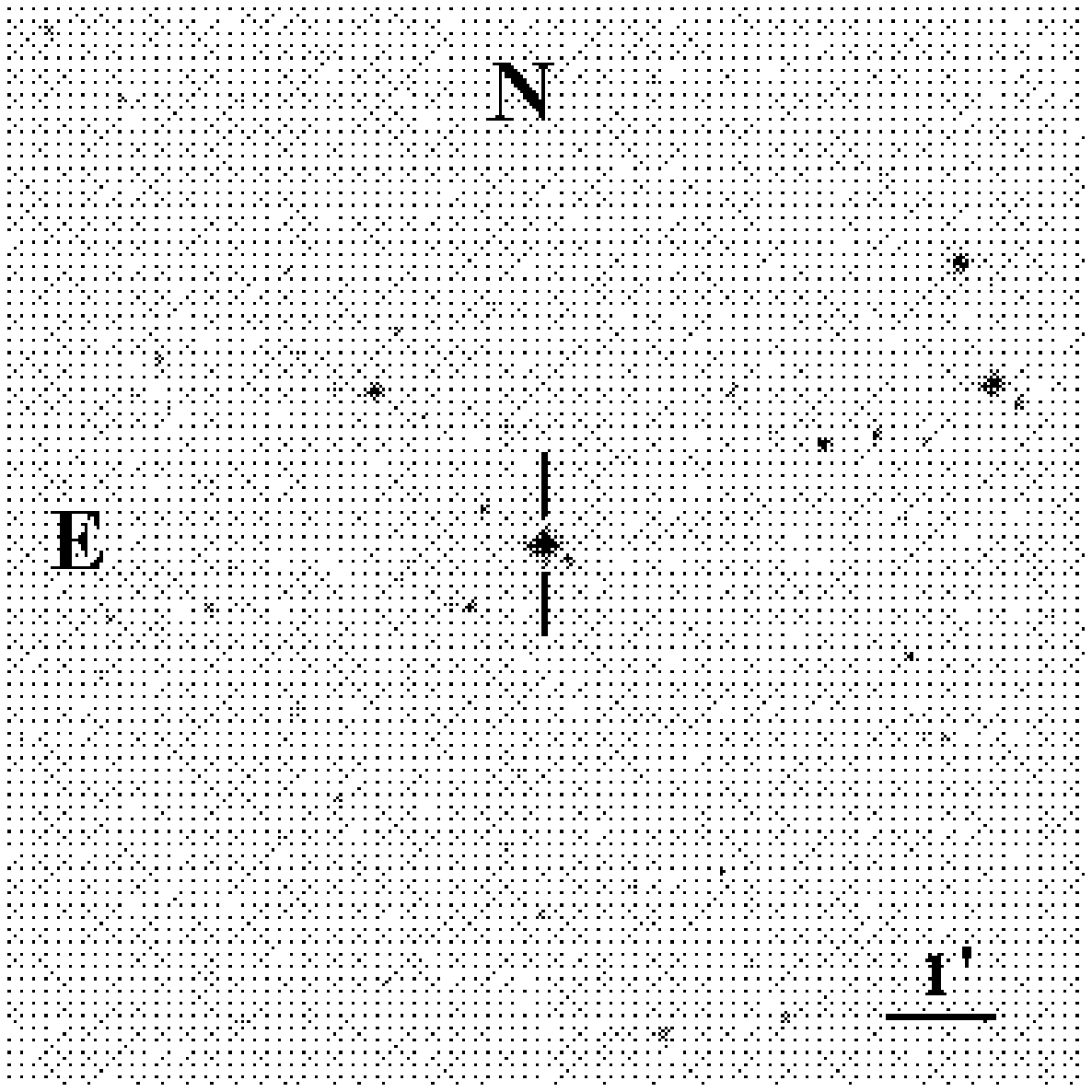}
\caption{Field, $10^{\prime}$ on a side, of the star JL~194.}
\label{fig:figure29}
\end{figure}

\begin{figure}
\plotone{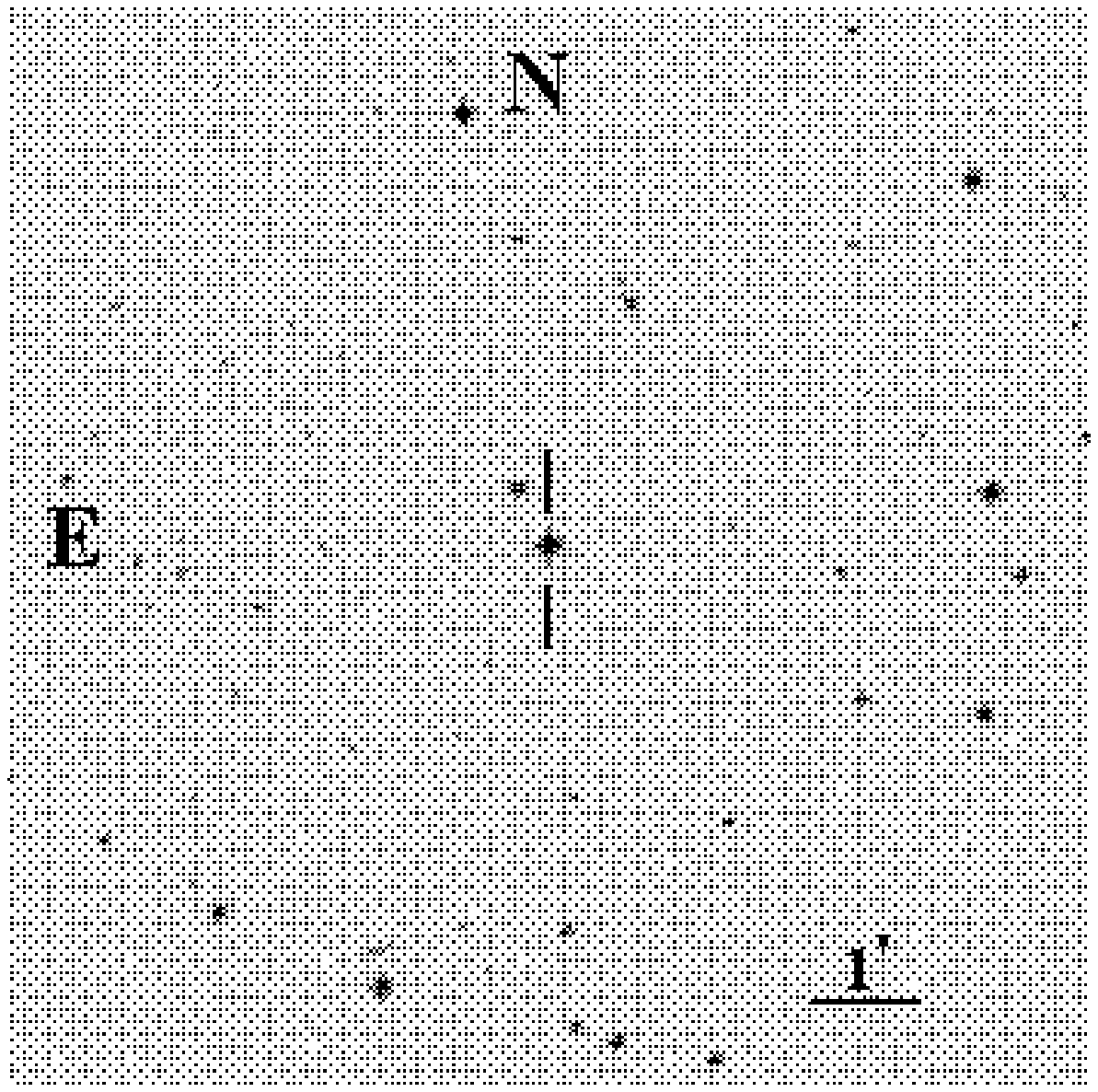}
\caption{Field, $10^{\prime}$ on a side, of the star JL~202.}
\label{fig:figure30}
\end{figure}

\begin{figure}
\plotone{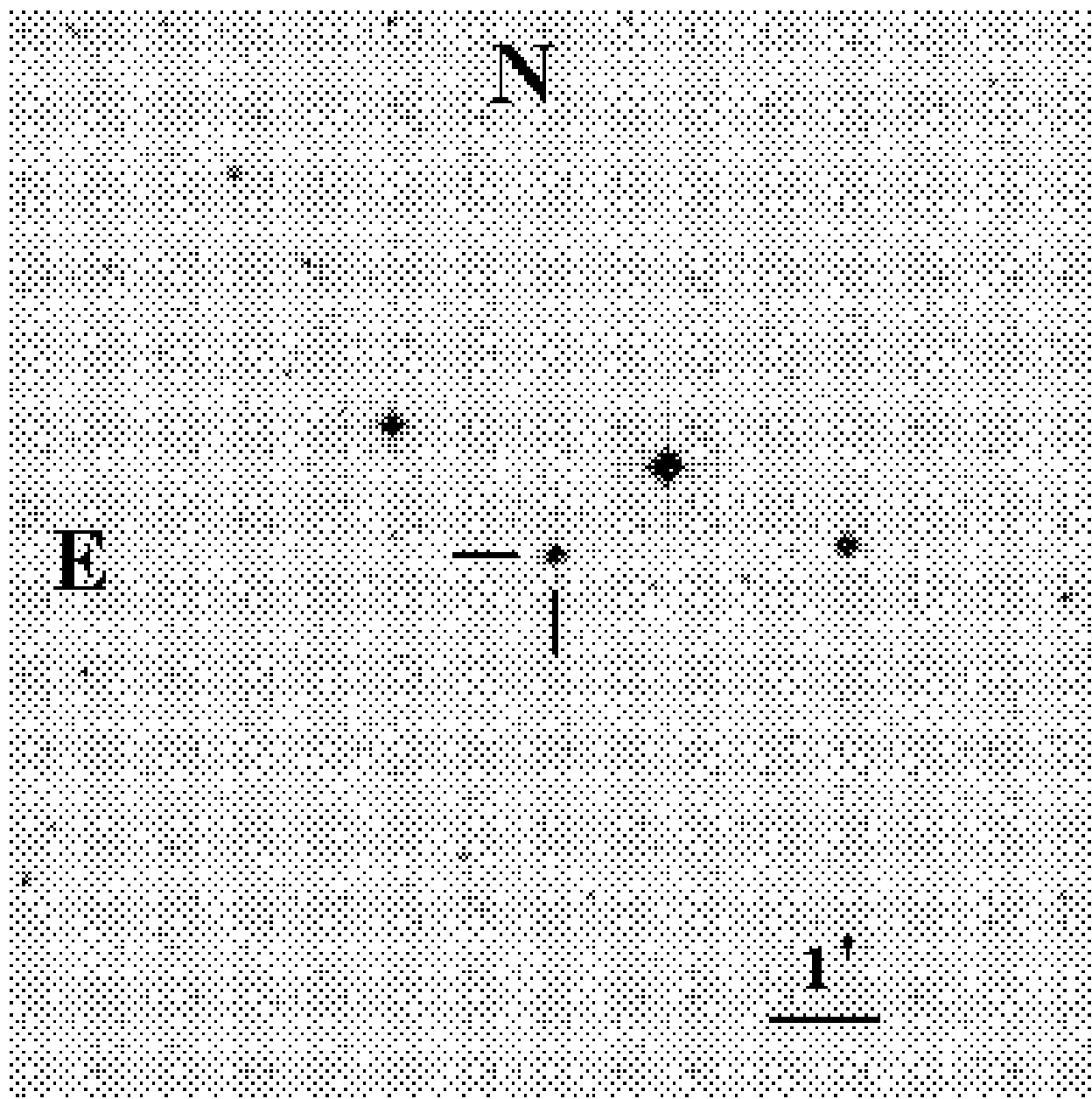}
\caption{Field, $10^{\prime}$ on a side, of the star GD~679.}
\label{fig:figure31}
\end{figure}

\begin{figure}
\plotone{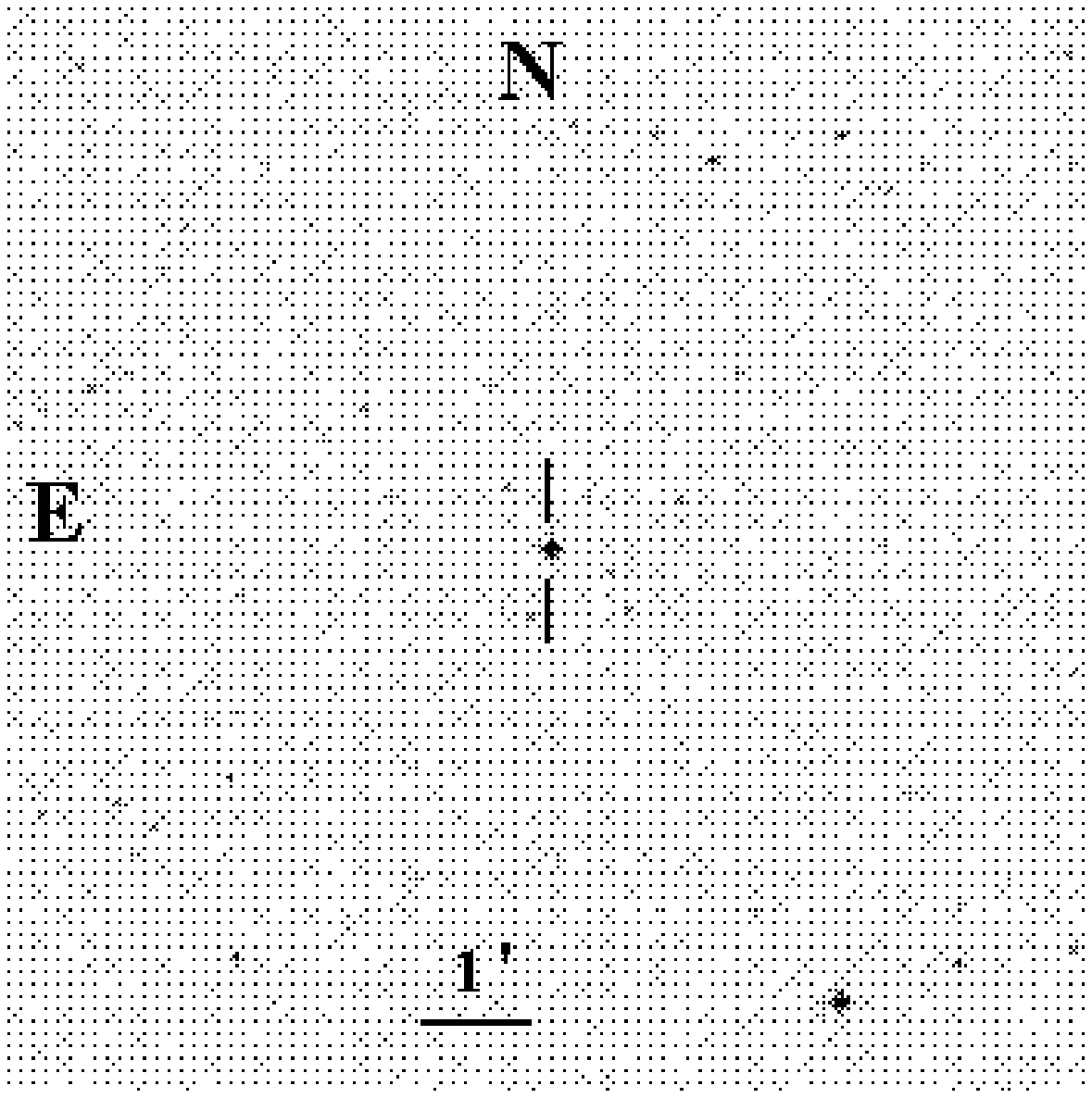}
\caption{Field, $10^{\prime}$ on a side, of the star JL~236.}
\label{fig:figure32}
\end{figure}

\begin{figure}
\plotone{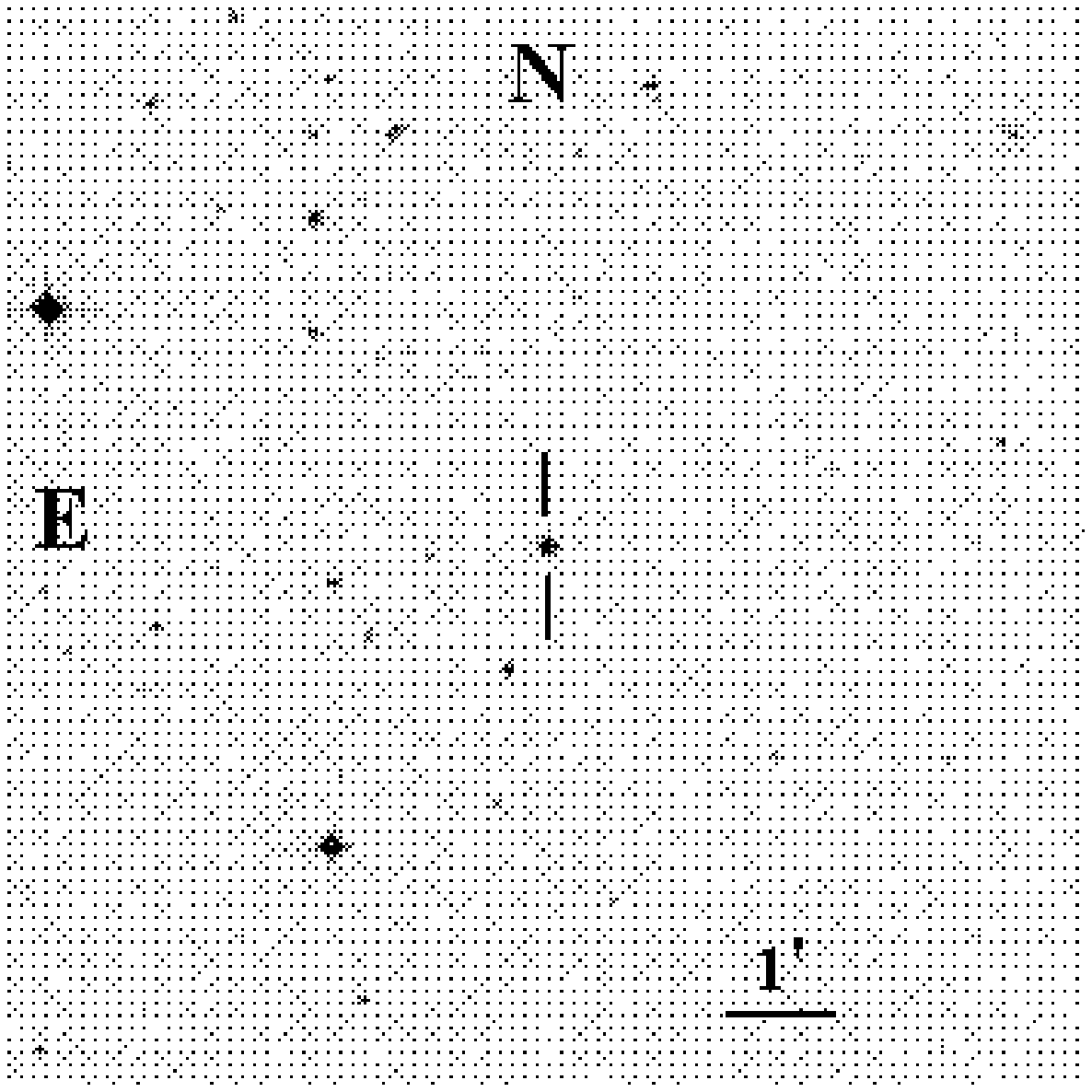}
\caption{Field, $10^{\prime}$ on a side, of the star JL~261.}
\label{fig:figure33}
\end{figure}

\begin{figure}
\plotone{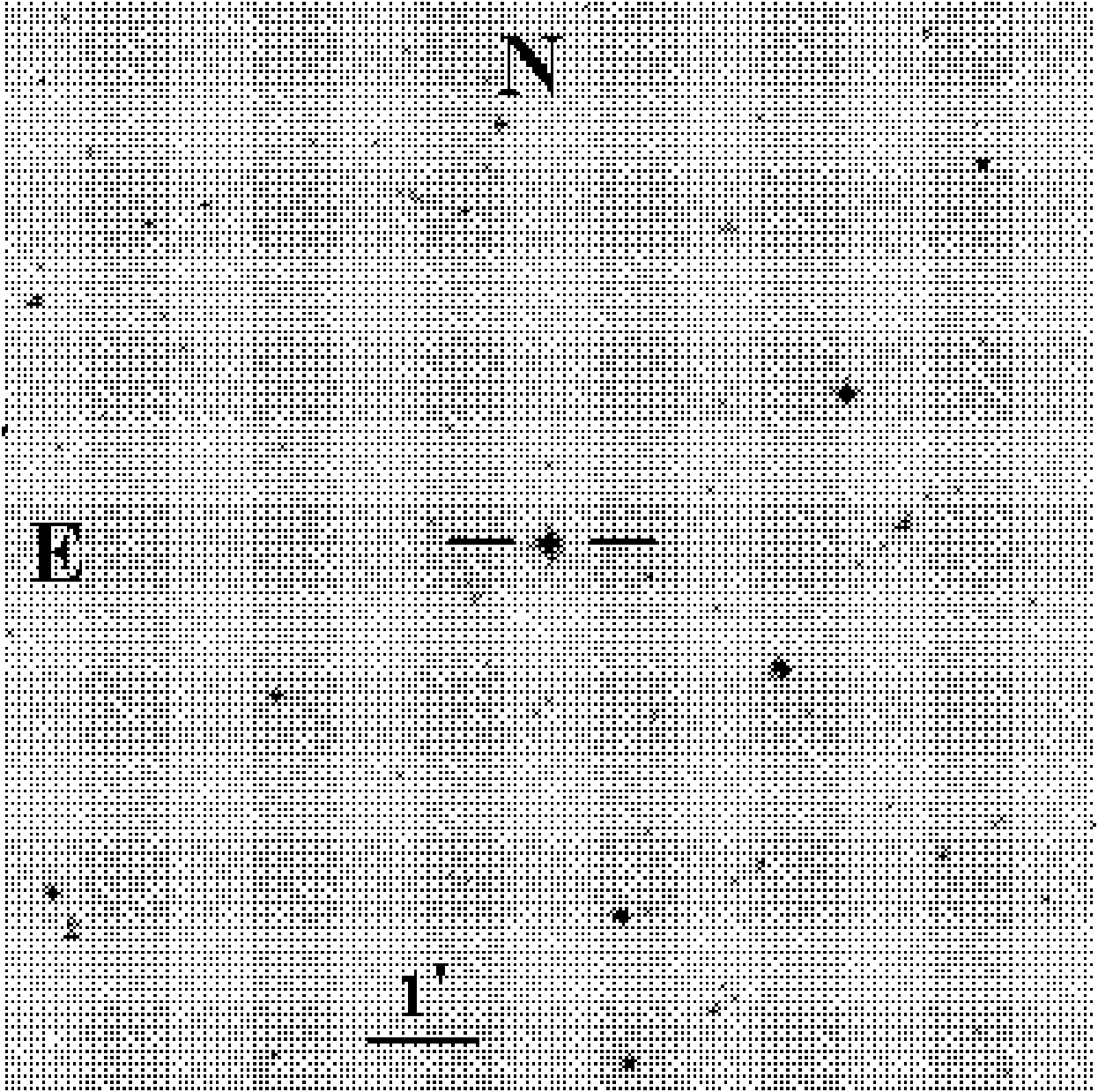}
\caption{Field, $10^{\prime}$ on a side, of the star LB~3241.}
\label{fig:figure34}
\end{figure}

\begin{figure}
\plotone{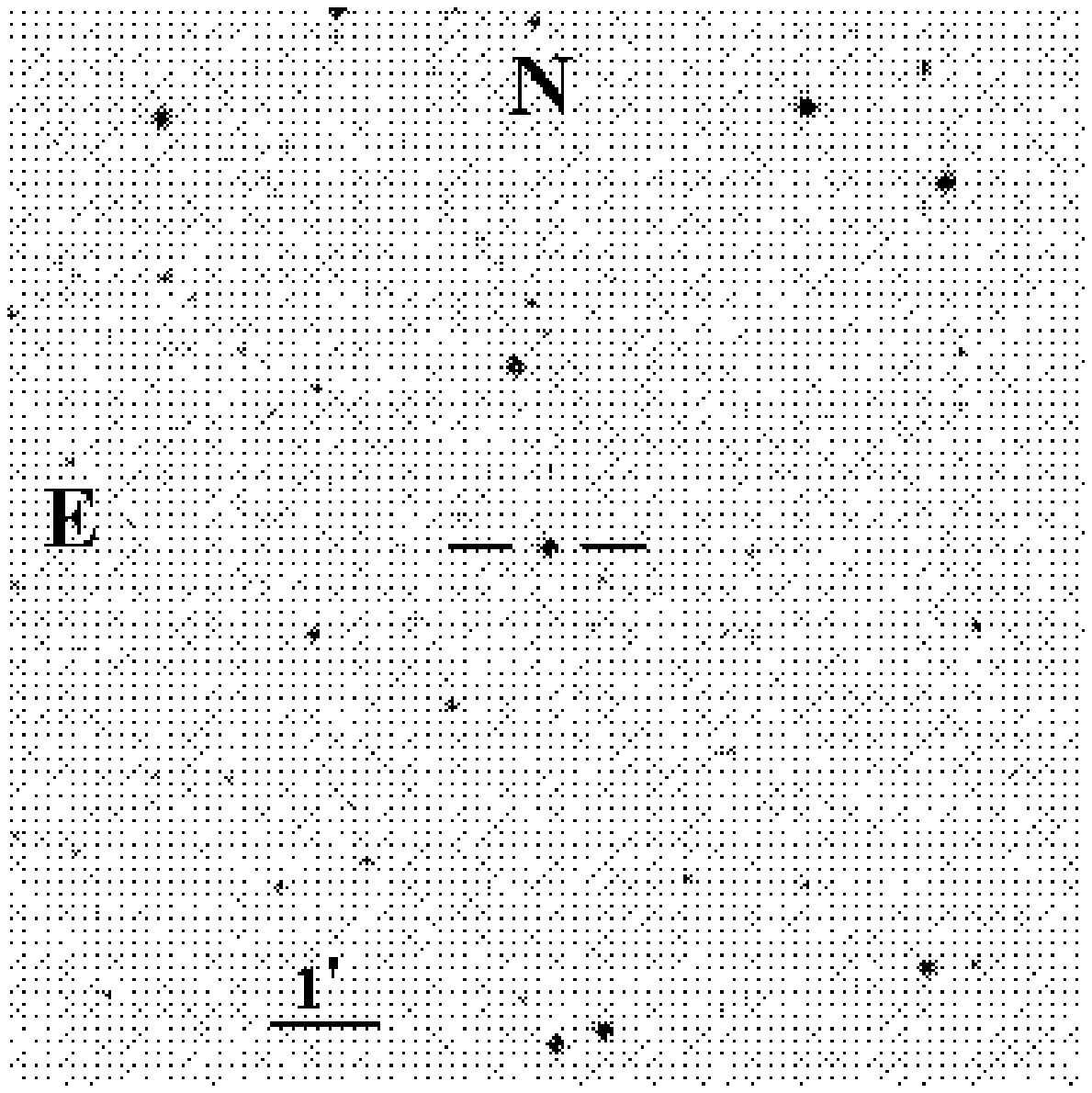}
\caption{Field, $10^{\prime}$ on a side, of the star JL~286.}
\label{fig:figure35}
\end{figure}

Isolated stars were measured on occasion as potential standard stars.  Several of them have just 
enough measures to be usable as standard stars, but a sequence never materialized in their 
vicinity.  Others have just too few data to be used as standards. However, their $UBVRI$ 
magnitudes and color indices may be of use for other purposes.  Final data for these stars are 
given in Table \ref{tab:table4}, where the column headings have the same explanation and their 
content the same basis as in Table \ref{tab:table1}.  Finding charts for these isolated stars 
may be found in Figures \ref{fig:figure28} - \ref{fig:figure46}.

The magnitude distribution of all the stars studied in this paper, that is, a combination of stars 
in Tables \ref{tab:table1} and \ref{tab:table4}, is plotted in Figure \ref{fig:figure47} in 0.25 
$V$ magnitude bins.  Figure \ref{fig:figure48} shows the $(B-V)$ color index range in 0.1 magnitude 
bins for all the stars studied in this paper.

Figures \ref{fig:figure49} - \ref{fig:figure55} have been plotted using data for the new standard 
stars in Table \ref{tab:table1}.  These figures show the mean error of a mean magnitude or color 
index, plotted as a function of magnitude or color index.

Figures \ref{fig:figure56} and \ref{fig:figure57} illustrate the $[(U-B), (B-V)]$ and $[(V-R), 
(R-I)]$ color-color plots for all stars measured in this paper, respectively, with stars from 
Table \ref{tab:table1} plotted as filled circles and stars from Table \ref{tab:table4} as open 
circles.

Cross-identifications are provided in Table \ref{tab:table5}.  On occasion the very best 
coordinate and proper motion information is needed for standard stars.  Hence, Table 
\ref{tab:table5} presents the most recent coordinates and proper motions for the new standard 
stars in Table \ref{tab:table1} and the isolated stars in Table \ref{tab:table4}.  All 
coordinates are for the epoch J2000.0. The 2MASS Point Source Catalog (PSC) positions come 
from The Two Micron All Sky Survey \citep[2MASS;][]{Skrutskie2006}.  The UCAC2 positions come 
from The Second USNO CCD Astrograph Catalog \citep{Zacharias2004}.  Spectral types are given 
where available.

\section{Comments on Individual Stars}
\label{sec:starcomments}

The numbering system for the stars labeled with LB (Luyten Blue) is described in 
\citet{Luyten1956}.

The numbering system for the stars labeled with JL arises from the observations of 
\citet{JaideeLynga1969}.  These authors provide coordinates and star charts resulting from their 
search for `faint violet stars.'

The LSE numbering system originated with \citet{Drilling1983}.

The LSS numbering system originated with the publication of the catalog Luminous Stars in the 
Southern Milky Way \citep{StephensonSanduleak1971}.

The MCT stars (Montreal-Cambridge-Tololo) initially were discussed by \citet{Demers1986} and 
\citet{Demers1987}.  CCD sequences used to calibrate the MCT~fields appeared in 
\citet{Demers1993a, Demers1993b}.  An initial set of photometry and charts for a selected 
subset of the MCT stars themselves appeared in \citet{Lamontagne2000}.  However, a final 
summary paper listing all the MCT stars discovered in the survey has not appeared, because a 
large number of these stars lack spectroscopic data \citep{Demers2007, Lamontagne2007}.

The nomenclature for the EC stars was defined in the Edinburgh-Cape survey \citep{Kilkenny1991}.

The nomenclature GJ pertains to stars in the \citet{GlieseJahreiss1979} catalog of nearby 
stars.

The {\it Hubble Space Telescope} Guide Star Catalog (GSC) acronym first appeared in 
\citet{Lasker1990}.

The NVS terminology began with the New Catalog of Suspected Variable Stars 
\citep{KukarkinKholopov1982}.  One most easily now can access variable and suspected variable star 
information by entering the Sternberg Astronomical Institute's webpage at {\url 
{http://www.sai.msu.edu}}, and clicking on the ``GCVS Research Group" (General Catalog of 
Variable Stars), and then going to the appropriate catalog.

The WD numbering system exists for white dwarf stars.  Excellent online sources of information for 
white dwarf stars include Jay Holberg's website at {\url {http://procyon.lpl.arizona.edu/WD/}} and 
G. McCook and E. Sion's website at {\url 
{http://www.astronomy.villanova.edu/WDCatalog/index.html}}.  A print description of the latter is 
in \citet{McCookSion1999}.

Expanded comments, in the sense of increasing right ascension, follow for individual stars 
which appear in both Tables \ref{tab:table1} and \ref{tab:table4}:

\noindent{JL~163B [= USNO-B1.0~0397-0001111 = USNO-A2.0~0375-00050408 = GSC2.2~S0120011284]}

\noindent{JL~163 [= GSC~08028-00524]}

\noindent{JL~166 [= GSC~08022-01020]}

\noindent{T$\,$Phe~B [= RW~Phe = AN~409.1929 = CD-47~128 = GSC~08024-00363] See note 1, Table 2 
in \citet{Landolt1992}.  Discovered by \citet{Dartayet1929}.  T$\,$Phe~B is sequence star ``h" 
for variable star T$\,$Phe in \citet{FlemingPickering1907} and \citet{CampbellPickering1913}.}

\noindent{T$\,$Phe~F [= GSC~08024-00830 = NSV~184] See comment by \citet{Dartayet1929} that the 
T$\,$Phe sequence star ``h$^1$" might be variable.  The sequence is in 
\citet{FlemingPickering1907} and \citet{CampbellPickering1913}.  The AAVSO (d) chart for 
T$\,$Phe (002546), plotted at a scale of $20^{\prime\prime}=1mm$, is based on 
\citet{FlemingPickering1907} and \citet{CampbellPickering1913}.  T$\,$Phe~F here is the star 
marked as 132 on the AAVSO chart.}

\noindent{JL~194 [= CD-48~106 = LB~1559 = GSC~08024-00844 = HIP~2499] Found to be a blue star by 
\citet{LuytenAnderson1958}.  The {\it Hipparcos} Catalog flagged JL~194 as a variable star, 
indicating a variability range between 0.06 and 0.60 magnitude. The flag indicates that the 
photometry quoted was ground-based.  The photometry of \citet{HillHill1966} differs from that 
herein (with their $V=12.36$), but one could say that their and the current photometry agrees, 
given Hill and Hill's quoted errors of 0.04 magnitude in $V$ and 0.03 magnitude in the color 
indices.  A more concordant value of $V=12.41$ is given in \citet{KilkennyHill1975}.  A ``near 
$V$" magnitude of 12.33 is quoted by \citet{Newell1973}.  \citet{NewellGraham1976} present a 
Str\"{o}mgren $y$ magnitude of 12.45.  $UBV$ photometry of \citet{Wegner1980} essentially agrees 
with the current results.  After a review of the literature, then, and given that the current 
photometry was obtained over a two year interval, JL~194 appears to be constant in brightness.}

\noindent{JL~202 [= CPD-55~142 = LB~1566 = GSC~08469-00387] Found to be a blue star by 
\citet{LuytenAnderson1958}.  A search for nebular emission around JL~202 \citep{Mendez1988} 
found none.  Nevertheless, \citet{Kohoutek1997} indicated that JL~202 might be a possible 
post-planetary nebulae, citing only Mendez et al. Photometry from \citet{HillHill1966} and from 
\citet{Wegner1980} bracket the new photometry herein, thereby indicating the likelihood of long 
term constancy in brightness for JL~202.}

\noindent{GD~679 [BPM~46934 = BPS~CS~30316-0011 = GSC~06999-02080 = WD~0104-33] Appeared in the 
``Bruce Proper Motion Survey" \citep{Luyten1963}.  Identified as a white dwarf suspect 
\citep{Giclas1972}. Appears in \citet{Lamontagne2000} as MCT~0104-3336.}

\noindent{JL~236 [= GSC~08474-00031]}

\noindent{JL~261 [= LB~3229 = GSC~08047-00420] Initially cataloged as a blue star by 
\citet{LuytenAnderson1959}.  \citet{Wegner1980} has photometry in fair agreement, considering his 
quoted errors, compared with the new photometry herein.  However, the photometry herein consists 
only of one measurement.}

\noindent{LB~3241 [= GSC~08045-00656 = JL~285 = NSV~759] Identified as a faint blue star by 
\citet{LuytenAnderson1959}.  The first measured proper motion for LB~3241 was recorded by 
\citet{Luyten1962}.  \citet{KilkennyHill1975} found a variation of 0.12~mag in $V$.}

\noindent{JL~286 [= GSC~08048-00322]}

\noindent{LB~1735 [= EC~04300-5341 = USNO-A1.0~0300-01365856] Identified as a faint blue star by 
\citet{LuytenAnderson1958}.  The first measured proper motion for LB~1735 was recorded by 
\citet{Luyten1962}.  \citet{Beers2001} classify LB~1735 as a horizontal-branch B-type (HBB) star.  
They report $UBV$ photometry of $V=13.66$, $(B-V)=-0.17$, and $(U-B)=-0.59$, with errors of two to 
three percent.}

\noindent{LB~1735~E  The mean error of a single observation in $V$ is 0.038, a bit high.  The 
star may bear watching for possible variability.}

\begin{figure}
\plotone{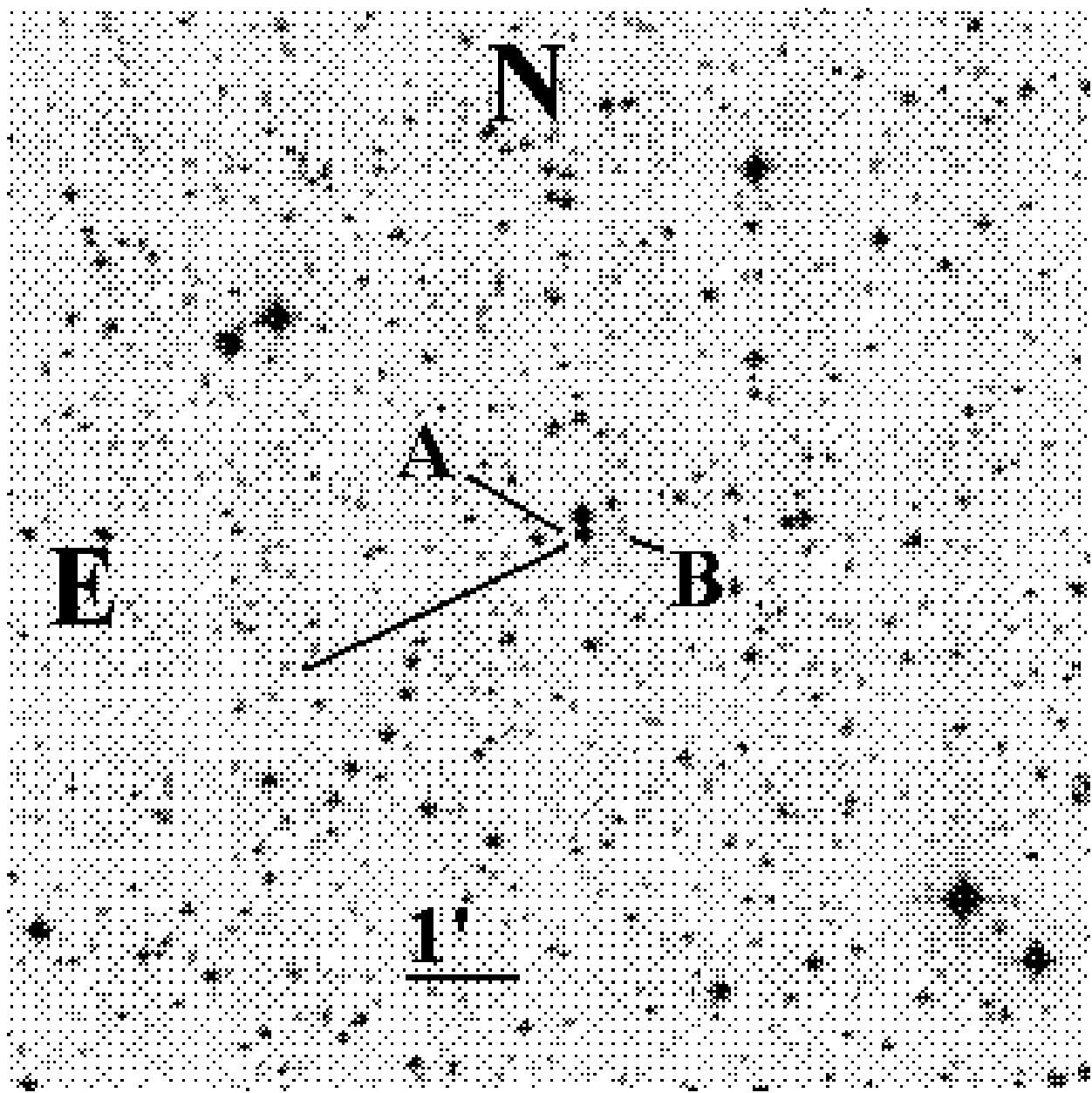}
\caption{Field, $10^{\prime}$ on a side, of the star L745-46A (see text, Section 4).}
\label{fig:figure36}
\end{figure}

\begin{figure}
\plotone{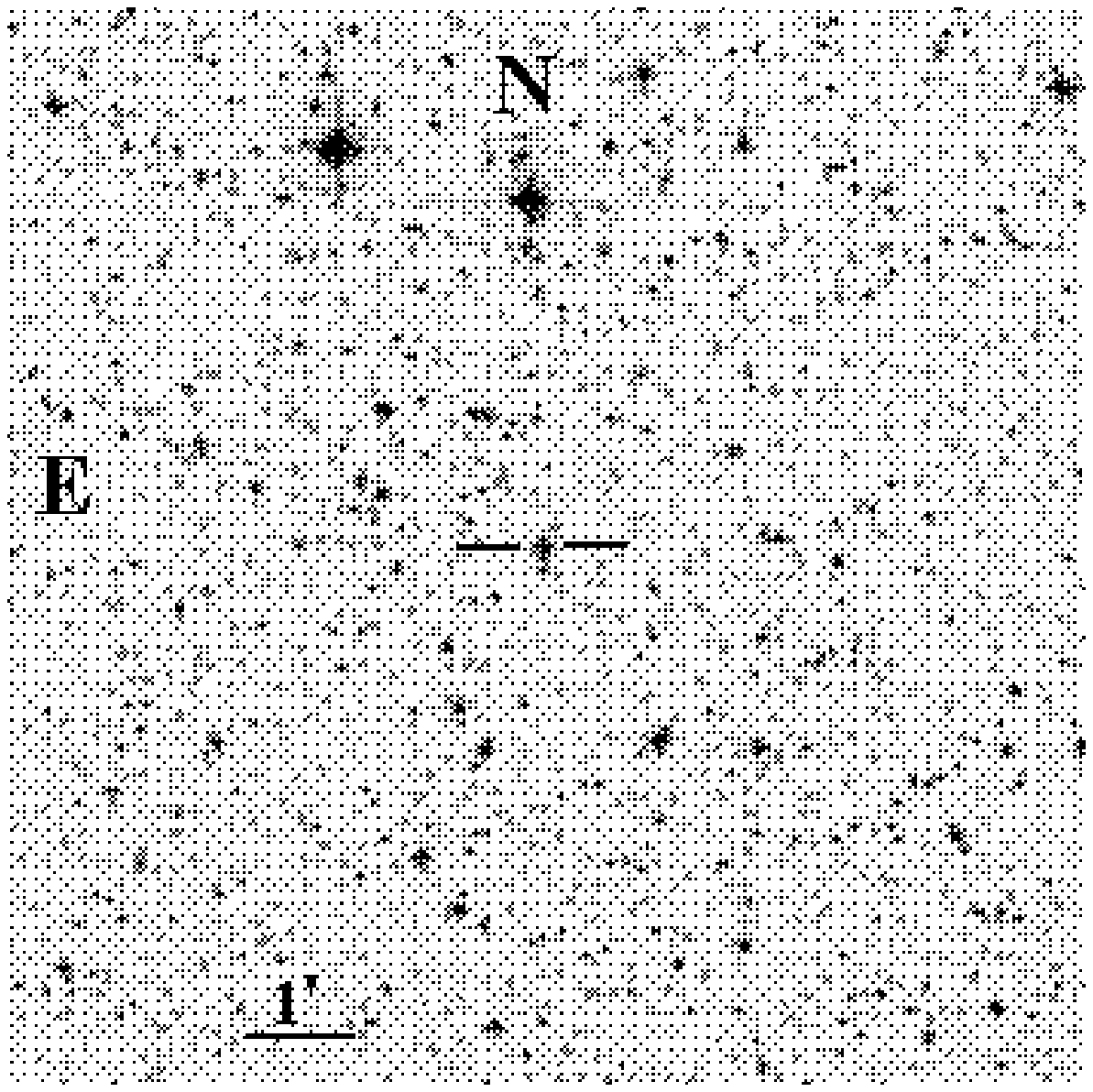}
\caption{Field, $10^{\prime}$ on a side, of the star LSS~1274.}
\label{fig:figure37}
\end{figure}

\begin{figure}
\plotone{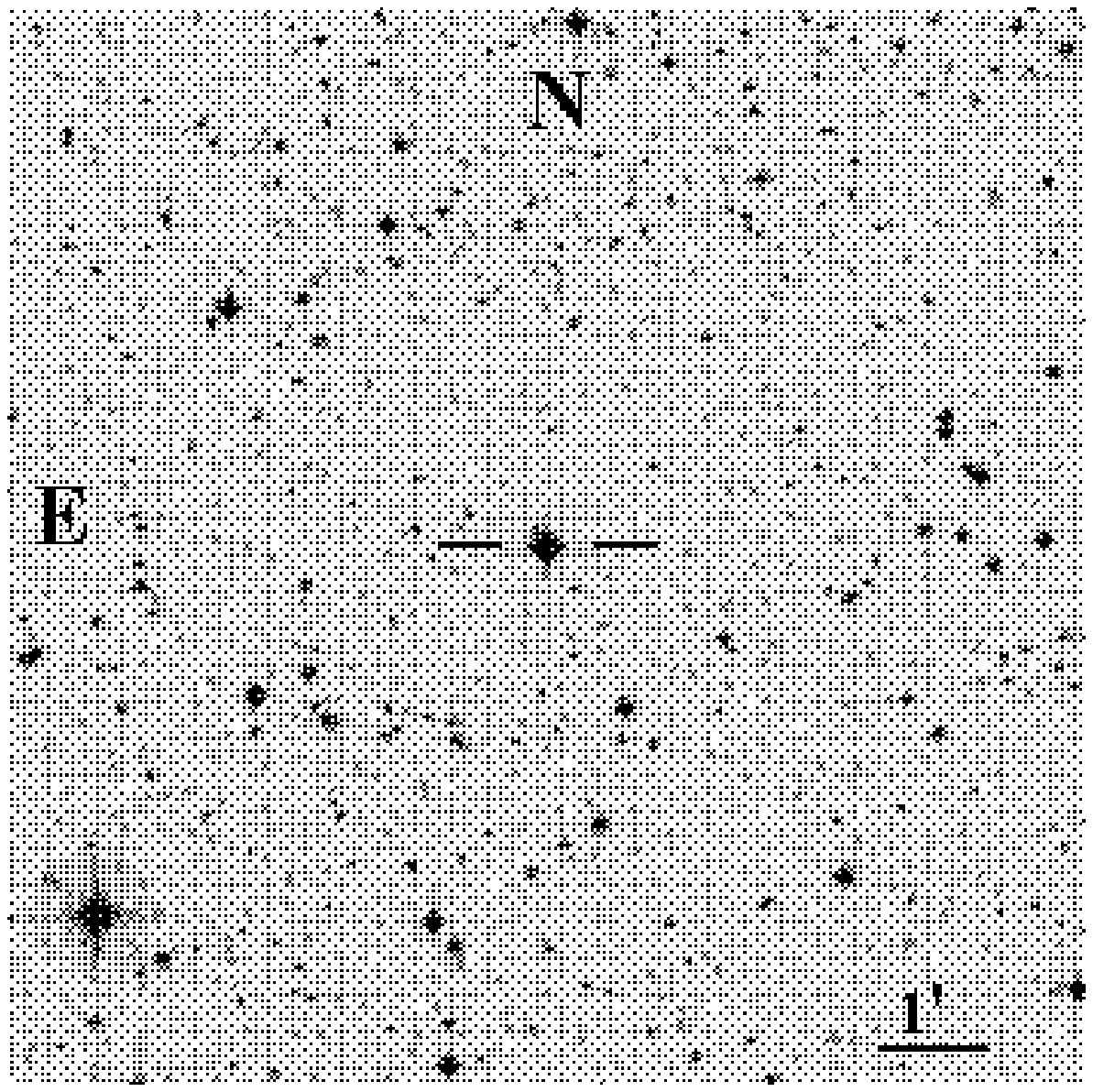}
\caption{Field, $10^{\prime}$ on a side, of the star LSS~1275.}
\label{fig:figure38}
\end{figure}

\begin{figure}
\plotone{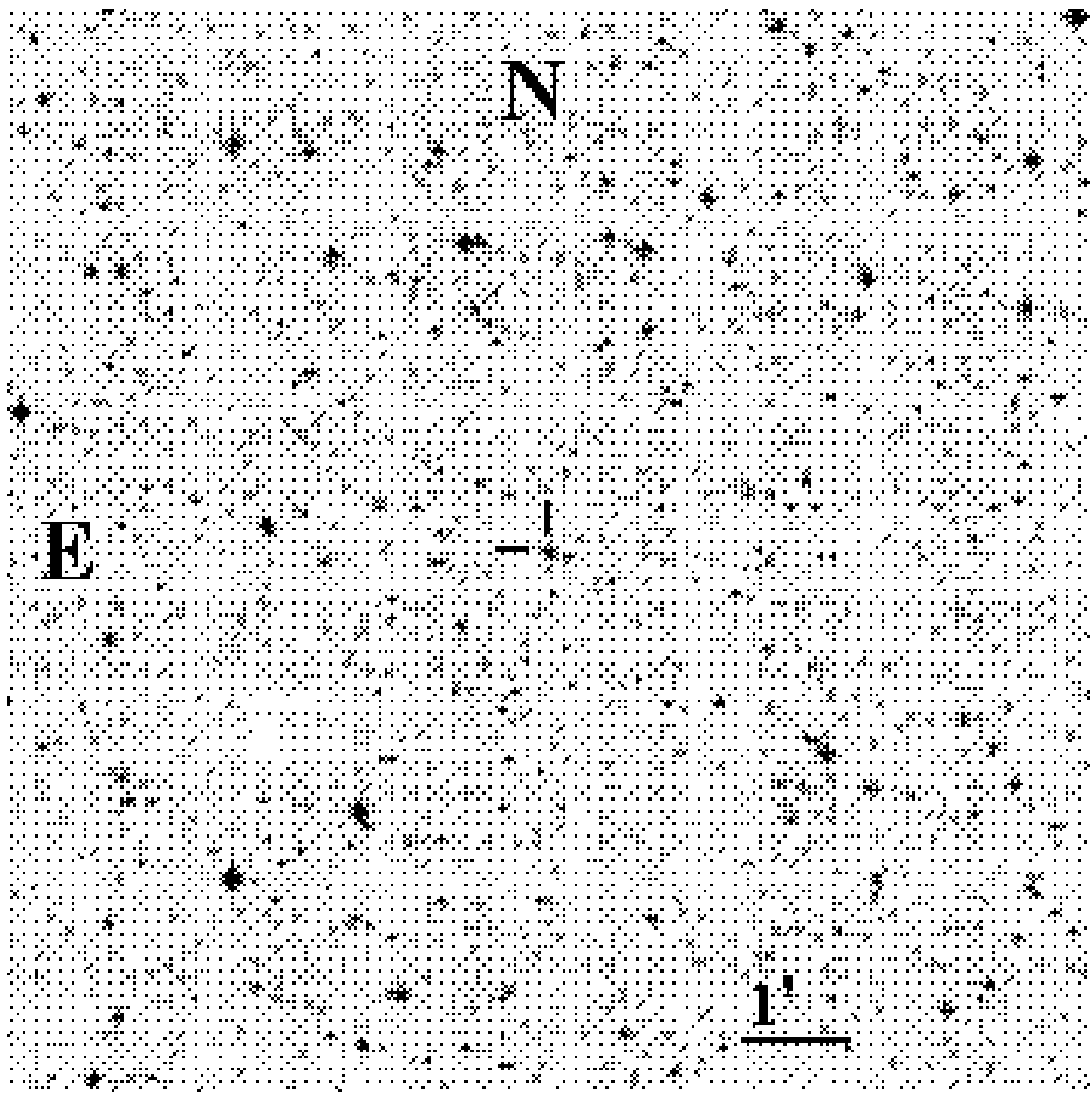}
\caption{Field, $10^{\prime}$ on a side, of the star LSS~1349.}
\label{fig:figure39}
\end{figure}

\begin{figure}
\plotone{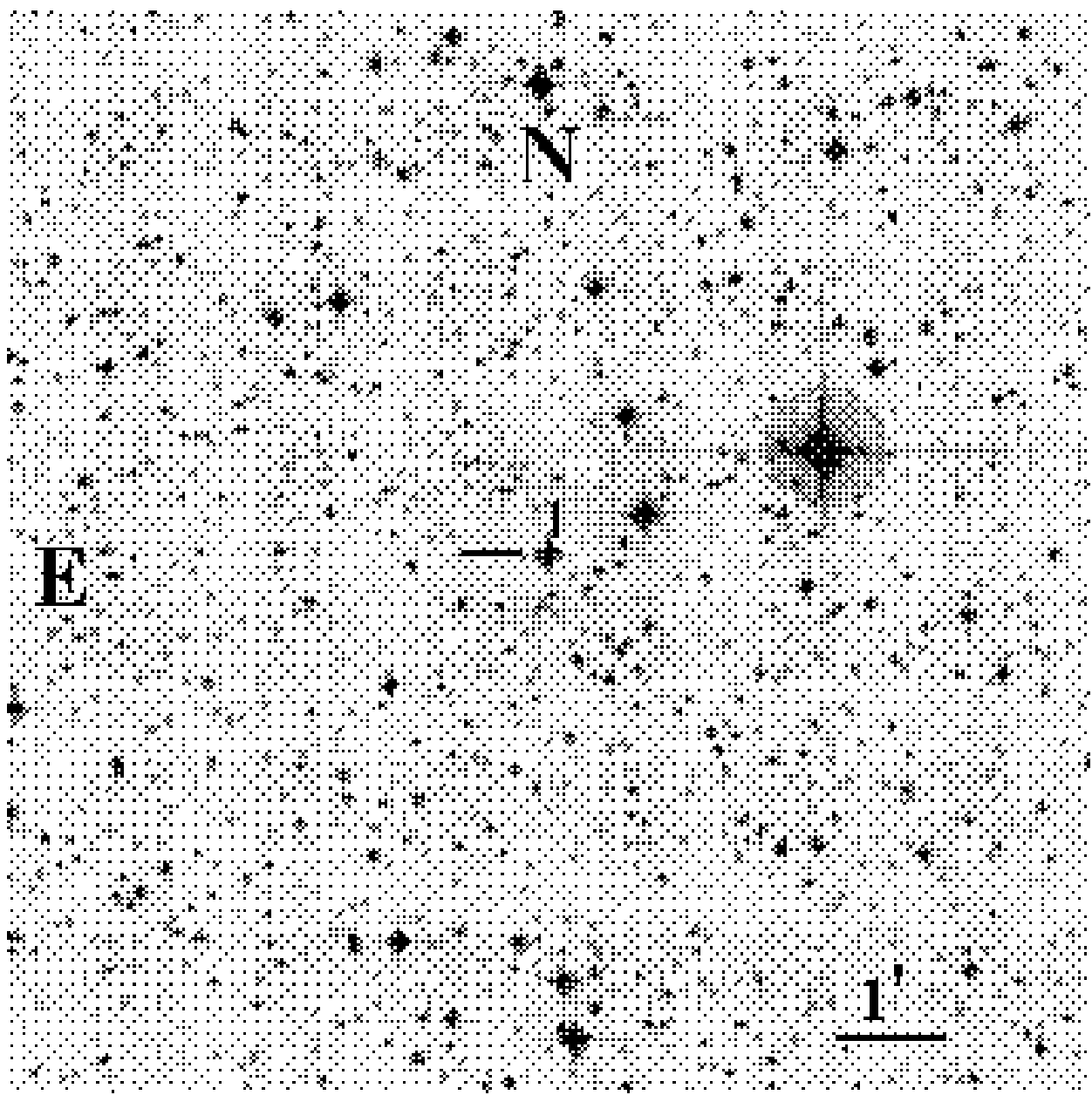}
\caption{Field, $10^{\prime}$ on a side, of the star LSS~1362.}
\label{fig:figure40}
\end{figure}

\begin{figure}
\plotone{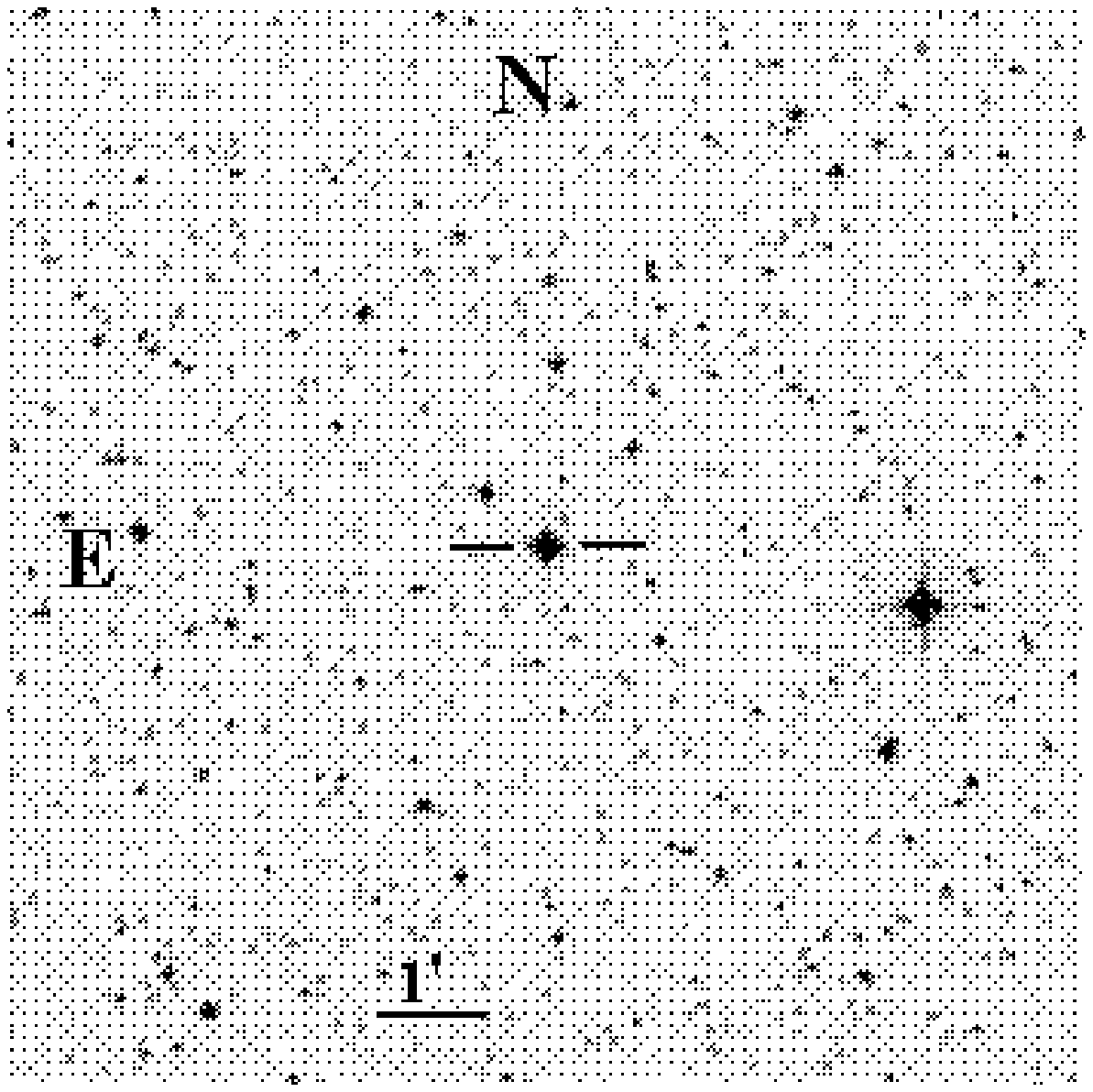}
\caption{Field, $10^{\prime}$ on a side, of the star LSE~153.}
\label{fig:figure41}
\end{figure}

\begin{figure}
\plotone{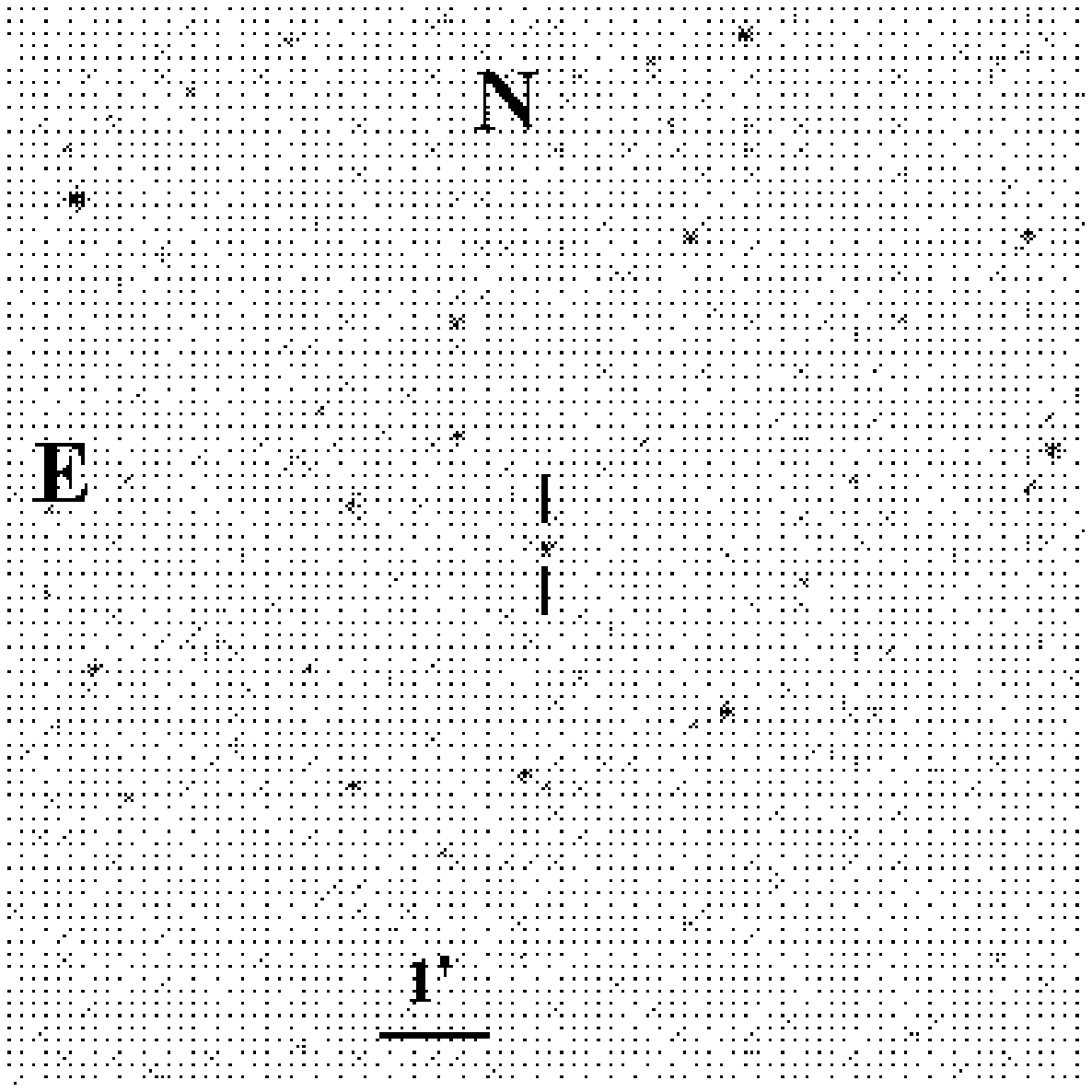}
\caption{Field, $10^{\prime}$ on a side, of the star LSE~125.}
\label{fig:figure42}
\end{figure}

\begin{figure}
\plotone{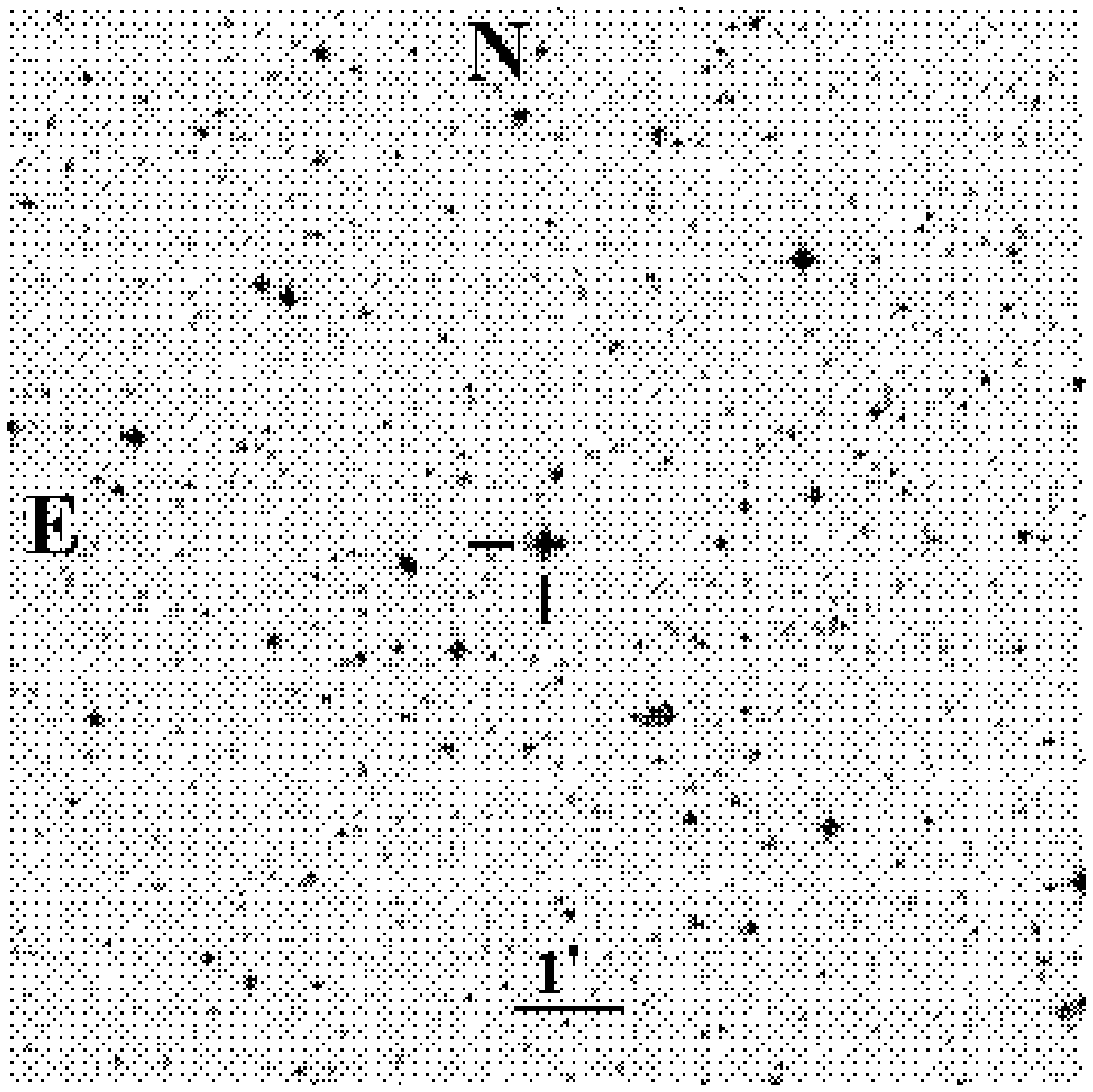}
\caption{Field, $10^{\prime}$ on a side, of the star LSE~234.}
\label{fig:figure43}
\end{figure}

\begin{figure}
\plotone{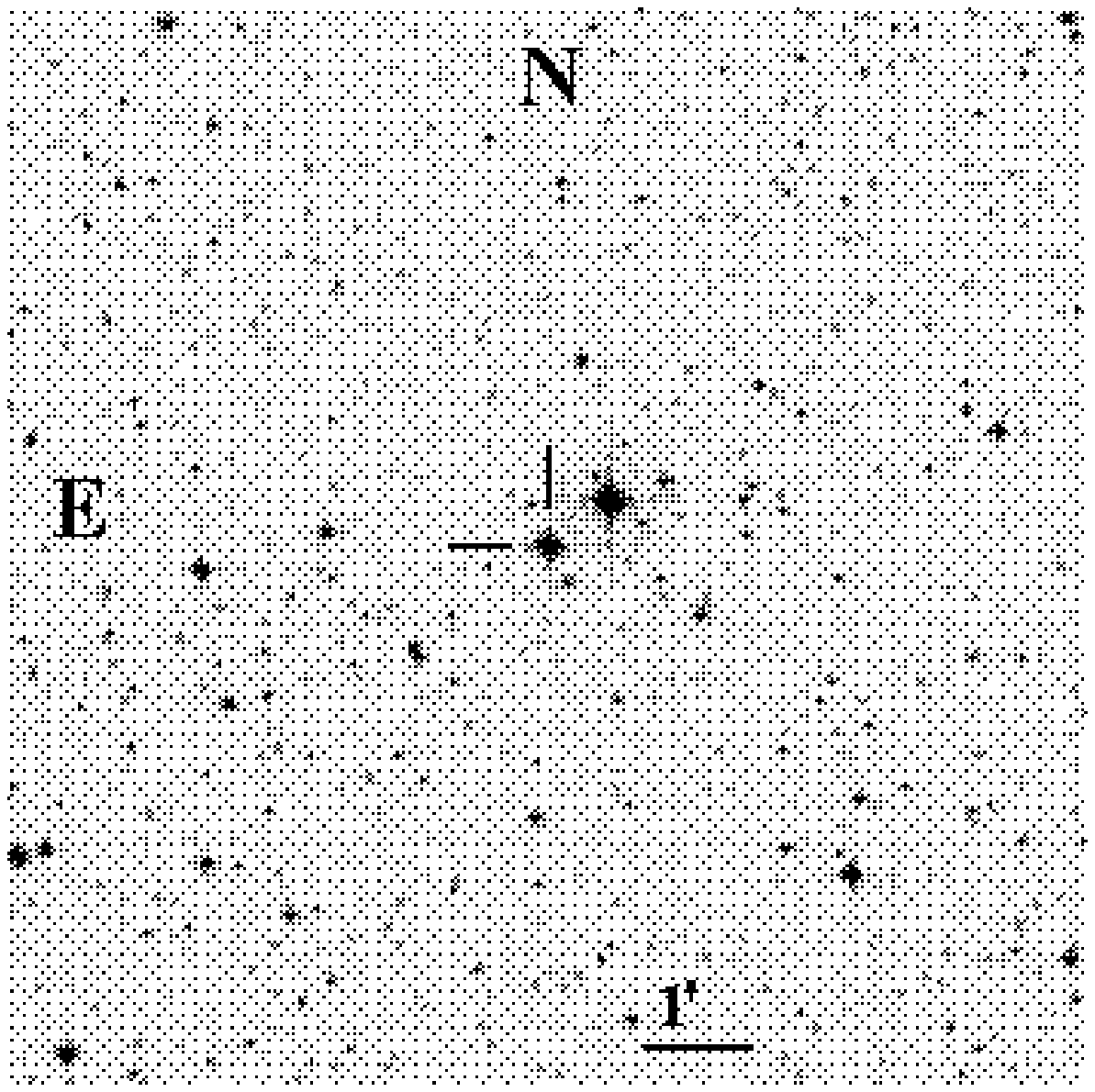}
\caption{Field, $10^{\prime}$ on a side, of the star LSE~263.}
\label{fig:figure44}
\end{figure}

\begin{figure}
\plotone{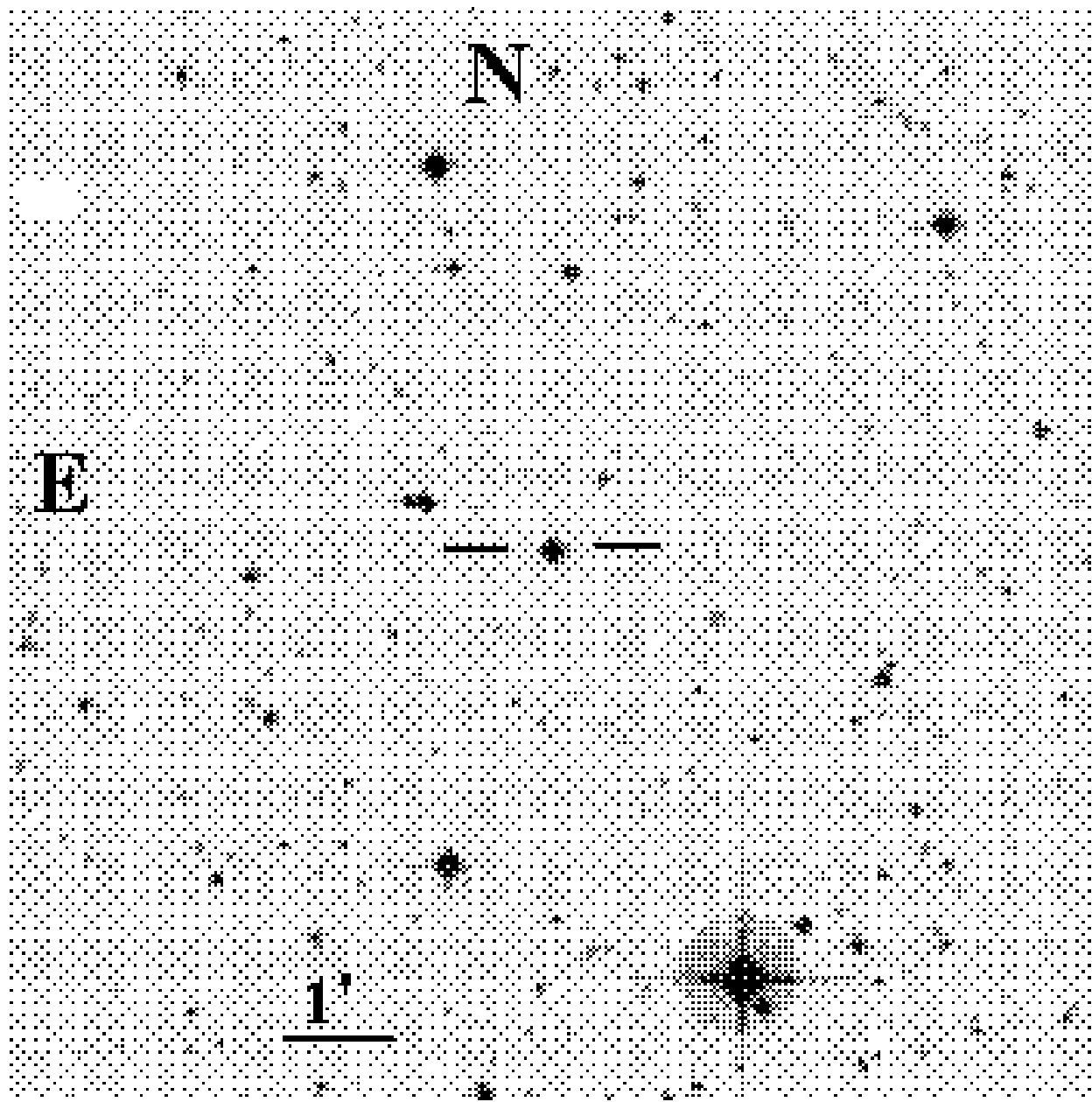}
\caption{Field, $10^{\prime}$ on a side, of the star JL~25.}
\label{fig:figure45}
\end{figure}

\begin{figure}
\plotone{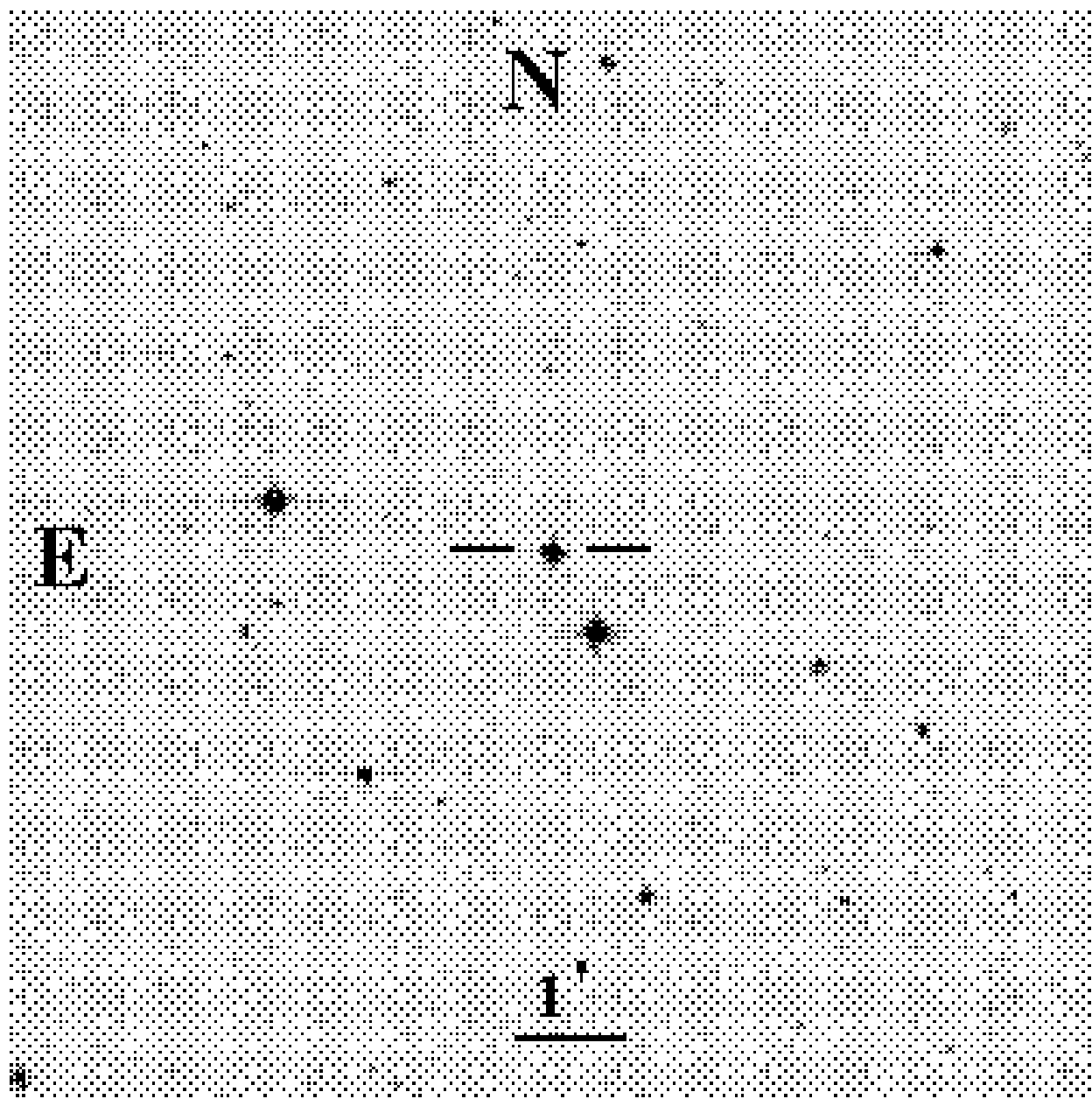}
\caption{Field, $10^{\prime}$ on a side, of the star LB~1516.}
\label{fig:figure46}
\end{figure}

\begin{figure}
\plotone{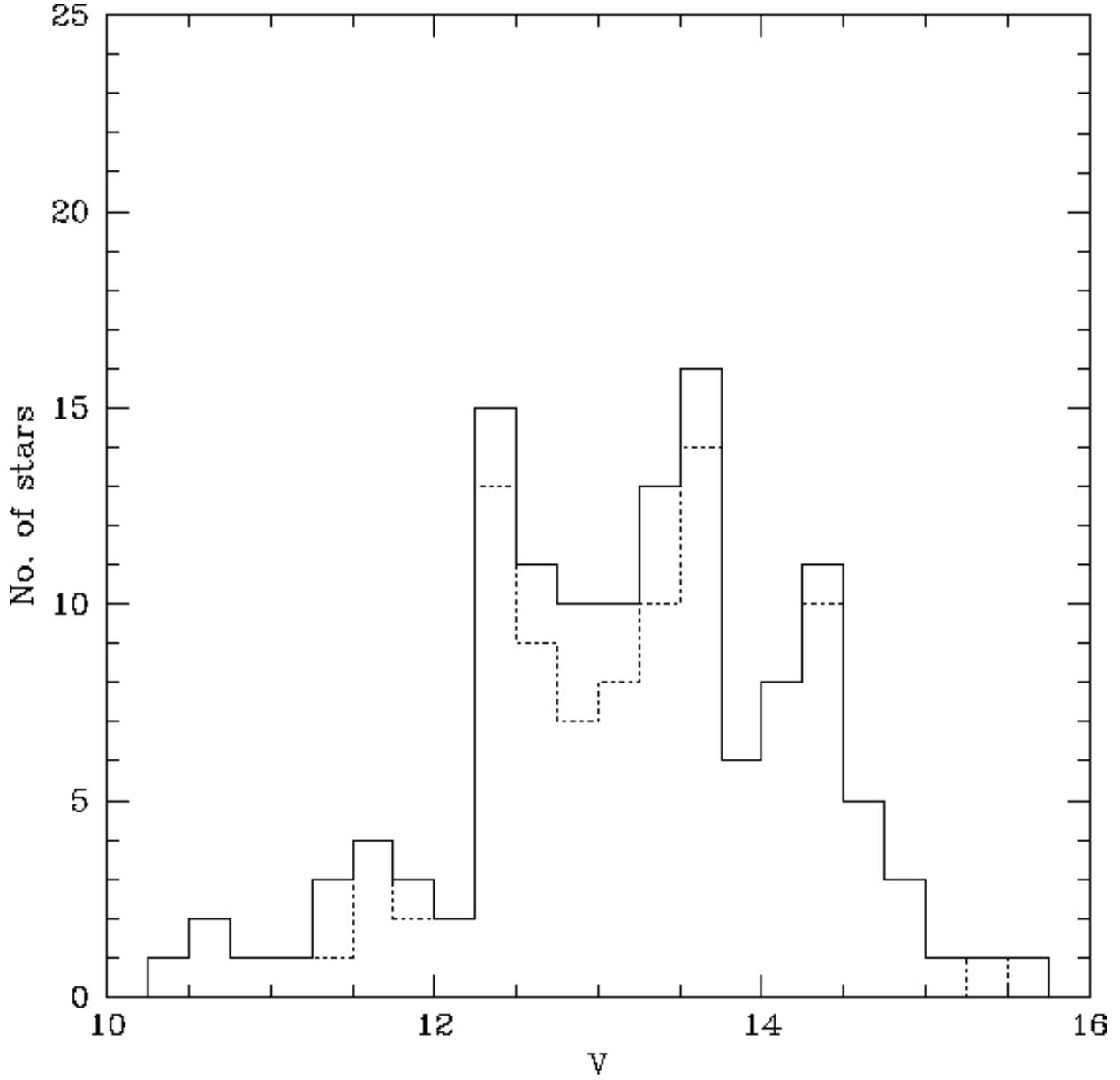}
\caption{The magnitude distribution for the new standards stars listed in Table \ref{tab:table1} 
($dotted~line$) and all stars in both Tables \ref{tab:table1} and \ref{tab:table4} ($solid~line$) 
in intervals of 0.25 $V$ mag.}
\label{fig:figure47}
\end{figure}

\begin{figure}
\plotone{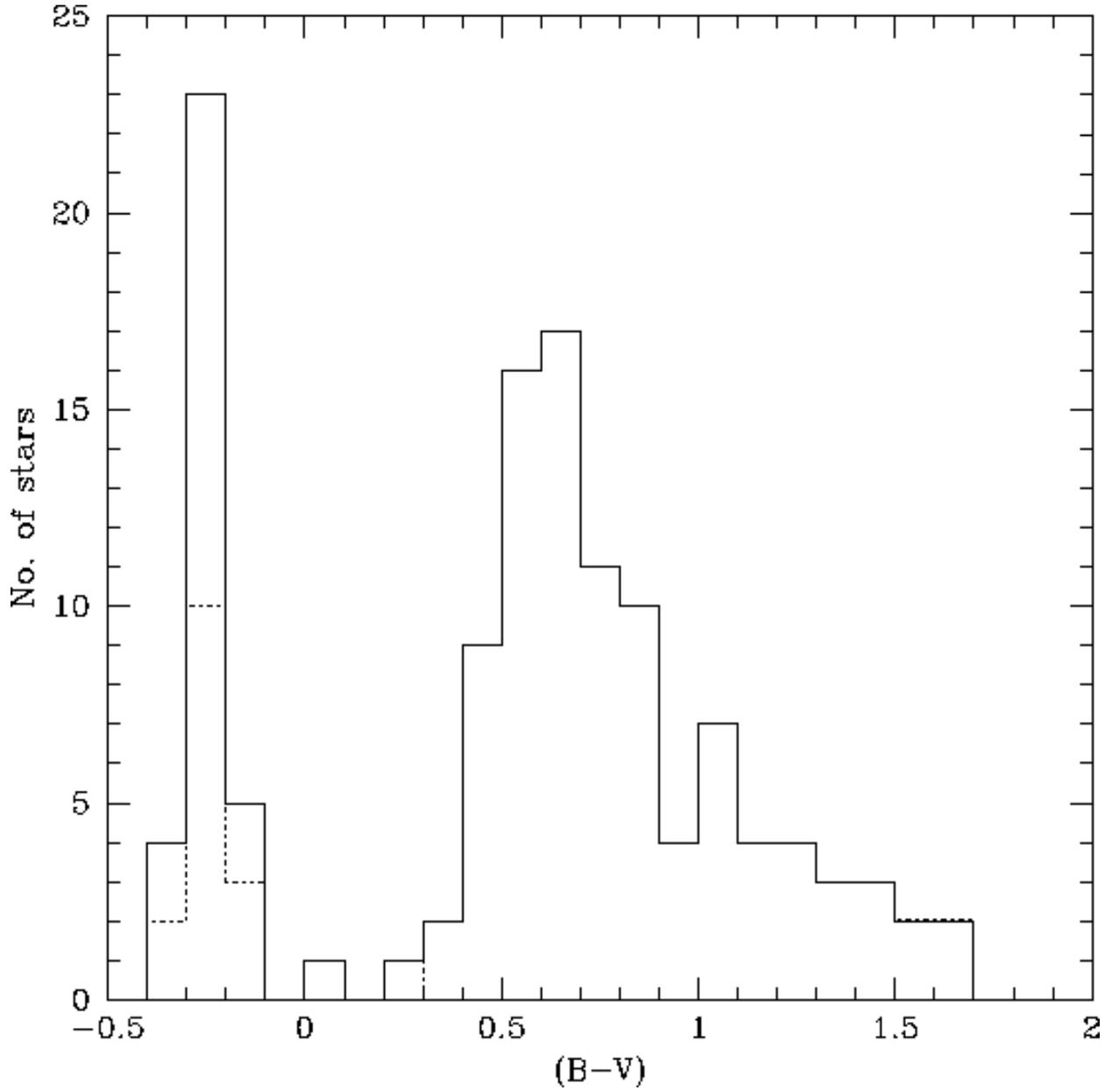}
\caption{The distribution in $(B-V)$ color index for the new standards stars listed in Table 
\ref{tab:table1} ($dotted~line$) and all stars in both Tables \ref{tab:table1} and \ref{tab:table4} 
($solid~line$) in intervals of 0.1 mag.}
\label{fig:figure48}
\end{figure}

\begin{figure}
\plotone{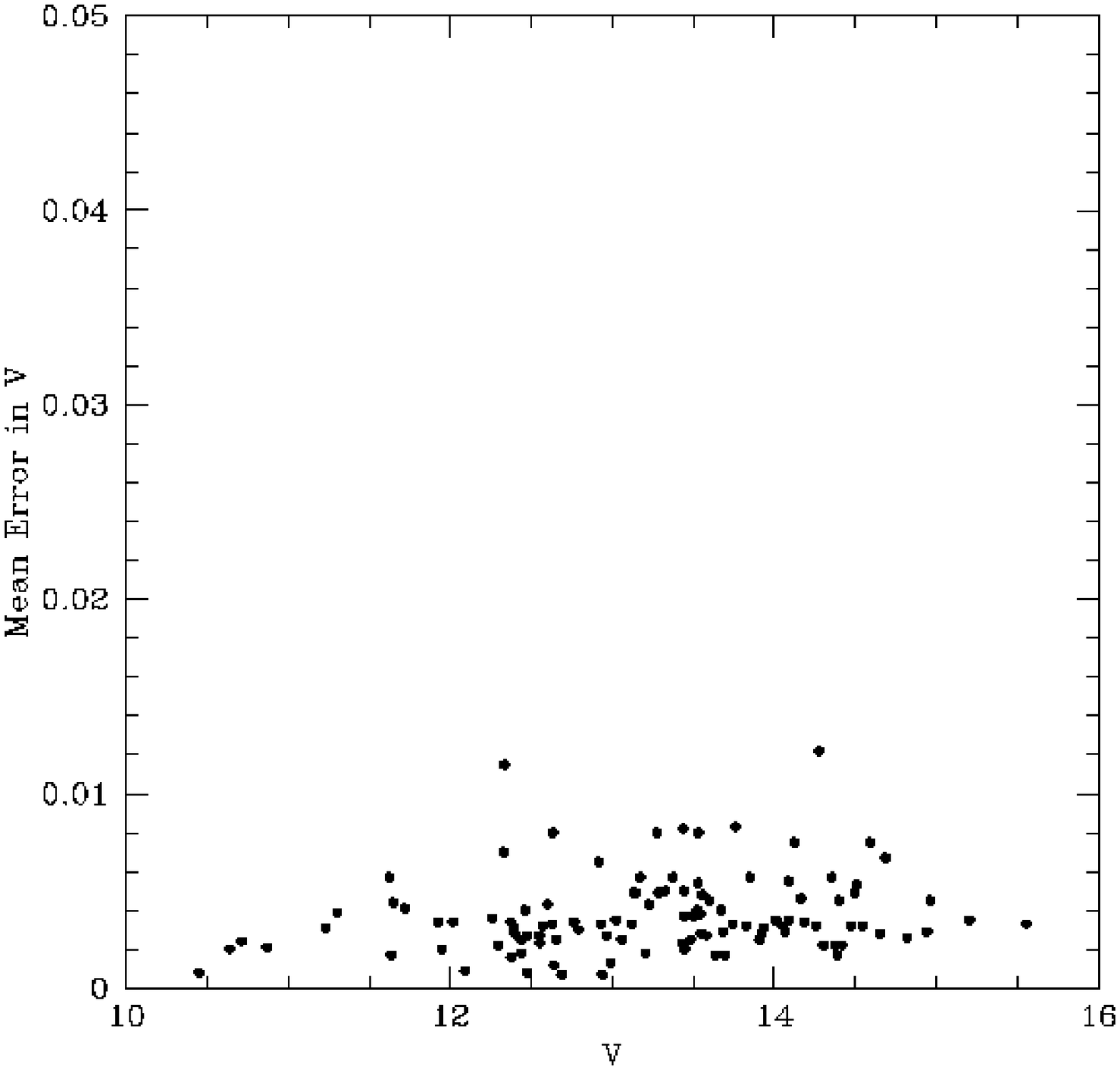}
\caption{The mean error of the mean of a single observation in $V$ for the new standard stars as a 
function of $V$.}
\label{fig:figure49}
\end{figure}

\begin{figure}
\plotone{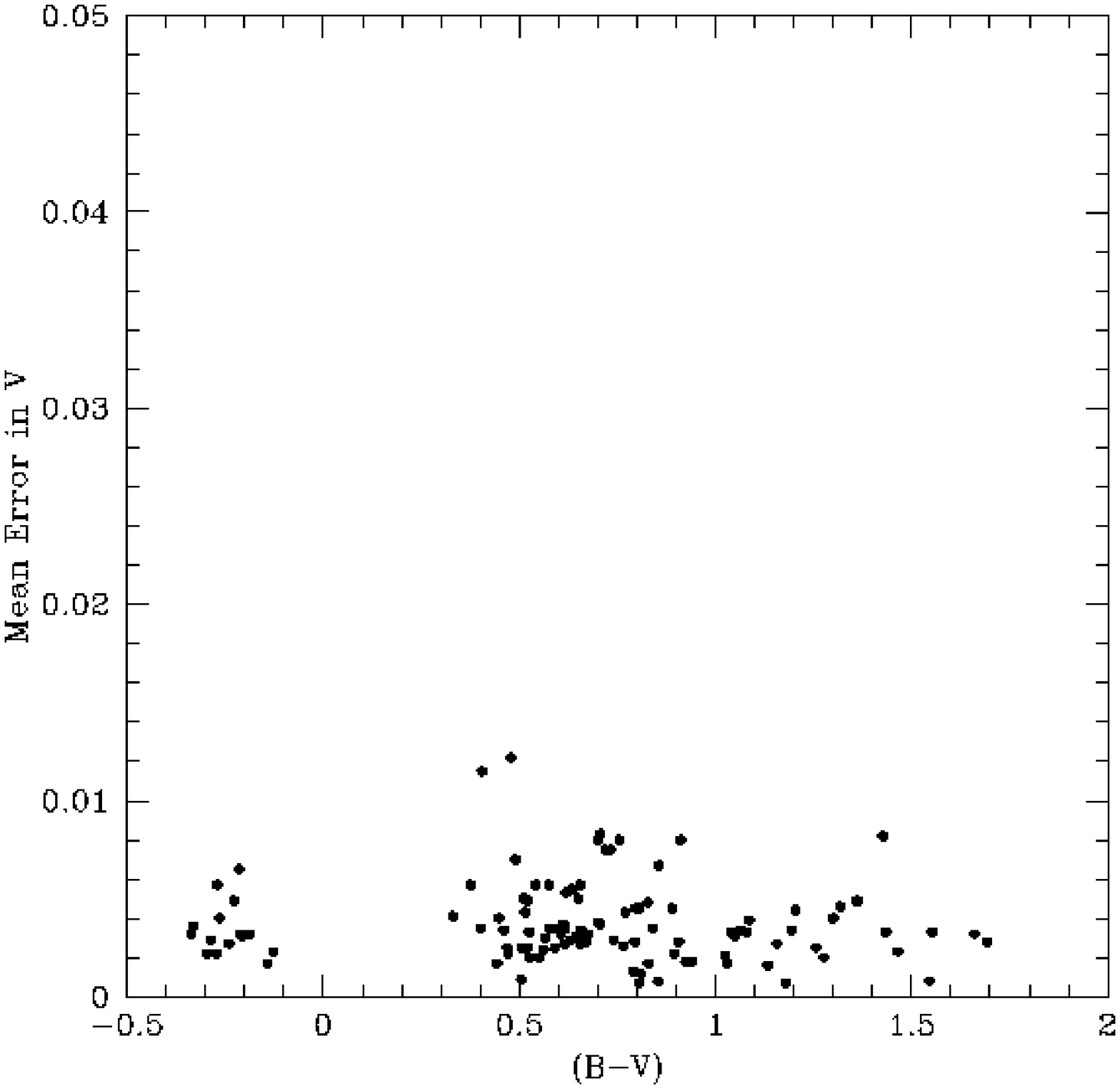}
\caption{The mean error of the mean of a single observation in $V$ for the new standard stars as a 
function of $(B-V)$.}
\label{fig:figure50}
\end{figure}

\begin{figure}
\plotone{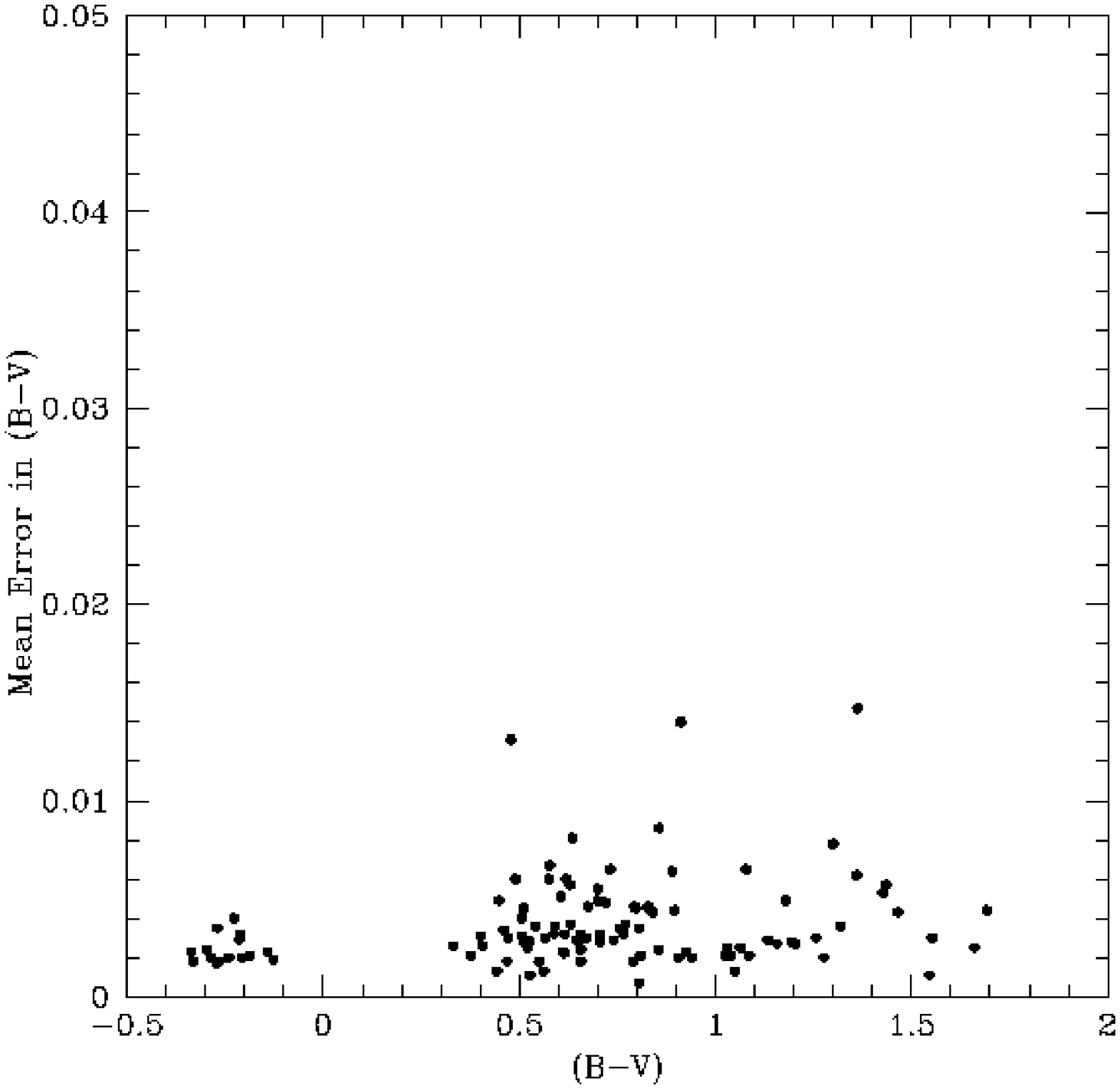}
\caption{The mean error of the mean of a single observation in $(B-V)$ for the new standard stars as a 
function of $(B-V)$.}
\label{fig:figure51}
\end{figure}

\begin{figure}
\plotone{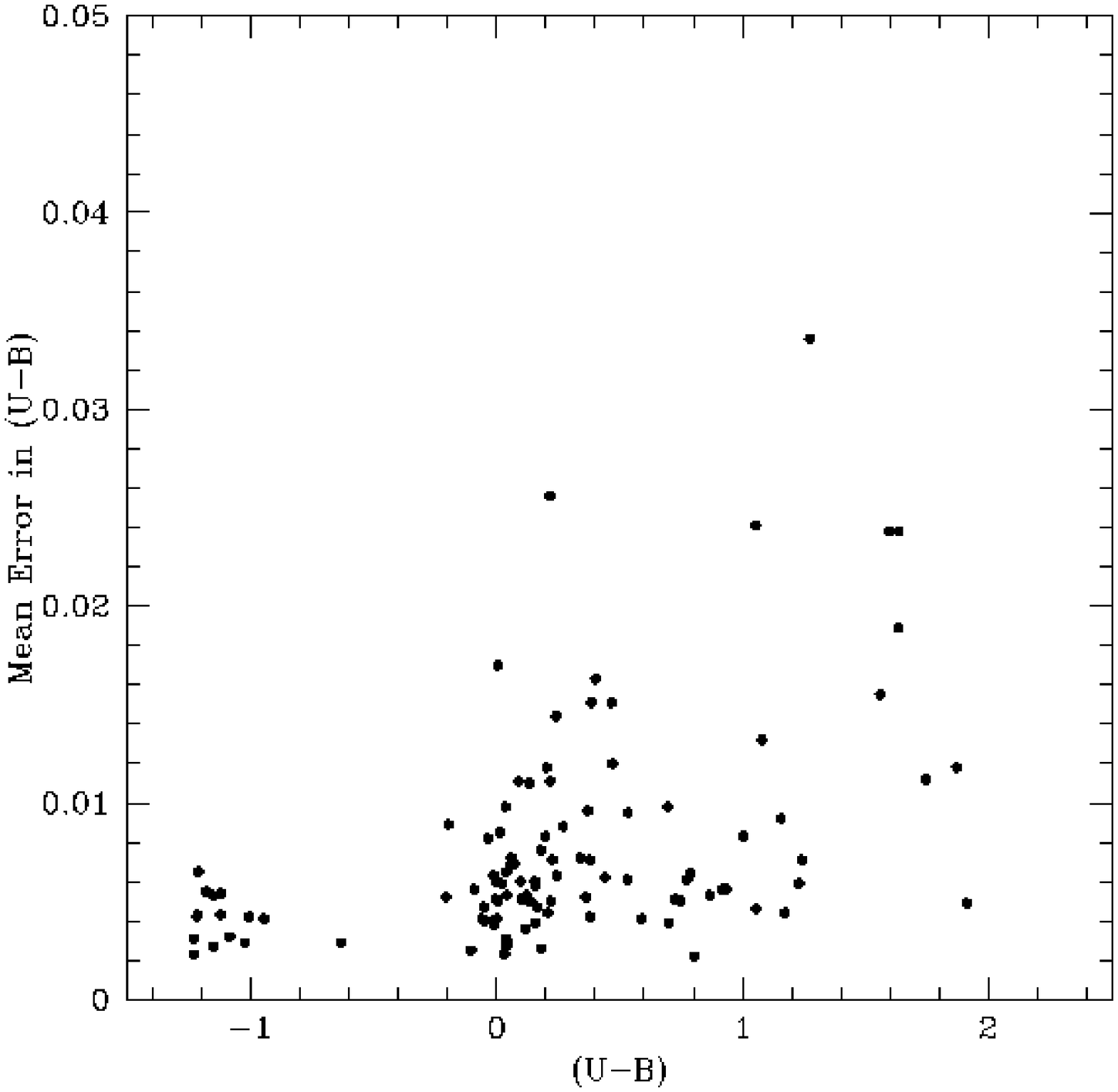}
\caption{The mean error of the mean of a single observation in $(U-B)$ for the new standard stars as a 
function of $(U-B)$.}
\label{fig:figure52}
\end{figure}

\begin{figure}
\plotone{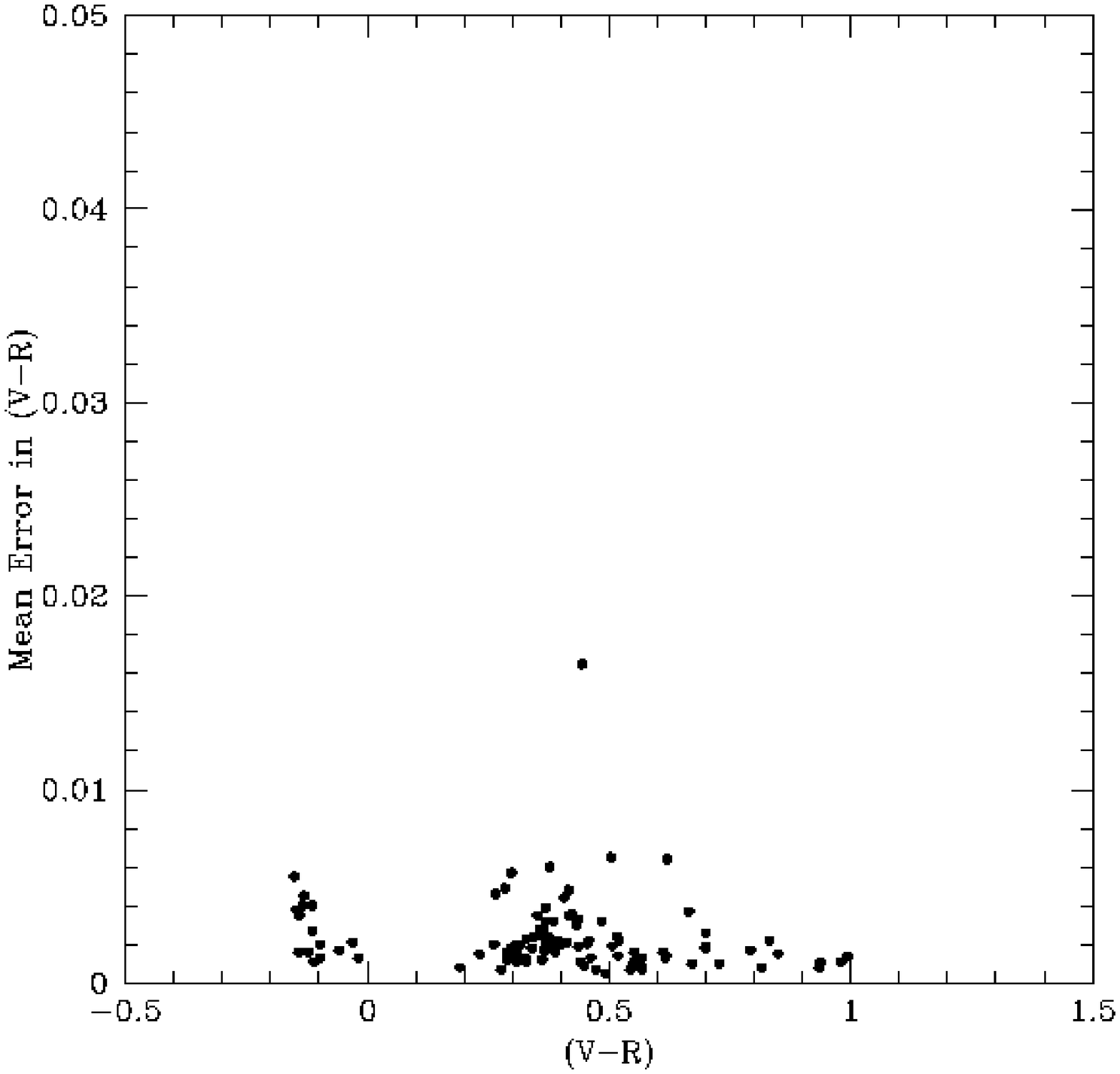}
\caption{The mean error of the mean of a single observation in $(V-R)$ for the new standard stars as a 
function of $(V-R)$.}
\label{fig:figure53}
\end{figure}

\begin{figure}
\plotone{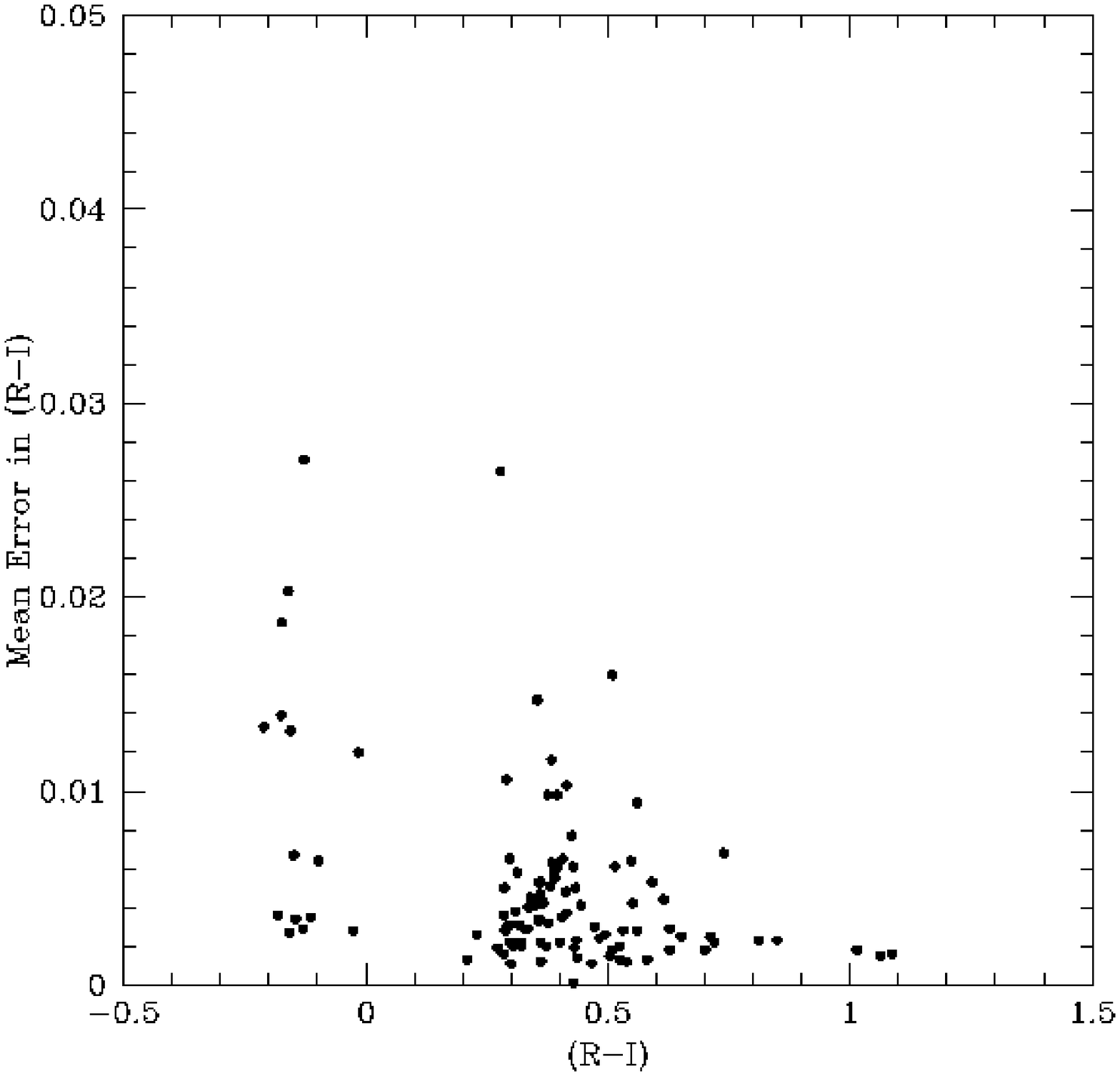}
\caption{The mean error of the mean of a single observation in $(R-I)$ for the new standard stars as a 
function of $(R-I)$.}
\label{fig:figure54}
\end{figure}

\begin{figure}
\plotone{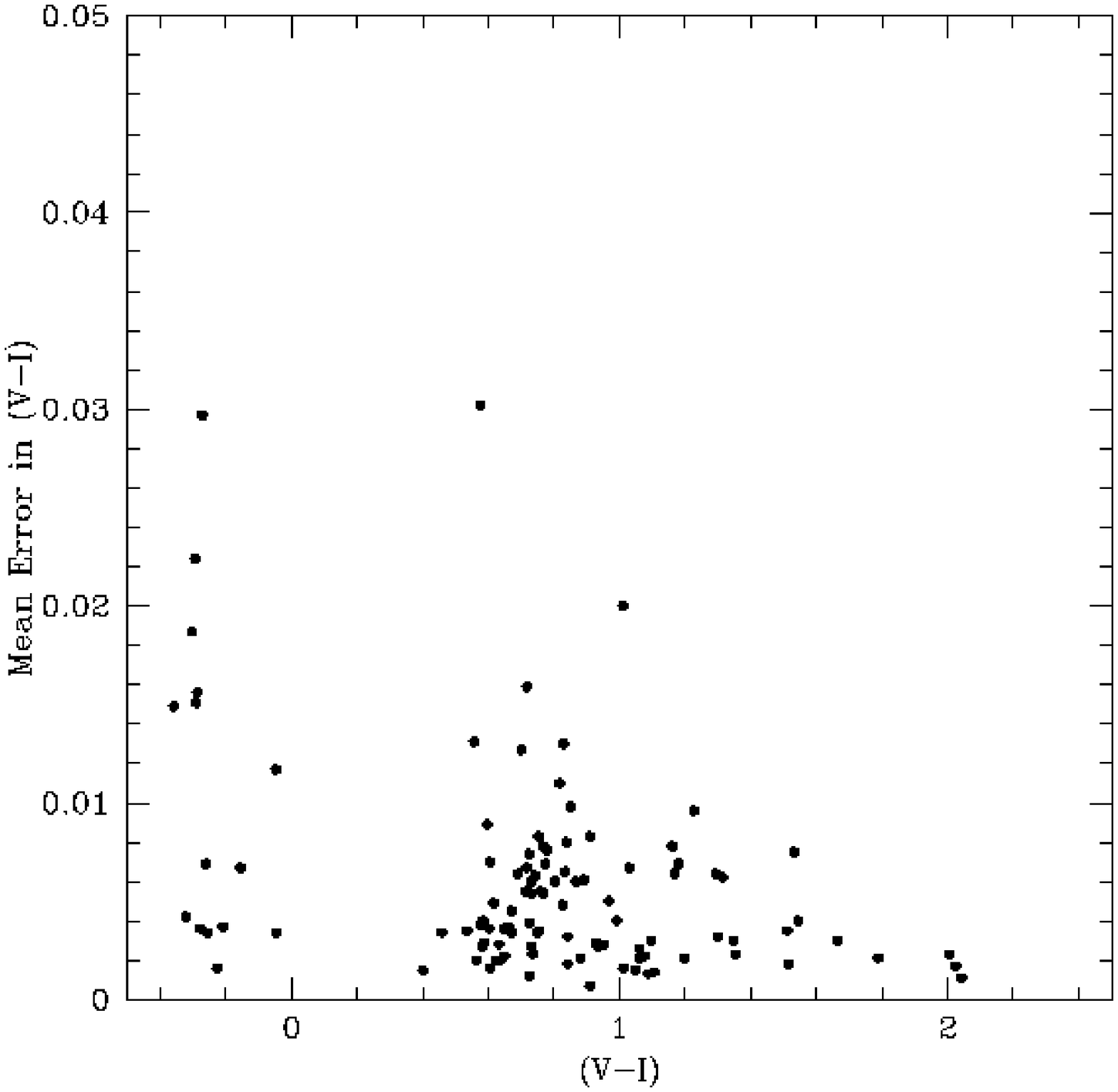}
\caption{The mean error of the mean of a single observation in $(V-I)$ for the new standard stars as a 
function of $(V-I)$.}
\label{fig:figure55}
\end{figure}

\begin{figure}
\plotone{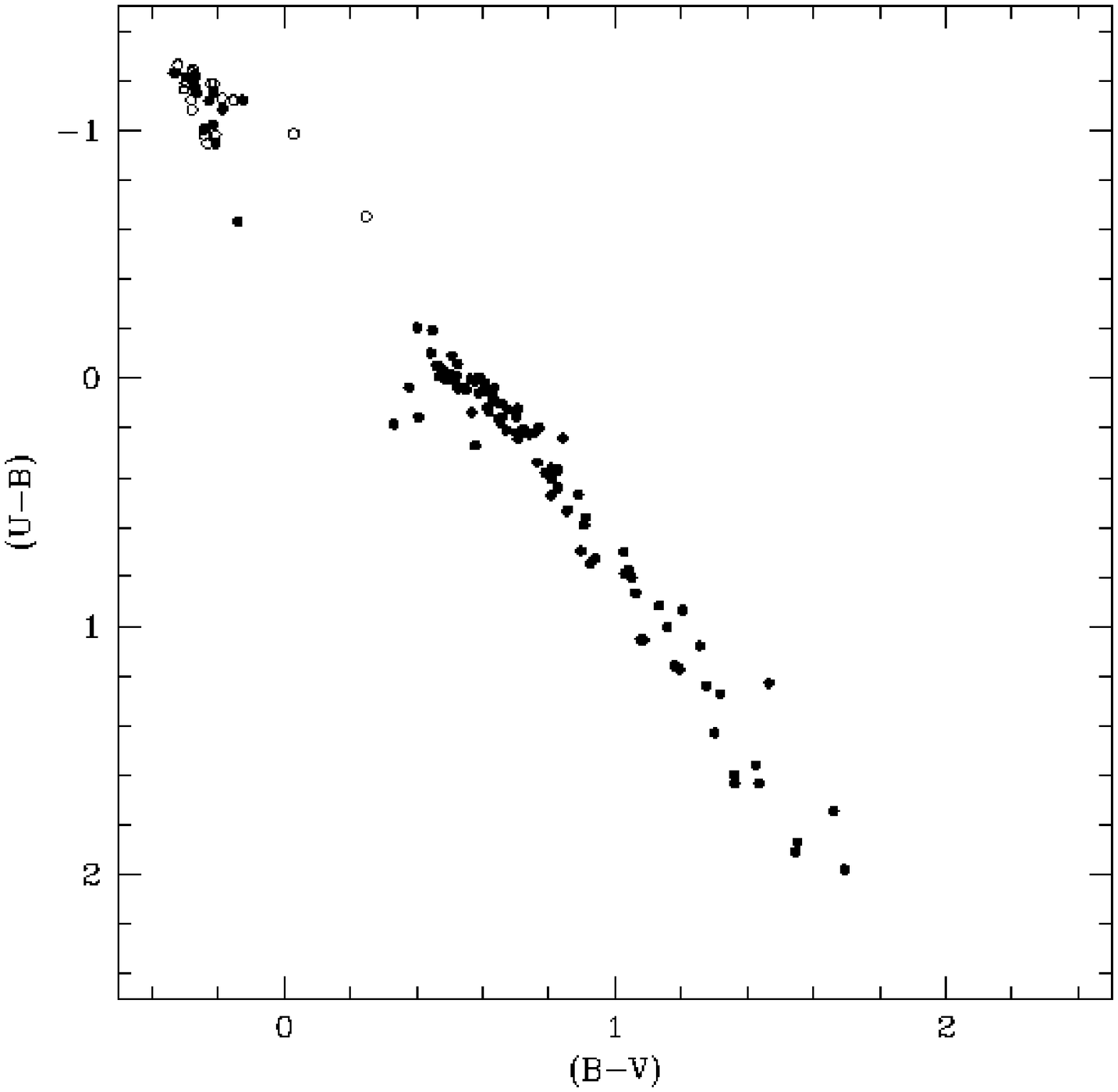}
\caption{The $[(U-B), (B-V)]$ color-color plot for all stars measured in this paper, filled 
circles from Table \ref{tab:table1} and open circles from Table \ref{tab:table4}.}
\label{fig:figure56}
\end{figure}

\begin{figure} 
\plotone{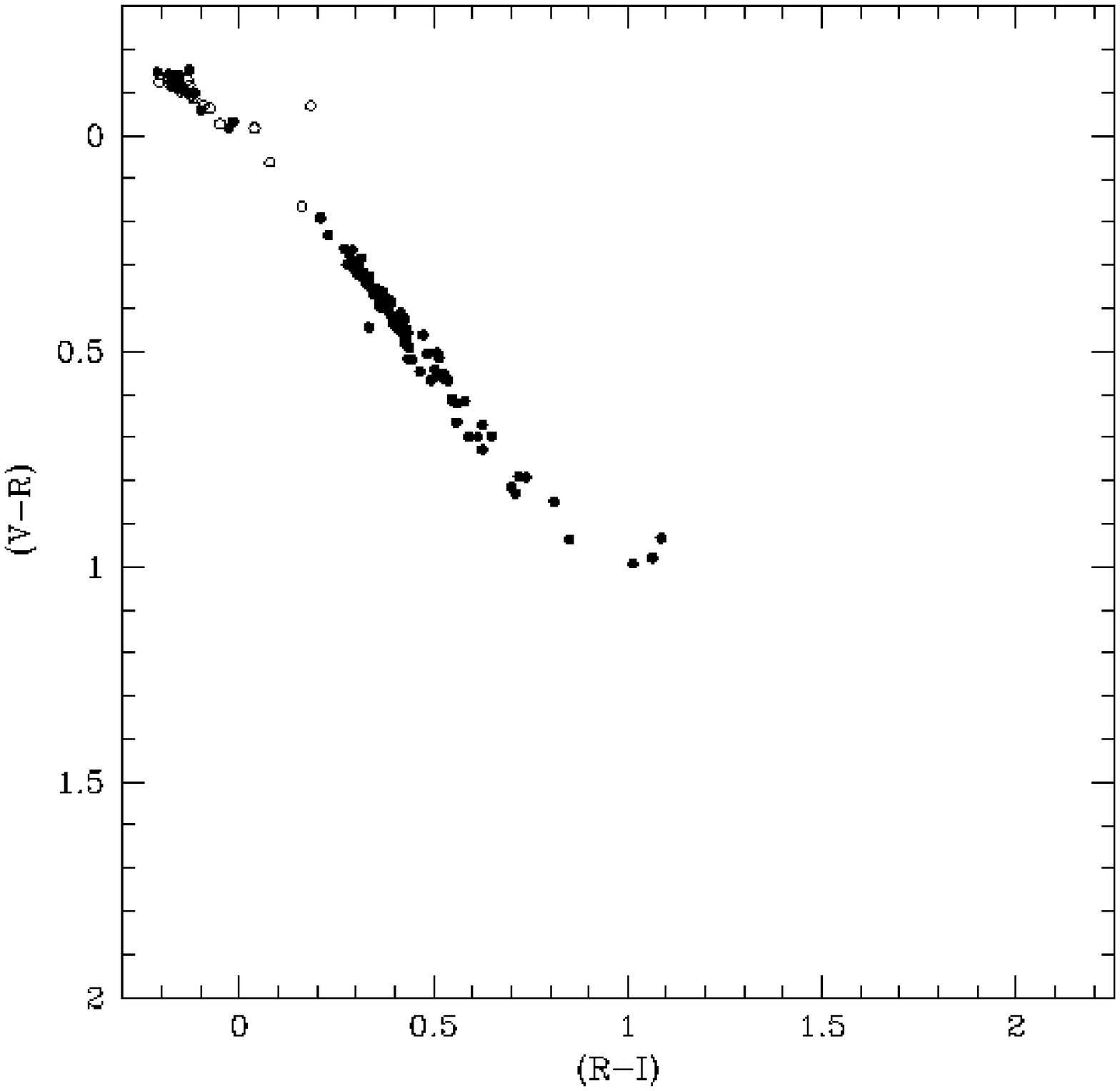} 
\caption{The $[(V-R), (R-I)]$ color-color plot for all stars measured in this paper, filled 
circles from Table \ref{tab:table1} and open circles from Table \ref{tab:table4}.}
\label{fig:figure57} 
\end{figure}

\noindent{L745-46A [= LPM~269 = GJ~283A] Discovered by \citet{Luyten1941} to be a high proper motion 
star, and cataloged by Luyten as L745-46 in the ``Bruce Proper Motion Survey", and as BPM~72393 
\citep{Luyten1963}.  Now known to be a common proper motion pair with modern astrometric information 
($\mu=1.2613 ^{\prime\prime}/yr$ at a position angle of $116.3^{\circ}$) published by 
\citet{Costa2005}.  Inspection showed that their $V$ magnitude is some 0.03~mag fainter than the 
author's in Table \ref{tab:table4} in this paper.  Another example of photometry in the literature is 
by \citet{Eggen1969}: $V=12.98$, $(B-V)=+0.29$, $(U-B)=-0.61$.  The fainter red member, L745-46B, was 
not measured in this program.  Both components are identified in Figure \ref{fig:figure36} with a 
straight line at a position angle of $116^{\circ}$ indicating their approximate direction of motion, 
but not the size of the motion.  L745-46A will move along this line a distance equal to the length of 
the scale marked at the bottom of the figure in 48 years.  The epoch of the image in Figure 
\ref{fig:figure36} is 1983.02 (The author is indebted to B. Skiff (2007) for ferreting this 
information from the literature.).}

\noindent{LSS~982 [= CD-40~3927 = CPD-40~2185] First was identified in a survey of luminous stars 
in the southern Milky Way \citep{StephensonSanduleak1971}.}

\noindent{WD~0830-535 [= GSC~08568-02947 = L~245-50 \citep{LuytenSmith1958} = BPM~19061 
\citep{Luyten1963}] \citet{Eggen1969} gave $V=14.47$, $(B-V)=-0.15$, and $(U-B)=-1.15$ in 
reasonable agreement with the values in Table \ref{tab:table1} herein.  \citet{Luyten1963} 
found proper motion components $\mu=0.167^{\prime\prime}$ and $\theta=166^{\circ}$.}

\noindent{LSS~1275 [= CD-45~5058 = CPD-45~3655 = HIP~45789 = GSC~08166-01456] This paper only has 
two photometric measurements made on one night for LSS~1275, but since the star is bright at 
$V=11.417$, those two measures should have value.  Also, the logbook noted that the night was 
photometric.  However, photometry in the literature indicates considerable differences as compared 
with that in Table \ref{tab:table4} herein.  \citet{KlareNeckel1977} found $V=11.33$, 
$(B-V)=-0.31$, $(U-B)=-1.21$; \citet{Denoyelle1977} found $V=11.36$, $(B-V)=-0.32$, $(U-B)=-1.24$; 
and, \citet{Schild1983} found $V=11.32$, $(B-V)=-0.32$, $(U-B)=-1.19$.  The {\it Hipparcos} Catalog 
(HIP~45789) provides a variability range of 0.2~mag.  A comparison of these various magnitudes 
indicates the possibility of variability.  One must remember, though, that each of the magnitudes 
quoted is based on only a couple measures.}

\noindent{LSS~1362 [= PN~G273.6+06.1] Found by \citet{Drilling1983} to be a subluminous O star.  
\citet{HeberDrilling1984} found it to be a planetary nebula, further studied by 
\citet{Mendez1988}.}

\noindent{WD~1056-384 [= GSC~07724-01874 = 2RE~J105818-384423 \citep{Mason1995} = 
EUVE~J1058-387 \citep{Vennes1996}] \citet{Pye1995} give a DA spectral type.  \citet{Mason1995} 
quote $V=13.5$ from the $HST$ Guide Star Catalog V.1.1, which \citet{Vennes1996} convert to a 
``photoelectric" $V=14.08$.}

\noindent{WD~1153-484 [= L~325-214 \citep{LuytenSmith1958} = BPM~36430 \citep{Luyten1963} = 
UCAC2~10852635] \citep{HintzenJensen1979} determined a DA spectral type.  \citet{Eggen1969} 
found $V=12.85$, $(B-V)=-0.205$, and $(U-B)=-1.01$.}

\noindent{LSE~44 [= GSC~08267-02418] Found by \citet{Drilling1983} to be a subluminous O star.}

\noindent{LSE~153 [= CD-46~8926 = CPD-46~6542 = GSC~08263-02213] Found by \citet{Drilling1983} to 
be a subluminous O star.}

\noindent{LSE~125 [= GSC~07837-01483 = PN~G335.5+12.4 (planetary nebula)] Found by 
\citet{Drilling1983} to be a subluminous O star.}

\noindent{LSE~259 [= GSC~08730-00028] Found by \citet{Drilling1983} to be a subluminous O star.}

\noindent{LSE~234 [= CPD-64~3829 = GSC~09063-01610 = NSV 24315] Found by \citet{Drilling1983} to 
be a subluminous O star.  With regard to NSV~24315, \citet{KukarkinKholopov1982} online at 
{\url {http://www.sai.msu.su/groups/cluster/gcvs/gcvs/nsvsup/ref.txt}} only lead to a 
preprint not locatable.}

\noindent{LSE~263 [= CD-51~11879 = GSC~08386-01370] Found by \citet{Drilling1983} to be a 
subluminous O star.}

\noindent{JL~25 [= GSC~09459-01325]}

\noindent{MCT~2019-4339 [= CS~22943-127 \citep{Beers1992}] \citet{Beers1992} give a spectral 
type of sdO.}

\noindent{JL~82 [= GSC~09331-00373 = EC~21313-7301]}

\noindent{JL~117 [= GSC~09340-00915]}

\noindent{LB~1516 [= GSC~08451-00403 = UCAC2~11226102] Identified as a faint blue star by 
\citet{LuytenAnderson1958}. The first measured proper motion for LB~1516 was recorded by 
\citet{Luyten1962}.}

\acknowledgements{It always is a pleasure to acknowledge the staff of the CTIO for their 
hospitality and assistance!  Individuals always available to help in anyway include A. Alvarez, E. 
Cosgrove, M. Fernandez, A. Gomez, A. Guerra, R. Leiton, D. Maturana, A. Perez, S. Pizarro, D. 
Rojas, O. Saa, N. Saavedra, E. Schmidt, H. Tirado, P. Ugarte, R. Venegas, and A. Zuniga.  Thanks go 
to John Drilling for calling the blue LSE stars to the author's attention, and to Serge Demers for 
providing preliminary charts for the MCT star fields.  Todd J. Henry and Wei-Chen Jao provided a 
modern finding chart for L745-46A.  Brian Skiff updated the author with techniques to ensure that 
the coordinates and proper motions are modern and accurate.  The appearance of this paper's figures 
and tables are due to the skills of James L. Clem.  The author is indebted to the referee, 
M.~S.~Bessell, for his helpful comments.  This observational program has been supported by NSF 
grants AST 9528177, AST-0097895, and AST-0503871.}

%***********************************************************************************************************

%**************************************************************************************************************

\newpage
\begin{deluxetable}{lcccccccccccccccc}
\rotate
\tablecaption{$UBVRI$ Photometry of Standard Stars Near $-50$ Degrees Declination}
\tabletypesize{\tiny}
\tablewidth{0pt}
\tablehead{\colhead{ }                                                       &
           \multicolumn{2}{c}{$\alpha$~~~~(2000)~~~~$\delta$}                &
           \multicolumn{8}{c}{ }                                             &
           \multicolumn{6}{c}{Mean Error of the Mean}                        \\
           \colhead{Star}                    &
           \colhead{\hms}                    &
           \colhead{\dms}                    &
           \colhead{$V$}                     &
           \colhead{$B-V$}                   &
           \colhead{$U-B$}                   &
           \colhead{$V-R$}                   &
           \colhead{$R-I$}                   &
           \colhead{$V-I$}                   &
           \colhead{$n$}                     &
           \colhead{$m$}                     &
           \colhead{$V$}                     &
           \colhead{$B-V$}                   &
           \colhead{$U-B$}                   &
           \colhead{$V-R$}                   &
           \colhead{$R-I$}                   &
           \colhead{$V-I$}                   }
\startdata
JL~163B          & 00 10 24.88  & $-$50 13 55.6  & 15.554 & $+$1.077 & $+$1.053 & $+$0.665 & $+$0.560 & $+$1.226 & 24 & 13 & 0.0033 &	0.0065  & 0.0241 & 0.0037 & 0.0094 & 0.0096 \\
JL~163A          & 00 10 26.309 & $-$50 15 03.52 & 12.927 & $+$0.524 & $-$0.058 & $+$0.317 & $+$0.315 & $+$0.632 & 21 & 12 & 0.0033 &	0.0028  & 0.0041 & 0.0020 & 0.0022 & 0.0028 \\
JL~163           & 00 10 33.221 & $-$50 15 24.37 & 12.963 & $-$0.240 & $-$1.006 & $-$0.122 & $-$0.158 & $-$0.278 & 20 & 11 & 0.0027 &	0.0020	& 0.0042 & 0.0016 & 0.0027 & 0.0036 \\
JL~163C          & 00 10 38.238 & $-$50 15 26.39 & 14.391 & $+$0.828 & $+$0.369 & $+$0.458 & $+$0.433 & $+$0.892 & 21 & 11 & 0.0017 &	0.0046	& 0.0096 & 0.0022 & 0.0050 & 0.0061 \\
JL~163D          & 00 10 41.62  & $-$50 13 45.6  & 14.300 & $+$0.896 & $+$0.695 & $+$0.520 & $+$0.444 & $+$0.967 & 21 & 11 & 0.0022 &	0.0044	& 0.0098 & 0.0022 & 0.0041 & 0.0050 \\
JL~163E          & 00 10 58.017 & $-$50 14 17.44 & 13.544 & $+$0.699 & $+$0.218 & $+$0.389 & $+$0.362 & $+$0.752 & 18 & 10 & 0.0038 &	0.0049	& 0.0111 & 0.0019 & 0.0042 & 0.0035 \\
JL~163F          & 00 11 00.225 & $-$50 12 53.81 & 12.638 & $+$0.808 & $+$0.361 & $+$0.448 & $+$0.430 & $+$0.881 & 18 &  9 & 0.0012 &	0.0021	& 0.0052 & 0.0009 & 0.0019 & 0.0021 \\
                 &              &                &        &          &          &          &          &          &    &    &        &           &        &        &        &        \\
TPhe~I           & 00 30 04.593 & $-$46 28 10.17 & 14.820 & $+$0.764 & $+$0.338 & $+$0.422 & $+$0.395 & $+$0.817 & 25 & 13 & 0.0026 &	0.0032	& 0.0072 & 0.0036 & 0.0098 & 0.0110 \\
TPhe~A           & 00 30 09.594 & $-$46 31 28.91 & 14.651 & $+$0.793 & $+$0.380 & $+$0.435 & $+$0.405 & $+$0.841 & 29 & 12 & 0.0028 &	0.0046	& 0.0071 & 0.0019 & 0.0035 & 0.0032 \\
TPhe~H           & 00 30 09.683 & $-$46 27 24.30 & 14.942 & $+$0.740 & $+$0.225 & $+$0.425 & $+$0.425 & $+$0.851 & 23 & 12 & 0.0029 &	0.0029	& 0.0071 & 0.0035 & 0.0077 & 0.0098 \\
TPhe~B           & 00 30 16.313 & $-$46 27 58.57 & 12.334 & $+$0.405 & $+$0.156 & $+$0.262 & $+$0.271 & $+$0.535 & 29 & 17 & 0.0115 &	0.0026	& 0.0039 & 0.0020 & 0.0019 & 0.0035 \\
TPhe~C           & 00 30 16.98  & $-$46 32 21.4  & 14.376 & $-$0.298 & $-$1.217 & $-$0.148 & $-$0.211 & $-$0.360 & 39 & 23 & 0.0022 &	0.0024	& 0.0043 & 0.0038 & 0.0133 & 0.0149 \\
TPhe~D           & 00 30 18.342 & $-$46 31 19.85 & 13.118 & $+$1.551 & $+$1.871 & $+$0.849 & $+$0.810 & $+$1.663 & 37 & 23 & 0.0033 &	0.0030	& 0.0118 & 0.0015 & 0.0023 & 0.0030 \\
TPhe~E           & 00 30 19.768 & $-$46 24 35.60 & 11.631 & $+$0.443 & $-$0.103 & $+$0.276 & $+$0.283 & $+$0.564 & 38 & 10 & 0.0017 &	0.0013	& 0.0025 & 0.0007 & 0.0016 & 0.0020 \\
TPhe~J           & 00 30 23.02  & $-$46 23 51.6  & 13.434 & $+$1.465 & $+$1.229 & $+$0.980 & $+$1.063 & $+$2.043 & 28 & 15 & 0.0023 &	0.0043	& 0.0059 & 0.0011 & 0.0015 & 0.0011 \\
TPhe~F           & 00 30 49.820 & $-$46 33 24.07 & 12.475 & $+$0.853 & $+$0.534 & $+$0.492 & $+$0.437 & $+$0.929 & 19 & 10 & 0.0008 &	0.0024	& 0.0095 & 0.0005 & 0.0014 & 0.0029 \\
TPhe~K           & 00 30 56.315 & $-$46 23 26.04 & 12.935 & $+$0.806 & $+$0.402 & $+$0.473 & $+$0.429 & $+$0.909 &  2 &  2 & 0.0007 &	0.0007	& 0.0163 & 0.0007 & 0.0001 & 0.0007 \\
TPhe~G           & 00 31 04.303 & $-$46 22 51.35 & 10.447 & $+$1.545 & $+$1.910 & $+$0.934 & $+$1.086 & $+$2.025 & 20 & 10 & 0.0008 &	0.0011	& 0.0049 & 0.0008 & 0.0016 & 0.0017 \\
                 &              &                &        &          &          &          &          &          &    &    &        &           &        &        &        &        \\
MCT~0401$-$4017E & 04 02 28.087 & $-$40 09 35.57 & 10.636 & $+$0.527 & $+$0.040 & $+$0.307 & $+$0.299 & $+$0.606 & 20 & 10 & 0.0020 &	0.0011	& 0.0027 & 0.0011 & 0.0011 & 0.0016 \\
MCT~0401$-$4017F & 04 02 29.559 & $-$40 10 21.19 & 12.990 & $+$0.790 & $+$0.380 & $+$0.441 & $+$0.400 & $+$0.841 & 20 &  9 & 0.0013 &	0.0018	& 0.0042 & 0.0011 & 0.0022 & 0.0018 \\
MCT~0401$-$4017D & 04 02 55.219 & $-$40 15 52.60 & 11.947 & $+$0.549 & $+$0.045 & $+$0.319 & $+$0.304 & $+$0.623 & 20 & 11 & 0.0020 &	0.0018	& 0.0029 & 0.0013 & 0.0020 & 0.0020 \\
MCT~0401$-$4017  & 04 03 04.540 & $-$40 09 41.05 & 14.418 & $-$0.272 & $-$1.180 & $-$0.141 & $-$0.156 & $-$0.291 & 21 & 11 & 0.0022 &	0.0017	& 0.0055 & 0.0035 & 0.0131 & 0.0151 \\
MCT~0401$-$4017A & 04 03 05.195 & $-$40 11 02.88 & 10.709 & $+$0.562 & $+$0.001 & $+$0.328 & $+$0.321 & $+$0.649 & 21 & 11 & 0.0024 &	0.0013	& 0.0041 & 0.0011 & 0.0022 & 0.0022 \\
MCT~0401$-$4017B & 04 03 29.846 & $-$40 10 58.46 & 12.654 & $+$0.506 & $-$0.092 & $+$0.315 & $+$0.319 & $+$0.633 & 20 & 11 & 0.0025 &	0.0031	& 0.0056 & 0.0013 & 0.0020 & 0.0020 \\
MCT~0401$-$4017C & 04 03 34.273 & $-$40 07 02.53 & 12.395 & $+$0.906 & $+$0.589 & $+$0.551 & $+$0.524 & $+$1.075 & 21 & 11 & 0.0028 &	0.0020	& 0.0041 & 0.0011 & 0.0020 & 0.0022 \\
                 &              &                &        &          &          &          &          &          &    &    &        &           &        &        &        &        \\
LB~1735          & 04 31 11.090 & $-$53 35 27.06 & 13.634 & $-$0.142 & $-$0.632 & $-$0.059 & $-$0.098 & $-$0.157 & 27 & 15 & 0.0017 &	0.0023	& 0.0029 & 0.0017 & 0.0064 & 0.0067 \\
LB~1735A         & 04 31 19.438 & $-$53 34 38.22 & 13.906 & $+$0.589 & $-$0.004 & $+$0.339 & $+$0.334 & $+$0.672 & 20 & 11 & 0.0025 &	0.0036	& 0.0051 & 0.0018 & 0.0029 & 0.0034 \\
LB~1735B         & 04 31 22.810 & $-$53 36 36.85 & 14.542 & $+$0.675 & $+$0.126 & $+$0.379 & $+$0.361 & $+$0.741 & 17 &  9 & 0.0032 &	0.0046	& 0.0051 & 0.0022 & 0.0053 & 0.0063 \\
LB~1735F         & 04 31 37.11  & $-$53 36 53.4  & 15.205 & $+$0.402 & $-$0.206 & $+$0.265 & $+$0.290 & $+$0.556 & 27 & 14 & 0.0035 &	0.0031	& 0.0052 & 0.0046 & 0.0106 & 0.0131 \\
LB~1735E         & 04 31 42.303 & $-$53 36 18.48 & 13.759 & $+$0.706 & $+$0.242 & $+$0.399 & $+$0.362 & $+$0.760 & 21 & 11 & 0.0083 &	0.0028	& 0.0063 & 0.0020 & 0.0044 & 0.0055 \\
LB~1735C         & 04 31 49.652 & $-$53 34 37.52 & 12.765 & $+$0.459 & $-$0.052 & $+$0.290 & $+$0.297 & $+$0.587 & 20 & 11 & 0.0034 &	0.0034	& 0.0047 & 0.0016 & 0.0022 & 0.0029 \\
LB~1735D         & 04 31 58.253 & $-$53 34 52.52 & 14.184 & $+$0.657 & $+$0.099 & $+$0.383 & $+$0.383 & $+$0.765 & 17 &  9 & 0.0034 &	0.0024	& 0.0051 & 0.0022 & 0.0063 & 0.0078 \\
LB~1735G         & 04 31 46.194 & $-$53 39 47.89 & 13.446 & $+$1.276 & $+$1.241 & $+$0.815 & $+$0.701 & $+$1.516 & 26 & 14 & 0.0020 &	0.0020	& 0.0071 & 0.0008 & 0.0018 & 0.0018 \\
                 &              &                &        &          &          &          &          &          &    &    &        &           &        &        &        &        \\
MCT~0436$-$4616A & 04 38 20.929 & $-$46 09 27.54 & 12.089 & $+$0.504 & $-$0.014 & $+$0.296 & $+$0.285 & $+$0.581 &  5 &  3 & 0.0009 &	0.0040	& 0.0063 & 0.0018 & 0.0036 & 0.0027 \\
MCT~0436$-$4616  & 04 38 27.320 & $-$46 10 52.69 & 13.823 & $-$0.213 & $-$1.153 & $-$0.032 & $-$0.015 & $-$0.048 &  8 &  5 & 0.0032 &	0.0032	& 0.0053 & 0.0021 & 0.0120 & 0.0117 \\
MCT~0436$-$4616B & 04 38 47.374 & $-$46 10 08.74 & 13.522 & $+$0.448 & $-$0.194 & $+$0.284 & $+$0.312 & $+$0.596 &  5 &  3 & 0.0040 &	0.0049	& 0.0089 & 0.0049 & 0.0058 & 0.0089 \\
                 &              &                &        &          &          &          &          &          &    &    &        &           &        &        &        &        \\
MCT~0550$-$4911E & 05 51 43.660 & $-$49 10 48.59 & 13.671 & $+$1.300 & $+$1.429 & $+$0.831 & $+$0.711 & $+$1.544 & 16 &  8 & 0.0040 &	0.0078	& 0.0935 & 0.0022 & 0.0025 & 0.0040 \\
MCT~0550$-$4911D & 05 51 49.207 & $-$49 11 18.95 & 14.592 & $+$0.732 & $+$0.217 & $+$0.406 & $+$0.383 & $+$0.701 & 15 &  9 & 0.0075 &	0.0065	& 0.0256 & 0.0044 & 0.0116 & 0.0127 \\
MCT~0550$-$4911B & 05 51 59.105 & $-$49 11 25.06 & 14.683 & $+$0.856 & $+$0.530 & $+$0.483 & $+$0.427 & $+$0.910 & 14 &  8 & 0.0067 &	0.0086	& 0.0061 & 0.0032 & 0.0061 & 0.0083 \\
MCT~0550$-$4911  & 05 52 02.712 & $-$49 11 22.38 & 14.355 & $-$0.270 & $-$1.219 & $-$0.114 & $-$0.175 & $-$0.288 & 18 & 11 & 0.0057 &	0.0035	& 0.0042 & 0.0040 & 0.0139 & 0.0156 \\
MCT~0550$-$4911C & 05 52 03.98  & $-$49 09 37.6  & 13.019 & $+$0.588 & $+$0.058 & $+$0.343 & $+$0.327 & $+$0.670 & 14 &  8 & 0.0035 &	0.0032	& 0.0072 & 0.0024 & 0.0029 & 0.0045 \\
MCT~0550$-$4911A & 05 52 04.104 & $-$49 10 48.55 & 14.255 & $+$0.606 & $+$0.020 & $+$0.351 & $+$0.340 & $+$0.690 & 14 &  9 & 0.0032 &	0.0051	& 0.0059 & 0.0035 & 0.0045 & 0.0064 \\
                 &              &                &        &          &          &          &          &          &    &    &        &           &        &        &        &        \\
LSS~982G         & 08 10 19.984 & $-$40 31 57.19 & 13.374 & $+$0.542 & $+$0.040 & $+$0.329 & $+$0.317 & $+$0.647 & 15 &  8 & 0.0057 &	0.0036	& 0.0065 & 0.0023 & 0.0031 & 0.0036 \\
LSS~982F         & 08 10 21.386 & $-$40 31 30.31 & 11.623 & $+$0.376 & $+$0.035 & $+$0.231 & $+$0.228 & $+$0.458 & 15 &  8 & 0.0057 &	0.0021	& 0.0031 & 0.0015 & 0.0026 & 0.0034 \\
LSS~982E         & 08 10 29.667 & $-$40 30 06.85 & 11.297 & $+$1.086 & $+$1.054 & $+$0.567 & $+$0.494 & $+$1.059 & 15 &  8 & 0.0039 &	0.0021	& 0.0046 & 0.0013 & 0.0026 & 0.0026 \\
LSS~982          & 08 10 31.719 & $-$40 32 47.07 & 12.258 & $-$0.331 & $-$1.233 & $-$0.143 & $-$0.181 & $-$0.323 & 20 & 11 & 0.0036 &	0.0018	& 0.0031 & 0.0016 & 0.0036 & 0.0042 \\
LSS~982A         & 08 10 32.719 & $-$40 31 14.52 & 12.292 & $+$0.470 & $-$0.009 & $+$0.290 & $+$0.287 & $+$0.577 & 16 &  9 & 0.0022 &	0.0030	& 0.0038 & 0.0012 & 0.0028 & 0.0038 \\
LSS~982B         & 08 10 34.325 & $-$40 30 49.16 & 12.435 & $+$0.468 & $-$0.049 & $+$0.297 & $+$0.291 & $+$0.587 & 16 &  9 & 0.0025 &	0.0018	& 0.0040 & 0.0018 & 0.0030 & 0.0038 \\
LSS~982D         & 08 10 38.089 & $-$40 28 20.84 & 11.648 & $+$1.204 & $+$0.934 & $+$0.671 & $+$0.626 & $+$1.298 & 17 &  9 & 0.0044 &	0.0027	& 0.0056 & 0.0010 & 0.0029 & 0.0032 \\
LSS~982C         & 08 10 38.222 & $-$40 30 09.95 & 12.571 & $+$1.660 & $+$1.745 & $+$0.937 & $+$0.849 & $+$1.787 & 19 & 11 & 0.0032 &	0.0025	& 0.0112 & 0.0011 & 0.0023 & 0.0021 \\
                 &              &                &        &          &          &          &          &          &    &    &        &           &        &        &        &        \\
WD~0830$-$535J   & 08 31 41.877 & $-$53 37 28.27 & 13.484 & $+$1.256 & $+$1.078 & $+$0.698 & $+$0.650 & $+$1.348 & 16 &  8 & 0.0025 &	0.0030	& 0.0132 & 0.0018 & 0.0025 & 0.0030 \\
WD~0830$-$535K   & 08 31 42.960 & $-$53 39 19.33 & 14.085 & $+$0.635 & $+$0.037 & $+$0.383 & $+$0.390 & $+$0.774 & 12 &  7 & 0.0055 &	0.0081	& 0.0098 & 0.0032 & 0.0055 & 0.0069 \\
WD~0830$-$535I   & 08 31 45.769 & $-$53 37 35.34 & 13.172 & $+$0.575 & $+$0.014 & $+$0.363 & $+$0.366 & $+$0.729 &  8 &  5 & 0.0057 &	0.0060	& 0.0085 & 0.0028 & 0.0042 & 0.0060 \\
WD~0830$-$535C   & 08 31 50.421 & $-$53 43 51.16 & 13.529 & $+$0.628 & $+$0.089 & $+$0.387 & $+$0.381 & $+$0.768 & 10 &  5 & 0.0054 &	0.0057	& 0.0111 & 0.0016 & 0.0051 & 0.0054 \\
WD~0830$-$535    & 08 31 51.91  & $-$53 40 32.5  & 14.497 & $-$0.227 & $-$1.119 & $-$0.134 & $-$0.161 & $-$0.293 & 18 & 12 & 0.0049 &	0.0040	& 0.0054 & 0.0040 & 0.0203 & 0.0224 \\
WD~0830$-$535B   & 08 31 54.356 & $-$53 42 37.91 & 13.847 & $+$0.654 & $+$0.181 & $+$0.392 & $+$0.389 & $+$0.780 & 10 &  7 & 0.0057 &	0.0032	& 0.0076 & 0.0022 & 0.0057 & 0.0076 \\
WD~0830$-$535A   & 08 31 54.454 & $-$53 40 48.08 & 14.123 & $+$0.719 & $+$0.205 & $+$0.415 & $+$0.414 & $+$0.829 & 11 &  6 & 0.0075 &	0.0048	& 0.0118 & 0.0048 & 0.0103 & 0.0130 \\
WD~0830$-$535H   & 08 32 03.822 & $-$53 37 50.36 & 12.790 & $+$0.566 & $+$0.136 & $+$0.364 & $+$0.371 & $+$0.734 &  9 &  5 & 0.0030 &	0.0030	& 0.0050 & 0.0017 & 0.0020 & 0.0023 \\
WD~0830$-$535D   & 08 32 04.970 & $-$53 44 14.80 & 12.599 & $+$0.514 & $+$0.006 & $+$0.327 & $+$0.335 & $+$0.661 &  9 &  5 & 0.0043 &	0.0027	& 0.0050 & 0.0013 & 0.0040 & 0.0037 \\
WD~0830$-$535G   & 08 32 08.190 & $-$53 40 30.15 & 14.009 & $+$0.577 & $+$0.270 & $+$0.368 & $+$0.357 & $+$0.725 &  8 &  5 & 0.0035 &	0.0067	& 0.0088 & 0.0039 & 0.0053 & 0.0074 \\
WD~0830$-$535F   & 08 32 08.639 & $-$53 40 44.59 & 14.509 & $+$0.618 & $+$0.134 & $+$0.377 & $+$0.376 & $+$0.752 &  7 &  4 & 0.0053 &	0.0060	& 0.0110 & 0.0060 & 0.0098 & 0.0083 \\
WD~0830$-$535E   & 08 32 11.137 & $-$53 41 58.54 & 13.225 & $+$0.769 & $+$0.199 & $+$0.462 & $+$0.473 & $+$0.932 &  9 &  5 & 0.0043 &	0.0037	& 0.0083 & 0.0013 & 0.0030 & 0.0027 \\
                 &              &                &        &          &          &          &          &          &    &    &        &           &        &        &        &        \\
WD~1056$-$384D   & 10 58 08.051 & $-$38 46 22.33 & 13.132 & $+$0.521 & $+$0.032 & $+$0.307 & $+$0.309 & $+$0.615 & 12 &  6 & 0.0049 &	0.0029	& 0.0023 & 0.0020 & 0.0038 & 0.0049 \\
WD~1056$-$384C   & 10 58 11.053 & $-$38 41 50.97 & 13.326 & $+$0.649 & $+$0.164 & $+$0.367 & $+$0.357 & $+$0.724 & 13 &  8 & 0.0050 &	0.0028	& 0.0047 & 0.0017 & 0.0033 & 0.0039 \\
WD~1056$-$384B   & 10 58 18.691 & $-$38 41 56.56 & 12.471 & $+$1.156 & $+$1.003 & $+$0.611 & $+$0.548 & $+$1.159 & 14 &  8 & 0.0027 &	0.0027	& 0.0083 & 0.0016 & 0.0064 & 0.0078 \\
WD~1056$-$384    & 10 58 20.11  & $-$38 44 25.1  & 14.047 & $-$0.187 & $-$1.085 & $-$0.132 & $-$0.174 & $-$0.303 & 14 &  8 & 0.0032 &	0.0021	& 0.0032 & 0.0045 & 0.0187 & 0.0187 \\
WD~1056$-$384A   & 10 58 23.720 & $-$38 44 01.19 & 12.375 & $+$1.132 & $+$0.916 & $+$0.616 & $+$0.581 & $+$1.197 & 14 &  8 & 0.0016 &	0.0029	& 0.0056 & 0.0013 & 0.0013 & 0.0021 \\
                 &              &                &        &          &          &          &          &          &    &    &        &           &        &        &        &        \\
WD~1153$-$484G   & 11 55 53.78  & $-$48 42 01.3  & 13.531 & $+$0.911 & $+$0.562 & $+$0.504 & $+$0.509 & $+$1.010 &  4 &  2 & 0.0080 &	0.0140	& 0.0755 & 0.0065 & 0.0160 & 0.0200 \\
WD~1153$-$484H   & 11 55 53.886 & $-$48 41 24.93 & 14.276 & $+$0.478 & $-$0.036 & $+$0.298 & $+$0.277 & $+$0.576 &  6 &  3 & 0.0122 &	0.0131	& 0.0082 & 0.0057 & 0.0265 & 0.0302 \\
WD~1153$-$484F   & 11 55 56.276 & $-$48 42 03.51 & 12.329 & $+$0.489 & $+$0.005 & $+$0.298 & $+$0.286 & $+$0.583 &  4 &  2 & 0.0070 &	0.0060	& 0.0170 & 0.0015 & 0.0050 & 0.0040 \\
WD~1153$-$484D   & 11 55 57.027 & $-$48 37 23.50 & 13.275 & $+$0.754 & $+$0.221 & $+$0.431 & $+$0.406 & $+$0.837 &  4 &  2 & 0.0080 &	0.0035	& 0.0050 & 0.0030 & 0.0065 & 0.0080 \\
WD~1153$-$484C   & 11 56 01.765 & $-$48 38 32.38 & 12.687 & $+$1.179 & $+$1.156 & $+$0.617 & $+$0.550 & $+$1.168 &  2 &  2 & 0.0007 &	0.0049	& 0.0092 & 0.0014 & 0.0042 & 0.0064 \\
WD~1153$-$484B   & 11 56 05.105 & $-$48 38 50.36 & 13.282 & $+$1.362 & $+$1.633 & $+$0.699 & $+$0.590 & $+$1.292 &  7 &  4 & 0.0049 &	0.0147	& 0.0189 & 0.0019 & 0.0053 & 0.0064 \\
WD~1153$-$484A   & 11 56 10.834 & $-$48 39 48.51 & 12.633 & $+$0.700 & $+$0.154 & $+$0.415 & $+$0.389 & $+$0.803 &  4 &  2 & 0.0080 &	0.0055	& 0.0060 & 0.0035 & 0.0060 & 0.0060 \\
WD~1153$-$484    & 11 56 11.427 & $-$48 40 03.30 & 12.915 & $-$0.215 & $-$1.023 & $-$0.098 & $-$0.130 & $-$0.228 &  6 &  3 & 0.0065 &	0.0029	& 0.0029 & 0.0020 & 0.0029 & 0.0016 \\
WD~1153$-$484E   & 11 56 13.505 & $-$48 42 16.00 & 13.440 & $+$0.511 &    0.000 & $+$0.308 & $+$0.297 & $+$0.605 &  4 &  2 & 0.0050 &	0.0045	& 0.0060 & 0.0020 & 0.0065 & 0.0070 \\
                 &              &                &        &          &          &          &          &          &    &    &        &           &        &        &        &        \\
LSE~44F          & 13 52 34.466 & $-$48 06 59.26 & 12.013 & $+$1.194 & $+$1.172 & $+$0.620 & $+$0.560 & $+$1.180 & 19 & 10 & 0.0034 &	0.0028	& 0.0044 & 0.0064 & 0.0028 & 0.0069 \\
LSE~44A          & 13 52 39.573 & $-$48 10 04.92 & 13.743 & $+$1.435 & $+$1.634 & $+$0.791 & $+$0.719 & $+$1.511 & 21 & 10 & 0.0033 &	0.0057	& 0.0238 & 0.0017 & 0.0022 & 0.0035 \\
LSE~44           & 13 52 40.779 & $-$48 08 22.75 & 12.459 & $-$0.265 & $-$1.152 & $-$0.112 & $-$0.145 & $-$0.255 & 20 & 11 & 0.0040 &	0.0018	& 0.0027 & 0.0011 & 0.0034 & 0.0034 \\
LSE~44B          & 13 52 43.784 & $-$48 11 34.04 & 12.632 & $+$1.039 & $+$0.774 & $+$0.567 & $+$0.538 & $+$1.105 & 18 &  9 & 0.0033 &	0.0021	& 0.0061 & 0.0007 & 0.0012 & 0.0014 \\
LSE~44C          & 13 52 51.850 & $-$48 09 56.32 & 13.437 & $+$1.426 & $+$1.559 & $+$0.793 & $+$0.738 & $+$1.531 & 17 &  9 & 0.0082 &	0.0053	& 0.0155 & 0.0017 & 0.0068 & 0.0075 \\
LSE~44D          & 13 52 52.115 & $-$48 08 38.43 & 13.140 & $+$1.360 & $+$1.598 & $+$0.728 & $+$0.626 & $+$1.354 & 15 &  8 & 0.0049 &	0.0062	& 0.0238 & 0.0010 & 0.0018 & 0.0023 \\
LSE~44E          & 13 52 55.514 & $-$48 08 19.58 & 13.554 & $+$0.827 & $+$0.440 & $+$0.454 & $+$0.413 & $+$0.866 & 16 &  8 & 0.0048 &	0.0045	& 0.0062 & 0.0020 & 0.0048 & 0.0060 \\
                 &              &                &        &          &          &          &          &          &    &    &        &           &        &        &        &        \\
LSE~259C         & 16 53 30.037 & $-$56 00 54.62 & 10.866 & $+$1.025 & $+$0.700 & $+$0.561 & $+$0.525 & $+$1.087 & 15 &  8 & 0.0021 &	0.0021	& 0.0039 & 0.0010 & 0.0013 & 0.0013 \\
LSE~259D         & 16 53 41.607 & $-$55 59 08.96 & 11.719 & $+$0.331 & $+$0.183 & $+$0.191 & $+$0.209 & $+$0.400 & 15 &  8 & 0.0041 &	0.0026	& 0.0026 & 0.0008 & 0.0013 & 0.0015 \\
LSE~259E         & 16 53 43.240 & $-$55 58 30.06 & 11.922 & $+$0.613 & $+$0.118 & $+$0.362 & $+$0.362 & $+$0.725 & 17 &  9 & 0.0034 &	0.0022	& 0.0036 & 0.0012 & 0.0012 & 0.0012 \\
LSE~259B         & 16 53 44.377 & $-$56 01 32.85 & 13.596 & $+$0.888 & $+$0.468 & $+$0.506 & $+$0.482 & $+$0.991 & 18 & 10 & 0.0045 &	0.0064	& 0.0151 & 0.0019 & 0.0024 & 0.0040 \\
LSE~259H         & 16 53 48.077 & $-$56 00 42.24 & 14.164 & $+$1.317 & $+$1.272 & $+$0.699 & $+$0.614 & $+$1.314 & 15 &  8 & 0.0046 &	0.0036	& 0.0336 & 0.0026 & 0.0044 & 0.0062 \\
LSE~259A         & 16 53 53.147 & $-$56 02 02.58 & 13.545 & $+$1.692 & $+$1.980 & $+$0.993 & $+$1.013 & $+$2.006 & 19 & 12 & 0.0028 &	0.0044	& 0.0695 & 0.0014 & 0.0018 & 0.0023 \\
LSE~259          & 16 53 54.573 & $-$56 01 54.76 & 12.551 & $-$0.127 & $-$1.123 & $-$0.019 & $-$0.026 & $-$0.046 & 22 & 14 & 0.0023 &	0.0019	& 0.0043 & 0.0013 & 0.0028 & 0.0034 \\
LSE~259F         & 16 53 55.538 & $-$55 59 26.16 & 13.580 & $+$0.615 & $+$0.052 & $+$0.374 & $+$0.375 & $+$0.749 & 17 &  9 & 0.0027 &	0.0032	& 0.0068 & 0.0024 & 0.0032 & 0.0034 \\
LSE~259G         & 16 54 00.237 & $-$55 59 20.05 & 14.092 & $+$0.841 & $+$0.241 & $+$0.515 & $+$0.514 & $+$1.029 & 14 &  7 & 0.0035 &	0.0043	& 0.0144 & 0.0024 & 0.0061 & 0.0067 \\
                 &              &                &        &          &          &          &          &          &    &    &        &           &        &        &        &        \\
MCT~2019$-$4339E & 20 22 38.910 & $-$43 31 17.07 & 13.693 & $+$1.029 & $+$0.788 & $+$0.566 & $+$0.530 & $+$1.096 & 28 & 15 & 0.0017 &	0.0025	& 0.0064 & 0.0009 & 0.0028 & 0.0030 \\
MCT~2019$-$4339D & 20 22 40.476 & $-$43 27 26.39 & 13.205 & $+$0.924 & $+$0.748 & $+$0.518 & $+$0.434 & $+$0.953 & 19 & 11 & 0.0018 &	0.0023	& 0.0050 & 0.0014 & 0.0023 & 0.0028 \\
MCT~2019$-$4339A & 20 22 45.332 & $-$43 29 43.33 & 13.055 & $+$0.521 & $-$0.011 & $+$0.307 & $+$0.295 & $+$0.602 & 20 & 11 & 0.0025 &	0.0025	& 0.0040 & 0.0016 & 0.0031 & 0.0036 \\
MCT~2019$-$4339B & 20 22 46.726 & $-$43 28 10.88 & 13.923 & $+$0.671 & $+$0.208 & $+$0.369 & $+$0.347 & $+$0.716 & 19 & 11 & 0.0028 &	0.0030	& 0.0044 & 0.0032 & 0.0041 & 0.0067 \\
MCT~2019$-$4339  & 20 22 49.056 & $-$43 30 11.53 & 13.685 & $-$0.288 & $-$1.212 & $-$0.115 & $-$0.149 & $-$0.261 & 24 & 13 & 0.0029 &	0.0020	& 0.0065 & 0.0027 & 0.0067 & 0.0069 \\
MCT~2019$-$4339C & 20 23 02.134 & $-$43 28 22.40 & 12.440 & $+$0.939 & $+$0.726 & $+$0.547 & $+$0.466 & $+$1.011 & 20 & 11 & 0.0018 &	0.0020	& 0.0051 & 0.0009 & 0.0011 & 0.0016 \\
MCT~2019$-$4339F & 20 23 03.99  & $-$43 31 21.9  & 13.936 & $+$0.647 & $+$0.097 & $+$0.369 & $+$0.360 & $+$0.729 & 20 & 11 & 0.0031 &	0.0029	& 0.0060 & 0.0022 & 0.0047 & 0.0054 \\
                 &              &                &        &          &          &          &          &          &    &    &        &           &        &        &        &        \\
JL~82C           & 21 35 45.005 & $-$72 50 12.76 & 13.440 & $+$0.612 & $+$0.041 & $+$0.357 & $+$0.358 & $+$0.715 & 19 & 10 & 0.0037 &	0.0023	& 0.0053 & 0.0028 & 0.0034 & 0.0055 \\
JL~82B           & 21 35 59.34  & $-$72 50 15.1  & 13.507 & $+$0.705 & $+$0.121 & $+$0.411 & $+$0.414 & $+$0.825 & 19 & 10 & 0.0037 &	0.0032	& 0.0053 & 0.0021 & 0.0037 & 0.0048 \\
JL~82            & 21 36 01.289 & $-$72 48 27.21 & 12.389 & $-$0.208 & $-$0.947 & $-$0.098 & $-$0.115 & $-$0.211 & 21 & 11 & 0.0031 &	0.0020	& 0.0041 & 0.0013 & 0.0035 & 0.0037 \\
JL~82D           & 21 36 15.362 & $-$72 45 27.21 & 12.371 & $+$1.062 & $+$0.865 & $+$0.551 & $+$0.509 & $+$1.061 & 19 & 10 & 0.0034 &	0.0025	& 0.0053 & 0.0016 & 0.0018 & 0.0021 \\
JL~82A           & 21 36 17.052 & $-$72 50 08.57 & 11.226 & $+$1.050 & $+$0.803 & $+$0.543 & $+$0.503 & $+$1.048 & 21 & 11 & 0.0031 &	0.0013	& 0.0022 & 0.0007 & 0.0015 & 0.0015 \\
                 &              &                &        &          &          &          &          &          &    &    &        &           &        &        &        &        \\
JL~117A          & 22 54 13.332 & $-$72 23 04.33 & 14.061 & $+$0.629 & $+$0.072 & $+$0.362 & $+$0.354 & $+$0.717 &  6 &  3 & 0.0029 &	0.0037	& 0.0069 & 0.0024 & 0.0147 & 0.0159 \\
JL~117           & 22 54 38.002 & $-$72 23 09.68 & 14.469 & $-$0.336 & $-$1.233 & $-$0.152 & $-$0.128 & $-$0.272 & 12 &  6 & 0.0032 &	0.0023	& 0.0023 & 0.0055 & 0.0271 & 0.0297 \\
JL~117B          & 22 54 51.021 & $-$72 23 23.61 & 14.961 & $+$0.806 & $+$0.471 & $+$0.444 & $+$0.334 & $+$0.779 &  4 &  2 & 0.0045 &	0.0035	& 0.0120 & 0.0165 & 0.9590 & 0.0755 \\
JL~117C          & 22 55 00.471 & $-$72 22 41.35 & 14.398 & $+$0.797 & $+$0.385 & $+$0.436 & $+$0.395 & $+$0.834 &  6 &  4 & 0.0045 &	0.0045	& 0.0151 & 0.0033 & 0.0061 & 0.0065 \\
JL~117D          & 22 55 37.574 & $-$72 22 34.28 & 12.550 & $+$0.657 & $+$0.159 & $+$0.369 & $+$0.360 & $+$0.731 &  5 &  2 & 0.0027 &	0.0018	& 0.0058 & 0.0018 & 0.0022 & 0.0027 \\
\enddata
\label{tab:table1}
\end{deluxetable}

\newpage
\begin{deluxetable}{lcc}
\tablecaption{Error Analysis for 109 New Standards Stars in Table \ref{tab:table1}.}
\tablewidth{0pt}
\tablehead{\colhead{ }                                    &
           \colhead{Mean Errors of a Single Observation}  &
           \colhead{Mean Errors of the Mean}              }
\startdata
$V$             & $0.0161\pm0.0101$ & $0.0035\pm0.0021$ \\
$B-V$           & $0.0137\pm0.0070$ & $0.0033\pm0.0024$ \\
$U-B$           & $0.0368\pm0.0488$ & $0.0091\pm0.0135$ \\
$V-R$           & $0.0091\pm0.0055$ & $0.0021\pm0.0019$ \\
$R-I$           & $0.0206\pm0.0202$ & $0.0048\pm0.0071$ \\
$V-I$           & $0.0240\pm0.0233$ & $0.0060\pm0.0093$ \\
\enddata
\label{tab:table2}
\end{deluxetable}

\newpage
\begin{deluxetable}{lcc}
\tablecaption{Field Centers for Sequences.}
\tablewidth{0pt}
\tablehead{\colhead{Field Name}                 &
           \colhead{$\alpha$(2000.0)}           &
           \colhead{$\delta$(2000.0)}           }
\startdata
JL~163          & 00 10 42.6 & $-$50 14 10 \\  
T~Phe           & 00 30 34.4 & $-$46 28 08 \\
MCT~0401-4017   & 04 03 01.2 & $-$40 11 28 \\
LB~1735         & 04 31 34.7 & $-$53 37 13 \\
MCT~0436-4616   & 04 38 34.2 & $-$46 10 10 \\
MCT~0550-4911   & 05 51 53.9 & $-$49 10 31 \\
LSS~982         & 08 10 29.1 & $-$40 30 34 \\
WD~0830-535     & 08 31 56.5 & $-$53 40 52 \\
WD~1056-384	& 10 58 15.9 & $-$38 44 07 \\
WD~1153-484     & 11 56 03.6 & $-$48 39 50 \\
LSE~44          & 13 52 45.0 & $-$48 09 17 \\
LSE~259         & 16 53 45.1 & $-$56 00 16 \\
MCT~2019-4339   & 20 22 51.5 & $-$43 29 24 \\
JL~82           & 21 36 01.0 & $-$72 47 51 \\
JL~117          & 22 54 55.5 & $-$72 22 59 \\
\enddata
\label{tab:table3}
\end{deluxetable}

\newpage
\begin{deluxetable}{lcccccccccccccccc}
\rotate
\tablecaption{$UBVRI$ Photometry of Isolated Stars}
\tabletypesize{\tiny}
\tablewidth{0pt}
\tablehead{\colhead{ }                                                       &
           \multicolumn{2}{c}{$\alpha$~~~~(2000)~~~~$\delta$}                &
           \multicolumn{8}{c}{ }                                             &
           \multicolumn{6}{c}{Mean Error of the Mean}                        \\
           \colhead{Star}                    &
           \colhead{\hms}                    &
           \colhead{\dms}                    &
           \colhead{$V$}                     &
           \colhead{$B-V$}                   &
           \colhead{$U-B$}                   &
           \colhead{$V-R$}                   &
           \colhead{$R-I$}                   &
           \colhead{$V-I$}                   &
           \colhead{$n$}                     &
           \colhead{$m$}                     &
           \colhead{$V$}                     &
           \colhead{$B-V$}                   &
           \colhead{$U-B$}                   &
           \colhead{$V-R$}                   &
           \colhead{$R-I$}                   &
           \colhead{$V-I$}                   }
\startdata
JL~166           & 00 12 43.023 & $-$45 51 15.48 & 15.341 & $-$0.281 & $-$1.122 & $-$0.069 & $+$0.184 & $+$0.116 &  2 &  1 &  0.0021 &  0.0014 &  0.0021 &  0.0021 &  0.0361 &  0.0339 \\
JL~194           & 00 31 41.651 & $-$47 25 20.03 & 12.397 & $-$0.234 & $-$0.947 & $-$0.122 & $-$0.157 & $-$0.279 &  5 &  3 &  0.0040 &  0.0018 &  0.0027 &  0.0022 &  0.0054 &  0.0076 \\
JL~202           & 00 40 13.303 & $-$55 01 52.20 & 13.087 & $-$0.296 & $-$1.195 & $-$0.138 & $-$0.184 & $-$0.321 &  5 &  3 &  0.0022 &  0.0009 &  0.0031 &  0.0027 &  0.0089 &  0.0098 \\
GD~679           & 01 06 51.07  & $-$33 20 31.3  & 13.537 & $-$0.278 & $-$1.083 & $-$0.137 & $-$0.156 & $-$0.289 &  6 &  3 &  0.0006 &  0.0004 &  0.0027 &  0.0007 &  0.0099 &  0.0117 \\
JL~236           & 01 14 06.692 & $-$52 44 02.19 & 13.376 & $-$0.242 & $-$1.005 & $-$0.129 & $-$0.132 & $-$0.264 &  1 &  1 & \nodata & \nodata & \nodata & \nodata & \nodata & \nodata \\
JL~261           & 01 47 17.465 & $-$51 33 39.18 & 13.589 & $-$0.299 & $-$1.191 & $-$0.123 & $-$0.206 & $-$0.329 &  1 &  1 & \nodata & \nodata & \nodata & \nodata & \nodata & \nodata \\
LB~3241          & 02 13 11.934 & $-$49 44 53.76 & 12.763 & $-$0.304 & $-$1.166 & $-$0.131 & $-$0.179 & $-$0.309 &  2 &  2 &  0.0014 &  0.0007 &  0.0078 &  0.0070 &  0.0127 &  0.0127 \\
JL~286           & 02 13 12.410 & $-$50 04 40.02 & 14.291 & $-$0.207 & $-$0.982 & $-$0.103 & $-$0.150 & $-$0.253 &  1 &  1 & \nodata & \nodata & \nodata & \nodata & \nodata & \nodata \\
L745-46A         & 07 40 20.794 & $-$17 24 49.20 & 13.034 & $+$0.248 & $-$0.652 & $+$0.165 & $+$0.161 & $+$0.326 &  5 &  3 &  0.0022 &  0.0022 &  0.0036 &  0.0013 &  0.0067 &  0.0063 \\
LSS~1274         & 09 18 56.013 & $-$57 04 25.42 & 12.959 & $-$0.212 & $-$1.187 & $-$0.070 & $-$0.092 & $-$0.166 &  2 &  1 &  0.0021 &  0.0007 &  0.0007 &  0.0014 &  0.0127 &  0.0113 \\
LSS~1275         & 09 20 10.130 & $-$45 31 54.96 & 11.417 & $-$0.322 & $-$1.263 & $-$0.135 & $-$0.170 & $-$0.306 &  2 &  1 &  0.0007 &  0.0007 &  0.0007 &  0.0001 &  0.0028 &  0.0028 \\
LSS~1349         & 09 46 57.109 & $-$50 12 16.05 & 13.379 & $+$0.028 & $-$0.986 & $+$0.062 & $+$0.079 & $+$0.140 &  2 &  1 &  0.0001 &  0.0014 &  0.0049 &  0.0028 &  0.0007 &  0.0042 \\
LSS~1362         & 09 52 44.524 & $-$46 16 47.47 & 12.533 & $-$0.222 & $-$1.189 & $-$0.064 & $-$0.075 & $-$0.141 &  2 &  1 &  0.0001 &  0.0014 &  0.0021 &  0.0007 &  0.0014 &  0.0007 \\
LSE~153          & 13 53 08.214 & $-$46 43 42.32 & 11.379 & $-$0.275 & $-$1.218 & $-$0.107 & $-$0.146 & $-$0.251 &  4 &  2 &  0.0100 &  0.0015 &  0.0030 &  0.0015 &  0.0020 &  0.0030 \\
LSE~125          & 15 43 05.042 & $-$39 18 14.59 & 12.372 & $-$0.153 & $-$1.119 & $-$0.027 & $-$0.049 & $-$0.074 &  4 &  2 &  0.0075 &  0.0040 &  0.0045 &  0.0030 &  0.0015 &  0.0040 \\
LSE~234          & 18 13 15.853 & $-$64 55 16.86 & 12.705 & $-$0.280 & $-$1.231 & $-$0.086 & $-$0.118 & $-$0.205 &  4 &  2 &  0.0100 &  0.0030 &  0.0095 &  0.0085 &  0.0070 &  0.0125 \\
LSE~263          & 19 02 11.740 & $-$51 30 09.54 & 11.771 & $-$0.278 & $-$1.245 & $-$0.091 & $-$0.122 & $-$0.211 &  2 &  1 &  0.0007 &  0.0007 &  0.0021 &  0.0001 &  0.0035 &  0.0035 \\
JL~25            & 19 39 38.960 & $-$76 01 17.25 & 13.321 & $-$0.188 & $-$1.129 & $-$0.017 & $+$0.040 & $+$0.018 &  4 &  2 &  0.0010 &  0.0015 &  0.0020 &  0.0035 &  0.0115 &  0.0085 \\
LB~1516          & 23 01 56.152 & $-$48 03 48.46 & 12.967 & $-$0.241 & $-$0.981 & $-$0.117 & $-$0.128 & $-$0.244 &  4 &  2 &  0.0035 &  0.0020 &  0.0020 &  0.0015 &  0.0055 &  0.0080 \\
\enddata
\label{tab:table4}
\end{deluxetable}

\newpage
\begin{deluxetable}{lccccrrccc}
\tablecaption{Accurate Coordinates and Proper Motions for the Stars in Tables \ref{tab:table1} and \ref{tab:table4}.}
\tabletypesize{\tiny}
\tablewidth{0pt}
\tablehead{\colhead{Star Name}                                &
           \multicolumn{2}{c}{$\alpha$~~~~(2000)~~~~$\delta$} &
           \colhead{UCAC2}                                    &
           \colhead{2MASS-PSC}                                &
           \colhead{$\mu_{\alpha}$}                           &
           \colhead{$\mu_{\delta}$}                           &
           \colhead{$\mu$ Ref.}                               &
           \colhead{Sp. Type}                                 &
           \colhead{Sp. Type Ref.}                            \\
           \colhead{ }                                        &
           \colhead{\hms}                                     &
           \colhead{\dms}                                     &
           \colhead{ }                                        &
           \colhead{ }                                        &
           \colhead{(mas/yr)}                                 &
           \colhead{(mas/yr)}                                 &
           \colhead{}                                         &
           \colhead{}                                         &
           \colhead{}                                         }
\tablenotetext{~}{1.  \citep{Monet2003}[USNO-B1.0 catalog]}
\tablenotetext{~}{2.  \citep{Zacharias2004}[UCAC2]}
\tablenotetext{~}{3.  \citep{Kilkenny1987}}
\tablenotetext{~}{4.  \citep{KilkennyMuller1989}}
\tablenotetext{~}{5.  \citep{Kilkenny1988}}
\tablenotetext{~}{6.  \citep{GreensteinSargent1974}}
\tablenotetext{~}{7.  \citep{KilkennyHill1975}}
\tablenotetext{~}{8.  \citep{Drilling1983}}
\tablenotetext{~}{9.  \citep{Platais1998}}
\tablenotetext{~}{10.  \citep{Platais2007}}
\startdata
JL~163B         & 00 10 24.88  & $-$50 13 55.6  &  \nodata  &   J00102488-5013556 &   $-$4   &  $-$20   &    1    & \nodata & \nodata \\ 
JL~163A         & 00 10 26.309 & $-$50 15 03.52 &  10092905 &   J00102631-5015034 &     13.3 &      9.1 &    2    & \nodata & \nodata \\ 
JL~163          & 00 10 33.221 & $-$50 15 24.37 &  10092907 &   J00103322-5015243 &      7.2 &      7.3 &    2    &     sdB &     3,4 \\ 
JL~163C         & 00 10 38.238 & $-$50 15 26.39 &  10092908 &   J00103824-5015263 &     11.4 &   $-$0.6 &    2    & \nodata & \nodata \\ 
JL~163D         & 00 10 41.62  & $-$50 13 45.6  &  \nodata  &   J00104162-5013456 &  \nodata &  \nodata & \nodata & \nodata & \nodata \\ 
JL~163E         & 00 10 58.017 & $-$50 14 17.44 &  10092916 &   J00105801-5014175 &      9.0 &   $-$0.6 &    2    & \nodata & \nodata \\ 
JL~163F         & 00 11 00.225 & $-$50 12 53.81 &  10092918 &   J00110022-5012539 &      7.7 &   $-$3.5 &    2    & \nodata & \nodata \\ 
JL~166          & 00 12 43.023 & $-$45 51 15.48 &  12134270 &   J00124302-4551153 &      2.0 &  $-$15.8 &    2    &    sdOB &       5 \\ 
TPhe~I          & 00 30 04.593 & $-$46 28 10.17 &  11909220 &   J00300459-4628102 &     13.6 &   $-$8.5 &    2    & \nodata & \nodata \\ 
TPhe~A          & 00 30 09.594 & $-$46 31 28.91 &  11679110 &   J00300959-4631289 &      8.3 &   $-$2.0 &    2    & \nodata & \nodata \\ 
TPhe~H          & 00 30 09.683 & $-$46 27 24.30 &  11909222 &   J00300968-4627243 &     11.2 &   $-$8.1 &    2    & \nodata & \nodata \\ 
TPhe~B          & 00 30 16.313 & $-$46 27 58.57 &  11909226 &   J00301631-4627586 &      1.0 &   $-$5.4 &    2    & \nodata & \nodata \\ 
TPhe~C          & 00 30 16.98  & $-$46 32 21.4  &  \nodata  &   J00301697-4632214 &  \nodata &  \nodata & \nodata & \nodata & \nodata \\ 
TPhe~D          & 00 30 18.342 & $-$46 31 19.85 &  11679116 &   J00301834-4631198 &      2.3 &      1.3 &    2    & \nodata & \nodata \\ 
TPhe~E          & 00 30 19.768 & $-$46 24 35.60 &  11909227 &   \nodata           &     51.4 &      1.6 &    2    & \nodata & \nodata \\ 
TPhe~J          & 00 30 23.02  & $-$46 23 51.6  &  \nodata  &   J00302301-4623516 &  \nodata &  \nodata & \nodata & \nodata & \nodata \\ 
TPhe~F          & 00 30 49.820 & $-$46 33 24.07 &  11679132 &   J00304980-4633239 &     80.6 &  $-$10.2 &    2    & \nodata & \nodata \\ 
TPhe~K          & 00 30 56.315 & $-$46 23 26.04 &  11909244 &   J00305632-4623260 &  $-$11.2 &   $-$1.6 &    2    & \nodata & \nodata \\ 
TPhe~G          & 00 31 04.303 & $-$46 22 51.35 &  11909246 &   J00310430-4622513 &     14.3 &      5.6 &    2    & \nodata & \nodata \\ 
JL~194          & 00 31 41.651 & $-$47 25 20.03 &  11454707 &   J00314165-4725200 &     21.2 &  $-$25.8 &    2    &     sdB &       5 \\ 
JL~202          & 00 40 13.303 & $-$55 01 52.20 &  07669947 &   J00401329-5501520 &     18.9 &      3.6 &    2    &     sdO &       5 \\ 
GD~679          & 01 06 51.07  & $-$33 20 31.3  &  \nodata  &   J01065106-3320313 &  \nodata &  \nodata & \nodata &     sdB &     6,5 \\ 
JL~236          & 01 14 06.692 & $-$52 44 02.19 &  08932516 &   J01140669-5244021 &     13.3 &   $-$7.7 &    2    &     sdB &       5 \\ 
JL~261          & 01 47 17.465 & $-$51 33 39.18 &  09398888 &   J01471747-5133391 &     16.2 &   $-$4.4 &    2    &  sdO:He &       4 \\ 
LB~3241         & 02 13 11.934 & $-$49 44 53.76 &  10323959 &   J02131193-4944537 &      1.1 &   $-$6.7 &    2    &    sdOB &       4 \\ 
JL~286          & 02 13 12.410 & $-$50 04 40.02 &  10095692 &   J02131241-5004399 &   $-$6.0 &      1.4 &    2    &     sdB &       4 \\ 
MCT~0401-4017E  & 04 02 28.087 & $-$40 09 35.57 &  14539866 &   J04022808-4009355 &     13.0 &     40.7 &    2    & \nodata & \nodata \\ 
MCT~0401-4017F  & 04 02 29.559 & $-$40 10 21.19 &  14539867 &   J04022955-4010211 &     24.3 &   $-$3.9 &    2    & \nodata & \nodata \\ 
MCT~0401-4017D  & 04 02 55.219 & $-$40 15 52.60 &  14539878 &   J04025522-4015525 &      5.9 &   $-$2.5 &    2    & \nodata & \nodata \\ 
MCT~0401-4017   & 04 03 04.540 & $-$40 09 41.05 &  14539887 &   J04030453-4009409 &  $-$15.6 &   $-$9.1 &    2    & \nodata & \nodata \\ 
MCT~0401-4017A  & 04 03 05.195 & $-$40 11 02.88 &  14539888 &   J04030519-4011029 &     19.2 &     19.9 &    2    & \nodata & \nodata \\ 
MCT~0401-4017B  & 04 03 29.846 & $-$40 10 58.46 &  14539898 &   J04032984-4010585 &  $-$11.0 &     12.3 &    2    & \nodata & \nodata \\ 
MCT~0401-4017C  & 04 03 34.273 & $-$40 07 02.53 &  14539899 &   J04033426-4007025 &  $-$12.5 & $-$100.7 &    2    & \nodata & \nodata \\ 
LB~1735         & 04 31 11.090 & $-$53 35 27.06 &  08436787 &   J04311106-5335270 &     11.6 &  $-$20.6 &    2    &    sdB? &   4,5,7 \\ 
LB~1735A        & 04 31 19.438 & $-$53 34 38.22 &  08436790 &   J04311942-5334382 &  $-$14.9 &     16.5 &    2    & \nodata & \nodata \\ 
LB~1735B        & 04 31 22.810 & $-$53 36 36.85 &  08436792 &   J04312279-5336369 &      0.3 &     26.9 &    2    & \nodata & \nodata \\ 
LB~1735F        & 04 31 37.11  & $-$53 36 53.4  &  \nodata  &   J04313710-5336534 &  \nodata &  \nodata & \nodata & \nodata & \nodata \\ 
LB~1735E        & 04 31 42.303 & $-$53 36 18.48 &  08436802 &   J04314228-5336185 &      4.2 &      5.6 &    2    & \nodata & \nodata \\ 
LB~1735G        & 04 31 46.194 & $-$53 39 47.89 &  08436803 &   J04314617-5339477 &      7.4 &     27.4 &    2    & \nodata & \nodata \\ 
LB~1735C        & 04 31 49.652 & $-$53 34 37.52 &  08436806 &   J04314963-5334375 &      3.1 &      8.3 &    2    & \nodata & \nodata \\ 
LB~1735D        & 04 31 58.253 & $-$53 34 52.52 &  08436811 &   J04315824-5334524 &   $-$5.0 &     16.6 &    2    & \nodata & \nodata \\ 
MCT~0436-4616A  & 04 38 20.929 & $-$46 09 27.54 &  11916012 &   J04382091-4609276 &      4.2 &      8.5 &    2    & \nodata & \nodata \\ 
MCT~0436-4616   & 04 38 27.320 & $-$46 10 52.69 &  11916019 &   J04382730-4610527 &      8.9 &      3.2 &    2    & \nodata & \nodata \\ 
MCT~0436-4616B  & 04 38 47.374 & $-$46 10 08.74 &  11916039 &   J04384737-4610087 &     42.4 &     52.6 &    2    & \nodata & \nodata \\ 
MCT~0550-4911E  & 05 51 43.660 & $-$49 10 48.59 &  10562675 &   J05514366-4910485 &     18.0 &     21.7 &    2    & \nodata & \nodata \\ 
MCT~0550-4911D  & 05 51 49.207 & $-$49 11 18.95 &  10562687 &   J05514921-4911188 &      8.8 &     10.4 &    2    & \nodata & \nodata \\ 
MCT~0550-4911B  & 05 51 59.105 & $-$49 11 25.06 &  10562695 &   J05515910-4911250 &     11.7 &      2.0 &    2    & \nodata & \nodata \\ 
MCT~0550-4911   & 05 52 02.712 & $-$49 11 22.38 &  10562698 &   J05520271-4911223 &   $-$2.9 &     11.7 &    2    & \nodata & \nodata \\ 
MCT~0550-4911C  & 05 52 03.98  & $-$49 09 37.6  &  \nodata  &   J05520398-4909376 &  \nodata &  \nodata & \nodata & \nodata & \nodata \\ 
MCT~0550-4911A  & 05 52 04.104 & $-$49 10 48.55 &  10562700 &   J05520411-4910485 &      0.0 &      4.0 &    2    & \nodata & \nodata \\ 
L745-46A        & 07 40 20.794 & $-$17 24 49.20 &  25277198 &   J07402064-1724481 &   1148.7 & $-$540.6 &    2    & \nodata & \nodata \\ 
LSS~982G        & 08 10 19.984 & $-$40 31 57.19 &  14358563 &   J08101998-4031571 &      1.1 &      9.4 &    2    & \nodata & \nodata \\ 
LSS~982F        & 08 10 21.386 & $-$40 31 30.31 &  14358565 &   J08102139-4031302 &   $-$7.0 &   $-$0.6 &    2    & \nodata & \nodata \\ 
LSS~982E        & 08 10 29.667 & $-$40 30 06.85 &  14358602 &   J08102966-4030068 &   $-$6.5 &      5.1 &    2    & \nodata & \nodata \\ 
LSS~982         & 08 10 31.719 & $-$40 32 47.07 &  14358605 &   J08103171-4032469 &     26.9 &  $-$39.9 &    2    &     sdO &       5 \\ 
LSS~982A        & 08 10 32.719 & $-$40 31 14.52 &  14358607 &   J08103271-4031145 &   $-$8.4 &     17.1 &    2    & \nodata & \nodata \\ 
LSS~982B        & 08 10 34.325 & $-$40 30 49.16 &  14358613 &   J08103432-4030491 &   $-$4.9 &      5.3 &    2    & \nodata & \nodata \\ 
LSS~982D        & 08 10 38.089 & $-$40 28 20.84 &  14569244 &   J08103808-4028209 &   $-$1.1 &      6.0 &    2    & \nodata & \nodata \\ 
LSS~982C        & 08 10 38.222 & $-$40 30 09.95 &  14358632 &   J08103822-4030100 &   $-$8.7 &      1.3 &    2    & \nodata & \nodata \\ 
WD0830-535J     & 08 31 41.877 & $-$53 37 28.27 &  08458552 &   J08314187-5337281 &   $-$3.5 &     11.0 &    2    & \nodata & \nodata \\ 
WD0830-535K     & 08 31 42.960 & $-$53 39 19.33 &  08458557 &   J08314296-5339192 &   $-$9.4 &      4.3 &    2    & \nodata & \nodata \\ 
WD0830-535I     & 08 31 45.769 & $-$53 37 35.34 &  08458569 &   J08314576-5337352 &      1.1 &      4.3 &    2    & \nodata & \nodata \\ 
WD0830-535C     & 08 31 50.421 & $-$53 43 51.16 &  08458589 &   J08315042-5343511 &   $-$9.1 &      5.1 &    2    & \nodata & \nodata \\ 
WD0830-535      & 08 31 51.91  & $-$53 40 32.5  &  \nodata  &   J08315191-5340324 &  $+$33.1 & $-$133.8 &   10    & \nodata & \nodata \\ 
WD0830-535B     & 08 31 54.356 & $-$53 42 37.91 &  08458600 &   J08315435-5342379 &  $-$11.2 &      6.6 &    2    & \nodata & \nodata \\ 
WD0830-535A     & 08 31 54.454 & $-$53 40 48.08 &  08458601 &   J08315445-5340481 &      1.1 &      8.5 &    2    & \nodata & \nodata \\ 
WD0830-535H     & 08 32 03.822 & $-$53 37 50.36 &  08458638 &   J08320382-5337504 &      1.4 &     11.0 &    2    & \nodata & \nodata \\ 
WD0830-535D     & 08 32 04.970 & $-$53 44 14.80 &  08458647 &   J08320497-5344147 &   $-$5.3 &   $-$3.0 &    2    & \nodata & \nodata \\ 
WD0830-535G     & 08 32 08.190 & $-$53 40 30.15 &  08458656 &   J08320818-5340301 &   $-$8.7 &      5.9 &    2    & \nodata & \nodata \\ 
WD0830-535F     & 08 32 08.639 & $-$53 40 44.59 &  08458658 &   J08320863-5340446 &   $-$0.2 &     16.7 &    2    & \nodata & \nodata \\ 
WD0830-535E     & 08 32 11.137 & $-$53 41 58.54 &  08458669 &   J08321113-5341585 &   $-$5.9 &      0.0 &    2    & \nodata & \nodata \\ 
LSS~1274        & 09 18 56.013 & $-$57 04 25.42 &  06750679 &   J09185601-5704254 &      2.6 &   $-$8.9 &    2    &     sdO &       5 \\ 
LSS~1275        & 09 20 10.130 & $-$45 31 54.96 &  12183510 &   J09201013-4531549 &  $-$19.5 &   $-$1.1 &    2    &     sdO &       5 \\ 
LSS~1349        & 09 46 57.109 & $-$50 12 16.05 &  10140464 &   J09465710-5012160 &  $-$10.9 &      8.7 &    2    &     sdO &       5 \\ 
LSS~1362        & 09 52 44.524 & $-$46 16 47.47 &  11964358 &   J09524453-4616474 &   $-$5.3 &     16.4 &    2    &     sdO &       8 \\ 
WD1056-384D     & 10 58 08.051 & $-$38 46 22.33 &  15265422 &   J10580804-3846222 &  $-$14.5 &  $-$11.1 &    2    & \nodata & \nodata \\ 
WD1056-384C     & 10 58 11.053 & $-$38 41 50.97 &  15265430 &   J10581105-3841509 &  $-$10.2 &      1.7 &    2    & \nodata & \nodata \\ 
WD1056-384B     & 10 58 18.691 & $-$38 41 56.56 &  15265449 &   J10581868-3841565 &   $-$7.9 &   $-$1.0 &    2    & \nodata & \nodata \\ 
WD1056-384      & 10 58 20.11  & $-$38 44 25.1  &  \nodata  &   J10582010-3844251 & $-$158.0 &     32.4 &    9    & \nodata & \nodata \\ 
WD1056-384A     & 10 58 23.720 & $-$38 44 01.19 &  15265461 &   J10582371-3844011 &  $-$16.1 &     14.5 &    2    & \nodata & \nodata \\ 
WD1153-484G     & 11 55 53.78  & $-$48 42 01.3  &  \nodata  &   J11555377-4842013 &  \nodata &  \nodata & \nodata & \nodata & \nodata \\ 
WD1153-484H     & 11 55 53.886 & $-$48 41 24.93 &  10852585 &   J11555388-4841248 &  $-$32.2 &     10.5 &    2    & \nodata & \nodata \\ 
WD1153-484F     & 11 55 56.276 & $-$48 42 03.51 &  10852592 &   J11555628-4842034 &      4.5 &   $-$2.9 &    2    & \nodata & \nodata \\ 
WD1153-484D     & 11 55 57.027 & $-$48 37 23.50 &  10852595 &   J11555702-4837234 &  $-$62.7 &   $-$8.3 &    2    & \nodata & \nodata \\ 
WD1153-484C     & 11 56 01.765 & $-$48 38 32.38 &  10852612 &   J11560176-4838322 &  $-$10.8 &      3.4 &    2    & \nodata & \nodata \\ 
WD1153-484B     & 11 56 05.105 & $-$48 38 50.36 &  10852620 &   J11560511-4838502 &   $-$7.5 &      3.5 &    2    & \nodata & \nodata \\ 
WD1153-484A     & 11 56 10.834 & $-$48 39 48.51 &  10852634 &   J11561083-4839484 &  $-$11.8 &   $-$7.5 &    2    & \nodata & \nodata \\ 
WD1153-484      & 11 56 11.427 & $-$48 40 03.30 &  10852635 &   J11561142-4840031 &  $-$57.9 &     14.2 &    2    & \nodata & \nodata \\ 
WD1153-484E     & 11 56 13.505 & $-$48 42 16.00 &  10852642 &   J11561350-4842158 &   $-$4.9 &      3.2 &    2    & \nodata & \nodata \\ 
LSE~44F         & 13 52 34.466 & $-$48 06 59.26 &  11101888 &   J13523446-4806592 &   $-$9.3 &   $-$6.5 &    2    & \nodata & \nodata \\ 
LSE~44A         & 13 52 39.573 & $-$48 10 04.92 &  11101903 &   J13523957-4810049 &   $-$4.4 &   $-$2.9 &    2    & \nodata & \nodata \\ 
LSE~44          & 13 52 40.779 & $-$48 08 22.75 &  11101909 &   J13524078-4808226 &  $-$27.3 &   $-$6.9 &    2    &     sdO &       8 \\ 
LSE~44B         & 13 52 43.784 & $-$48 11 34.04 &  11101921 &   J13524378-4811340 &   $-$9.7 &      1.8 &    2    & \nodata & \nodata \\ 
LSE~44C         & 13 52 51.850 & $-$48 09 56.32 &  11101947 &   J13525185-4809563 &   $-$9.6 &   $-$0.2 &    2    & \nodata & \nodata \\ 
LSE~44D         & 13 52 52.115 & $-$48 08 38.43 &  11101949 &   J13525211-4808384 &   $-$8.9 &   $-$0.9 &    2    & \nodata & \nodata \\ 
LSE~44E         & 13 52 55.514 & $-$48 08 19.58 &  11101960 &   J13525550-4808195 &   $-$0.3 &   $-$1.8 &    2    & \nodata & \nodata \\ 
LSE~153         & 13 53 08.214 & $-$46 43 42.32 &  11776514 &   J13530821-4643422 &  $-$24.4 &  $-$20.6 &    2    &     sdO &       8 \\ 
LSE~125         & 15 43 05.042 & $-$39 18 14.59 &  15080788 &   J15430504-3918145 &  $-$12.1 &  $-$14.2 &    2    & \nodata & \nodata \\ 
LSE~259C        & 16 53 30.037 & $-$56 00 54.62 &  07384133 &   J16533002-5600544 &      1.6 &   $-$5.9 &    2    & \nodata & \nodata \\ 
LSE~259D        & 16 53 41.607 & $-$55 59 08.96 &  07623191 &   J16534160-5559089 &   $-$6.3 &   $-$0.6 &    2    & \nodata & \nodata \\ 
LSE~259E        & 16 53 43.240 & $-$55 58 30.06 &  07623206 &   J16534323-5558300 &   $-$8.2 &      0.5 &    2    & \nodata & \nodata \\ 
LSE~259B        & 16 53 44.377 & $-$56 01 32.85 &  07384228 &   J16534436-5601328 &      5.3 &      3.7 &    2    & \nodata & \nodata \\ 
LSE~259H        & 16 53 48.007 & $-$56 00 42.24 &  07384247 &   J16534807-5600422 &   $-$8.7 &   $-$2.2 &    2    & \nodata & \nodata \\ 
LSE~259A        & 16 53 53.147 & $-$56 02 02.58 &  07384280 &   J16535314-5602025 &      0.8 &   $-$6.8 &    2    & \nodata & \nodata \\ 
LSE~259         & 16 53 54.573 & $-$56 01 54.76 &  07384295 &   J16535458-5601547 &      0.3 &  $-$14.4 &    2    &     sdO &       8 \\ 
LSE~259F        & 16 53 55.538 & $-$55 59 26.16 &  07623288 &   J16535554-5559261 &  $-$16.3 &   $-$6.4 &    2    & \nodata & \nodata \\ 
LSE~259G        & 16 54 00.237 & $-$55 59 20.05 &  07623325 &   J16540027-5559199 &   $-$3.4 &   $-$2.4 &    2    & \nodata & \nodata \\ 
LSE~234         & 18 13 15.853 & $-$64 55 16.86 &  03401203 &   J18131583-6455168 &   $-$5.0 &  $-$29.2 &    2    &     sdO &       8 \\ 
LSE~263         & 19 02 11.740 & $-$51 30 09.54 &  09610773 &   J19021174-5130094 &  $-$11.8 &      0.0 &    2    &     sdO &       8 \\ 
JL~25           & 19 39 38.960 & $-$76 01 17.25 &  00667985 &   J19393895-7601172 &      2.5 &  $-$22.3 &    2    &    sdOB &       4 \\ 
MCT~2019-4339E  & 20 22 38.910 & $-$43 31 17.07 &  13255341 &   J20223890-4331171 &      9.7 &   $-$7.3 &    2    & \nodata & \nodata \\ 
MCT~2019-4339D  & 20 22 40.476 & $-$43 27 26.39 &  13472188 &   J20224047-4327264 &     30.6 &  $-$39.1 &    2    & \nodata & \nodata \\ 
MCT~2019-4339A  & 20 22 45.332 & $-$43 29 43.33 &  13472194 &   J20224532-4329433 &      3.9 &  $-$23.4 &    2    & \nodata & \nodata \\ 
MCT~2019-4339B  & 20 22 46.726 & $-$43 28 10.88 &  13472198 &   J20224672-4328108 &      5.2 &   $-$4.1 &    2    & \nodata & \nodata \\ 
MCT~2019-4339   & 20 22 49.056 & $-$43 30 11.53 &  13255356 &   J20224905-4330116 &   $-$3.7 &  $-$12.3 &    2    & \nodata & \nodata \\ 
MCT~2019-4339C  & 20 23 02.134 & $-$43 28 22.40 &  13472215 &   J20230213-4328224 &     47.8 &  $-$11.4 &    2    & \nodata & \nodata \\ 
MCT~2019-4339F  & 20 23 03.99  & $-$43 31 21.9  &  \nodata  &   J20230398-4331219 &  \nodata &  \nodata & \nodata & \nodata & \nodata \\ 
JL~82C          & 21 35 45.005 & $-$72 50 12.76 &  01156348 &   J21354498-7250127 &   $-$3.5 &      9.1 &    2    & \nodata & \nodata \\ 
JL~82B          & 21 35 59.34  & $-$72 50 15.1  &  \nodata  &   J21355933-7250150 &  \nodata &  \nodata & \nodata & \nodata & \nodata \\ 
JL~82           & 21 36 01.289 & $-$72 48 27.21 &  01156353 &   J21360128-7248272 &     13.1 &  $-$15.7 &    2    &     sdB &       4 \\ 
JL~82D          & 21 36 15.362 & $-$72 45 27.21 &  01156358 &   J21361534-7245272 &   $-$2.6 &      3.5 &    2    & \nodata & \nodata \\ 
JL~82A          & 21 36 17.052 & $-$72 50 08.57 &  01156361 &   J21361703-7250084 &      2.4 &      4.0 &    2    & \nodata & \nodata \\ 
JL~117A         & 22 54 13.332 & $-$72 23 04.33 &  01245946 &   J22541335-7223043 &     15.2 &  $-$10.4 &    2    & \nodata & \nodata \\ 
JL~117          & 22 54 38.002 & $-$72 23 09.68 &  01245961 &   J22543801-7223097 &      0.8 &     10.7 &    2    &     sdO &       5 \\ 
JL~117B         & 22 54 51.021 & $-$72 23 23.61 &  01245972 &   J22545102-7223236 &     16.4 &  $-$22.2 &    2    & \nodata & \nodata \\ 
JL~117C         & 22 55 00.471 & $-$72 22 41.35 &  01245977 &   J22550047-7222413 &      7.7 &   $-$7.2 &    2    & \nodata & \nodata \\ 
JL~117D         & 22 55 37.574 & $-$72 22 34.28 &  01245986 &   J22553759-7222343 &     22.5 &   $-$2.1 &    2    & \nodata & \nodata \\ 
LB~1516         & 23 01 56.152 & $-$48 03 48.46 &  11226102 &   J23015616-4803484 &      3.2 &      3.5 &    2    &     sdB &       5 \\ 
\enddata
\label{tab:table5}
\end{deluxetable}

\clearpage 

\clearpage 

\clearpage 

\clearpage 

\clearpage 

\clearpage 

\clearpage

\clearpage

\clearpage

\clearpage

\clearpage

\clearpage

\clearpage

\clearpage

\clearpage

\clearpage

\clearpage

\clearpage

\clearpage

\clearpage

\clearpage

\clearpage

\clearpage

\clearpage

\clearpage

\clearpage

\clearpage

\clearpage

\clearpage

\clearpage

\clearpage

\clearpage

\clearpage

\clearpage

\clearpage

\clearpage

\clearpage

\clearpage

\clearpage

\clearpage

\clearpage

\clearpage

\clearpage

\clearpage

\clearpage

\clearpage

\clearpage

\clearpage

\clearpage

\clearpage

\clearpage

\clearpage

\clearpage

\clearpage

\clearpage

\clearpage

\clearpage 

\end{document}